\documentclass[nolinenum]{jfp}

\usepackage{hyperref,wrapfig, amsthm, pict2e, booktabs, subcaption,
  subfiles, rotating, pifont, indentfirst, amsmath,
  amsfonts, mathrsfs, adjustbox, mathtools,booktabs, microtype,
  ebproof, enumitem, multirow, multicol, setspace, xcolor, lineno,
  listings, stmaryrd, xspace, stackengine}

\usepackage{empheq}

\usepackage[tableposition=top]{caption}
\usepackage{tikz-cd}
\usetikzlibrary{decorations.pathmorphing}
\usepackage[ruled,linesnumbered,vlined]{algorithm2e}

\SetCommentSty{mycommfont}
\usepackage{graphicx}
\usepackage{minted}
\usepackage{tikz}
\newcounter{markeq}
\usepackage{spacingtricks}

\usepackage[T1]{fontenc}

\usepackage{mdframed}

\usepackage[xcolor]{changebar}
\cbcolor{blue}

\usepackage{spacingtricks, commands, refinementtydef}

\theoremstyle{definition}
\newtheorem{theorem}{Theorem}[section]

\newtheorem{lemma}[theorem]{Lemma}

\newtheorem{example}[theorem]{Example}
\newtheorem{definition}[theorem]{Definition}

\theoremstyle{definition}

\renewcommand{\paragraph}[1]{\textbf{#1}.\ }
\renewcommand{\cite}[1]{\citep{#1}}

\begin{document}

\journaltitle{JFP}
\cpr{Cambridge University Press}
\doival{10.1017/xxxxx}

\lefttitle{Polymorphic Coverage Types}
\righttitle{Polymorphic Coverage Types}

\totalpg{\pageref{lastpage01}}
\jnlDoiYr{2025}

\title{Polymorphic Coverage Types\footnote{Portions of this work
    previously appeared in the proceedings of the 2023 ACM Conference
    on Programming Language Design and Implementation.}}


\begin{authgrp}
\author{Zhe Zhou}
\affiliation{Department of Computer Science, Purdue University, USA\\
(\email{zhou956@purdue.edu})}

\author{Ashish Mishra}
\affiliation{Department of Computer Science, Indian Institute of Technology Hyderabad, India\\
(\email{mishraashish@cse.iith.ac.in})}

\author{Benjamin Delaware}
\affiliation{Department of Computer Science, Purdue University, USA\\
(\email{bendy@purdue.edu})}

\author{Suresh Jagannathan}
\affiliation{Department of Computer Science, Purdue University, USA\\
(\email{suresh@cs.purdue.edu})}
\end{authgrp}


\begin{abstract}
  Test input generators are an important part of property-based
  testing (PBT) frameworks. Because PBT is intended to test deep
  semantic and structural properties of a program, the outputs
  produced by these generators can be complex data structures,
  constrained to satisfy properties the developer believes is most
  relevant to testing the function of interest.  An important feature
  expected of these generators is that they be capable of producing
  \emph{all} acceptable elements that satisfy the function's input
  type and generator-provided constraints.  However, it is not readily
  apparent how we might validate whether a particular generator's
  output satisfies this \emph{coverage} requirement.  Typically,
  developers must rely on manual inspection and post-mortem analysis
  of test runs to determine if the generator is providing sufficient
  coverage; these approaches are error-prone and difficult to scale as
  generators become more complex.  To address this important concern,
  we present a new refinement type-based verification procedure for
  validating the coverage provided by input test generators, based on
  a novel interpretation of types that embeds ``\emph{must}-style''
  underapproximate reasoning principles as a fundamental part of the
  type system.  The types associated with expressions now capture the
  set of values \emph{guaranteed} to be produced by the expression,
  rather than the typical formulation that uses types to represent the
  set of values an expression \emph{may} produce. We have formalized
  the notion of \emph{coverage types} in a rich core language that
  supports higher-order functions and inductive datatypes.
  \cbnewadding{To better support checking coverage properties of
    real-world test generators, we extend this type system with type
    and qualifier polymorphism. These extensions enable our system to
    statically verify coverage guarantees of test input generators
    constructed using the sort of monadic combinators found in most
    PBT frameworks. We have implemented a coverage type checker for
    OCaml programs based on this core calculus, and present a detailed
    evaluation of the utility of our ideas using a corpus of
    benchmarks drawn from both the PBT literature and open source
    projects}.
\end{abstract}


\maketitle

\section{Introduction}
\label{sec:intro}

Property-based testing (PBT) is a popular technique for automatically
testing deep semantic and structural properties of
programs. Originally pioneered by the QuickCheck~\cite{quickCheck}
library for Haskell, PBT frameworks now exist for many programming
languages, including QCheck~\cite{QCheck}, JavaScript~\cite{fastCheck},
Rust~\cite{rustCheck}, Python~\cite{hypothesis},
Scala~\cite{scalaCheck}, and Rocq/Coq~\cite{quickchick}. The PBT
methodology rests on two key components: \emph{executable properties}
that capture the expected input-output behaviors of the program under
test, and \emph{test input generators} that generate random values of
the input types needed to validate these behaviors. In contrast to
unit tests, which rely on single examples of inputs and outputs,
generators are meant to provide a \emph{family} of inputs against
which programs can be tested, with the goal of ensuring the set of
generated tests provide good coverage of all possible inputs. In order to
prune out irrelevant inputs, PBT frameworks allow users to define
custom generators that reflect the specific shape of data that the
developer believes is most likely to trigger interesting (aka faulty)
behavior. As one simple example, to test a tree compression or balancing
function, the developer may want to use a generator that produces
$n$-ary trees with randomly chosen height and arity but whose leaves
are ordered according to a user-provided ordering relation.

%
Given the critical role they play in the assurance case provided by
PBT frameworks, it is reasonable to ask what constitutes a ``good''
specification for a test generator. For our example, one answer could
be that it should only produce ordered trees. Of course, this is not a
very satisfactory characterization of the behavior we desire: the
``constant'' generator that always produces trees of height one
trivially meets this specification, but it is unlikely to produce
useful tests for a compression function!  Ideally, we would like a
generator to intelligently enumerate the space of \emph{all} possible
ordered trees, thereby helping to maximize the likelihood of finding
bugs in the function under test.  Because defining such an enumeration
procedure for arbitrary datatypes can be hard, even when
complete enumeration is computationally feasible, PBT frameworks instead
give developers the ability to assemble generators for complex data
structures \emph{compositionally}, building on generators for simpler
types where randomly sampling elements of the type is straightforward
and sufficient.
For example, we could implement an ordered tree
generator in terms of a primitive random number generator that is used
to non-deterministically select the height, arity, and elements of a
candidate tree, checking (or enforcing) the orderness of the tree
before returning it as a feasible test input.
Although the
random number generator might provide a guarantee that its underlying
probability density function (PDF) is always non-zero on all elements
in its sample space, determining that a tree generator that is built
using it can actually enumerate all the ordered trees desired is a
substantially harder problem.  Even if we know the generator is
capable of eventually yielding all trees, constraints imposed by the
function's precondition might require the generator to perform further
filtering or transformations over generated trees.  However, proving
that any filtering operations the generator uses do not mistakenly
prune out valid ordered trees or that any transformations the
generator performs over candidate trees preserve the elements of the
random tree being transformed, pose additional challenges.  In other
words, verifying that the generator is \emph{complete} with respect to
our desired orderness property entails reasoning that is independent
of the behavior of the primitive generators used to build the tree.
Consequently, we require some alternative mechanism to help qualify
the part of the target function's input space the generator is
actually guaranteed to cover.  Devising such a mechanism is
challenging precisely because the properties that need to be tested
may impose complex structural and semantic constraints on the
generated output (e.g., requiring that an output tree be a binary
search tree, or that it satisfies a red-black property, etc.); the
complexity of these constraints is directly correlated to the
sparseness of the function's input space preconditions.

\begin{figure}[!ht]
\begin{minted}[xleftmargin=5pt, numbersep=4pt, fontsize = \normalsize, escapeinside=<>, linenos = true]{ocaml}
type 'a tree =
  | Leaf
  | Node of ('a * 'a tree * 'a tree)
let rec bst_gen (lo: int) (hi: int): int tree =
  if lo + 1 >= hi then Leaf else
  (* <\textcolor{red}{Leaf $\oplus$}> *)
  (let (x: int) = int_range (lo + 1, hi - 1) in
      Node (x, bst_gen lo x, bst_gen x hi))
\end{minted}
\caption{ A BST generator.  Failing to uncomment line 6
  results in the generator never producing trees that contain only a
  subset of the elements in the interval between \Code{lo} and
  \Code{hi}, which is inconsistent with the developer's intent.}
\label{fig:gen-ex}
\end{figure}

To illustrate this distinction more concretely, consider the input
test generator shown in Figure~\ref{fig:gen-ex} that is intended to
generate all binary search trees (BSTs) whose elements are between the
interval \Code{lo} and \Code{hi}.  If we ignore the comment on line 6, we
can conclude this generator always produces a non-empty BST whenever
\Code{lo} $<$ \Code{hi}.  While the generator is correct - it always
generates a well-formed BST - it is also incomplete; the call
\Code{bst\_gen 0 10}, for example, will never produce a tree
containing just \Code{Leaf} or a tree with a shape like
\Code{Node(1,Leaf,Leaf)}, even though these instances are valid trees
consistent with the constraints imposed by the generator's argument
bounds.  In fact, this implementation \emph{never} generates a BST
that only contains a proper subset of the elements that reside within
the interval defined by \Code{lo} and \Code{hi}.   By uncommenting line 6, however, we allow the generator to
non-deterministically choose (via operator $\oplus$) to either return
a \Code{Leaf} or left and right BST subtrees based on value returned
by the \Code{int\_range} generator, enabling it to potentially produce
BSTs containing all valid subsets of the provided interval, thus
satisfying our desired desired completeness behavior.  The subtleties involved in reasoning
about such coverage properties is clearly non-trivial.  We reiterate
that recognizing the distinction between these two implementations is
not merely a matter of providing a precise output type capturing the
desired sortedness property of a BST: the incomplete
implementation clearly satisfies such a type!  Furthermore, simply
knowing that the underlying \Code{int\_range} generator used in the
implementation samples all elements within the range of the arguments
it is provided is also insufficient to conclude that the BST generator
can yield \emph{all} possible BSTs within the supplied interval.
Similar observations have led prior work to consider ways to improve a
generator's coverage through mechanisms such as
fuzzing~\cite{quickcheck-coverage-guided, crowbar}, or to
automatically generate complete-by-construction generators for certain
classes of datatypes~\cite{LPP18}.

In contrast to these approaches, this paper embeds the notion of
coverage as an integral part of a test input generator's \emph{type}
specification.  By doing so, a generator's type now specifies the set
of behaviors the generator is \emph{guaranteed} to exhibit; a well-typed
generator is thus guaranteed to produce \emph{every possible} value
satisfying a desired structural property, e.g., that the repaired
(complete) version of \Code{bst\_gen} is capable of producing every
valid BST.  By framing the notion of coverage in type-theoretic terms,
our approach neither requires instrumentation of the target program to
assess the coverage effectiveness of a candidate generator (as in
~\citet{quickcheck-coverage-guided}) nor does it depend on a specific
compilation strategy for producing generators (as in~\citet{LPP18}).
Instead, our approach can automatically verify the coverage properties of an
\emph{arbitrary} test input generator, regardless of whether it was
hand-written or automatically synthesized.

Key to our approach is a novel formulation of a \emph{must}-style
analysis~\cite{Smash+Godefroid,GM21,JT+98} of a test input generator's
behavior in type-theoretic terms. In our proposed type system, we say
an expression $e$ has \emph{coverage type}
$\textcolor{CobaltBlue}{\tau}$ if every value contained in
$\textcolor{CobaltBlue}{\tau}$ \emph{must} be producible by $e$.  Note
how this definition differs from our usual notion of what a type
represents: ordinarily, if $e$ has type $\textcolor{magenta}{\tau}$
then we are allowed to conclude only that any value contained in
$\textcolor{magenta}{\tau}$ \emph{may} be produced by $e$.
Informally, types interpreted in this usual way define an
\emph{overapproximation} of the values an expression $e$ can yield,
without obligating $e$ to produce any specific such value.  In
contrast, coverage types define an \emph{underapproximation} - they
characterize the values an expression $e$ has to produce, potentially
eliding other values that $e$ may also evaluate to. When the set of
elements denoted by a generator's (underapproximate) coverage
type matches that of its (overapproximate) normal type, however, we
can soundly assert that the generator is complete.  As we illustrate
in the remainder of the paper, this characterization allows us to
reason about a program's coverage behavior on the same formal footing
as its safety properties.

In this sense, our solution can be seen a type-theoretic
interpretation of recently proposed Incorrectness Logics
(IL)~\cite{IncorrectnessLogic, RBD+20, LR+22}, in much the same way
that refinement-type systems like Liquid Types~\cite{LiquidHaskell,JV21}
relate to traditional program
logics~\cite{AxiomaticSemantics}. Despite the philosophical
similarities with IL, however, we use underapproximate reasoning for a
very different goal. While IL has been primarily used to precisely
capture the conditions that will lead a program to fault, this work
explores how type-based underapproximate reasoning can be used to
verify the completeness properties of a test generator in the context
of PBT.

This interpretation leads to a fundamental recasting of how types
relate to one another: ordinarily, we are always allowed to assert
that $\textcolor{magenta}{\tau <: \top}$.  This means that any typing
context that admits an expression with type
$\textcolor{magenta}{\tau}$ can also admit that expression at a type
with a logically weaker structure.  In contrast, the subtyping
relation for coverage types inverts this relation, so that
$\textcolor{CobaltBlue}{\top} <: \textcolor{CobaltBlue}{\tau}$.
Intuitively, $\top$ represents the coverage type that obligates an
expression ascribed this type to be capable of producing \emph{all}
elements in $\tau$. But, any context that requires an expression to
produce all such elements can always guarantee that the expression
will also produce a subset of these elements.  In other words, we are
always allowed to weaken an overapproximation (i.e., grow the set of
values an expression may evaluate to), and strengthen an
underapproximation (i.e., shrink the set of values an expression must
evaluate to).  Thus, in our setting, a random number generator over
the integers has coverage type
$\textcolor{CobaltBlue}{\top}$ under the mild
assumption that its underlying PDF provides a non-zero likelihood of
returning every integer.  In contrast, a faulty computation like
\Code{1 div 0} has coverage type $\textcolor{CobaltBlue}{\bot}$ since
there are no guarantees provided by the computation on the value(s) it
must return.  Here, $\bot$ represents a type that defines a degenerate
underapproximation, imposing no constraints on the values an
expression ascribed this type must produce.




\begin{figure}
    \centering
    \begin{minted}[xleftmargin=5pt, numbersep=4pt, linenos = true, fontsize = \small, escapeinside=??]{OCaml}
    val return: 'a -> 'a Gen.t
    val bind: 'a -> ('a -> 'b Gen.t) -> 'b Gen.t
    val frequency: (int -> (int * 'a Gen.t)) -> 'a Gen.t
    val fix: (int -> (int -> 'a -> 'a Gen.t) -> 'a Gen.t) -> int -> 'a Gen.t

    type 'a tree = Leaf of 'a | Node of 'a tree * 'a tree
    let tree_gen : int -> int tree Gen.t = fix
      (fun n self -> match n with
        | 0 -> return err
        | 1 -> bind nat_gen (fun x -> return (Leaf x))
        | n ->
          frequency
            (function
            | 0 -> (1, bind nat_gen (fun x -> return (Leaf x)))
            | _ -> (2, bind (self n/2) (fun tr1 ->
                         bind (self n/2) (fun tr2 ->
                           return (Node (tr1, tr2))))))
      )
    \end{minted}
    \caption{A tree generator expressed using \Code{QCheck}.}
    \label{fig:qcheck-example}
\end{figure}


\cbnewadding{In practice, developers often implement input generators
  using \emph{test generator combinators} using combinators provided
  by the PBT framework. In this setting, generators are defined in
  terms of a monad that facilitates the compositional construction of
  complex generators from simpler ones in a modular way.  The OCaml
  PBT framework \Code{QCheck}~\cite{QCheck}, for example, ascribes the
  type $\tvar{a}\TM$ to such generators, as can be seen in the signature
  for the binary tree generator $\Code{tree\_gen}$ shown
  in \autoref{fig:qcheck-example}. This generator is implemented by $4$
  combinators: in addition to the basic monad operators
  ($\Code{return}$ and $\Code{bind}$), $\Code{tree\_gen}$ also uses
  the $\Code{frequency}$ combinator to sample from multiple generators
  according to the given weights, as well as the $\Code{fix}$
  combinator to recursively construct trees of greater height from
  smaller ones.  Extending coverage types to support monadic
  combinators and type polymorphism (on both types and qualifiers) in
  requires significant extensions over the core monomorphic,
  non-monadic formulation.}



In summary, this article makes the following contributions:

\begin{enumerate}
\item Introduces the notion of \emph{coverage types}, types that characterize the values that an input test generator is guaranteed to
produce.

\item Formalizes the semantics of coverage types in an ML-like
  functional language with support for higher-order functions and
  inductive datatypes.\footnote{A Rocq formalization of this calculus,
    its type system, and its metatheory (i.e.,
    Theorem~\ref{theorem:type-sound}) is provided on
    Zenodo~\cite{artifact}. }

\item Develops a bi-directional type-checking algorithm for
  coverage types in this language.

\item \cbnewadding{Extends this core calculus and typing algorithm to
    support coverage type polymorphism and monadic composition.}

\item Incorporates these ideas in a tool (\name{}) that operates over
  OCaml programs equipped with input generators and typed using
  coverage types.

\item Presents an extensive empirical evaluation justifying the
  utility of coverage types, by using \name{} to verify the coverage
  properties of both \cbnewadding{hand-written generators drawn from
    real-world open source PBT applications} and automatically
  synthesized generators, covering a rich class of datatypes and their
  structural properties.
\end{enumerate}

\noindent The remainder of this article is structured as follows.  In
the next section, we present an informal overview of the key features
of our type system.  Section~\ref{sec:lang} presents the syntax and
semantics for a core call-by-value higher-order functional language
with inductive datatypes that we use to formalize our approach.
Section \ref{sec:rules} describes our coverage type system, and its
metatheory is given in Section \ref{sec:denotation}. A bidirectional
typing algorithm is then given in Section~\ref{sec:algorithm}.  We
describe details about the implementation of \name{} and provide
benchmark results in Section~\ref{sec:evaluation}.  Related work and
conclusions are given in Sections~\ref{sec:related}
and~\ref{sec:conc}.

\section{Overview} 
\label{sec:motivation}

Before presenting the full details of our type system, we begin with
an informal overview of its key features.


\paragraph{Base types} In the following, we write $\urt{b}{\phi}$ to denote the coverage
type that qualifies the base type $b$ using the predicate $\phi$.  As
described in the previous section, an application of the primitive
built-in generator for random numbers:
$\Code{int\_gen : unit\,{\shortrightarrow}\, int}$ has the coverage
type $\Code{int\_gen}~(): \nuut{int}{\top}$.  We use
brackets
$\textcolor{CobaltBlue}{\textbf{[}}...\textcolor{CobaltBlue}{\textbf{]}}$
to emphasize that a coverage type has a different meaning from the
types typically found in other refinement type
systems~\cite{JV21, LiquidHaskell} where a qualified type $b$, written
as $\ort{b}{\phi}$, uses a predicate $\phi$ to constrain the set of
values a program \emph{might} evaluate to. To illustrate this
distinction, consider the combinations of expressions and
types shown in \autoref{tab:examples}.
These examples demonstrate the previous observation that it is
always possible to strengthen the refinement predicate used in an
underapproximate type and weaken such a predicate in an overapproximate
type.
A similar phenomena appears in IL's rule of consequence, which inverts
the direction of the implications on pre- and postconditions in the
overapproximate version of the rule.
As a result, the bottom type $\nuut{int}{\bot}$ is the universal
supertype in our type hierarchy, as it places no restrictions on the
values a term \emph{must} produce. Thus, we sometimes abbreviate
$\nuut{int}{\bot}$ as \Code{int}, since the information provided by
both types is the same. Importantly, the coverage type for the error
term (\Code{err}) can \emph{only} be qualified with $\bot$, since an
erroneous computation is unconstrained with the respect to the values
it is obligated to produce.

\begin{table}[t!]
\renewcommand{\arraystretch}{0.8}
\caption{Examples of overapproximate and underapproximate
  (coverage) typings. We use $\goodtype$ and $\badtype$ to identify
  whether a term can or cannot be assigned the corresponding type,
  resp. The constant $\Code{err}$ represents a special error value,
  which causes the program to halt when encountered. }
\footnotesize
\begin{tabular}{l|llll}
\toprule
  \ocamlinline{int_gen ()}
  & $\goodtype \nuut{int}{\top}$
  & $\goodtype \nuut{int}{\nu = \Code{1 \lor 2}}$
  & $\goodtype \nuut{int}{\nu = \Code{1}}$
  & $\goodtype \nuut{int}{\bot}$ \\
  & $\goodtype \nuot{int}{\top}$
  & $\badtype \nuot{int}{\nu = \Code{1 \lor 2}}$
  & $\badtype \nuot{int}{\nu = \Code{1}}$
  & $\badtype\nuot{int}{\bot}$ \\
\midrule
  \ocamlinline{1}
  & $\goodtype \nuut{int}{\nu = \Code{1}}$
  & $\goodtype \nuut{int}{\bot}$
  & $\goodtype \nuot{int}{\top}$
  & $\goodtype \nuot{int}{\nu = \Code{1 \lor 2}}$ \\
  & $\goodtype \nuot{int}{\nu = \Code{1}}$
  & $\badtype \nuut{int}{\top}$
  & $\badtype \nuut{int}{\nu = \Code{1 \lor 2}}$
  & $\badtype\nuot{int}{\bot}$ \\
\midrule
  \ocamlinline{err}
  & $\goodtype \nuut{int}{\bot}$
  & $\badtype \nuut{int}{\top}$
  & $\badtype \nuut{int}{\nu = \Code{1 \lor 2}}$
  & $\badtype \nuut{int}{\nu = \Code{1}}$ \\
   & $\badtype \nuot{int}{\top}$
   & $\badtype \nuot{int}{\nu = \Code{1 \lor 2}}$
   & $\badtype \nuot{int}{\nu = \Code{1}}$
   & $\badtype\nuot{int}{\bot}$ \\
\bottomrule
\end{tabular}
\label{tab:examples}
\end{table}



Coverage types can also qualify inductive datatypes, like lists and
trees.  In particular, the complete generator for BSTs presented in
the introduction can be successfully type-checked using the following
result type:
\begin{align*}
    \nuut{int\ tree}{\I{bst}(\nu) \land \forall u, \mem(\nu, u) \impl \Code{lo} < u < \Code{hi}  }
\end{align*}\noindent
where $\I{bst}(\nu)$ and $\I{mem}(\nu, u)$ are \emph{uninterpreted predicates} used to encode semantic
properties of the datatype.
In the type given above, the qualifier
requires that \Code{bst\_gen}'s result is a BST (encoded by the
predicate $\I{bst}(\nu)$) and that every element $u$ stored in the tree
(encoded by the predicate $\mem(\nu, u)$) is between \Code{lo} and
\Code{hi}; the coverage type thus constrains the implementation to
produce \emph{all} trees that satisfy this qualifier predicate.  In
contrast, the incomplete version of the generator (i.e., the
implementation that does not allow prematurely terminating tree
generation with a \Code{Leaf} node) could only be type-checked using
the following (stronger) type:
\begin{align*}
    \nuut{int\ tree}{\I{bst}(\nu) \land \forall u, \mem(\nu, u) \iff \Code{lo} < u < \Code{hi}  }
\end{align*}\noindent
This signature asserts that all trees produced by the generator are
BSTs, that any element contained in the tree is within the interval
bounded by \Code{lo} and \Code{hi}, \emph{and} moreover, any element
in that interval \emph{must} be included in the tree.  The subtle
difference between the two implementations, reflected in the different
implication constraints expressed in their respective refinements,
precisely captures how their coverage properties differ.




\begin{figure}[t!]
  \begin{minted}[fontsize = \normalsize,xleftmargin=1pt]{ocaml}
 let even_gen () =
   let (n: int) = int_gen () in
   let (b: bool) = n mod 2 == 0 in
   if b then n else err
\end{minted}
\caption{An even number generator defined in terms of an integer number generator. }
\label{fig:even}
\end{figure}

\paragraph{Control Flow} Just as underapproximate coverage types
invert the standard overapproximate subtyping relationship, they also
invert the standard relationship between a control flow construct and
its subexpressions. To see how, consider the simple generator for even
numbers shown in \autoref{fig:even}. When the integer generator,
$\Code{int\_gen ()}$, yields an odd number, $\Code{even\_gen}$ faults;
otherwise it simply returns the generated number. Consider the
following type judgment that arises when type checking this program:
\begin{align}
  &\Code{n}{:}\nuut{int}{\top},
  \Code{b}{:}\nuut{bool}{\nu{\iff} n\; \Code{mod}\; 2 = 0}\nonumber
  \\&\qquad\vdash \ite{\Code{b}}{\Code{n}}{\Code{err}}
  : \nuut{int}{\nu\; \Code{mod}\; 2 = 0}
  \label{frm:if+type}
\end{align}\noindent Intuitively, this judgment asserts that the \Code{if}
expression covers all even numbers (i.e., has the type
$\nuut{int}{\nu\; \Code{mod}\; 2 = 0}$) assuming that the \emph{local
  variable} \Code{n} can be instantiated with an arbitrary number, and
that the variable \Code{b} is true precisely when $n$ is even. Notice
how the typing context encodes the potential control-flow path that
\emph{must} reach the non-faulting branch of the conditional
expression.  Enforcing the requirement that the conditional be able to
return all even numbers does \emph{not} require each of its branches
to be a subtype of the expected type, in contrast to standard type
systems.  Our type system must instead establish that, \emph{in
  total}, the values produced by each of the branches cover the even
numbers.  Because the \Code{false} branch of the conditional faults,
it is only typeable at the universal supertype, i.e.,
$\nuut{int}{\bot}$. Thus, if the standard subtyping relationship
between this conditional and its branches held, it could only be typed
at $\nuut{int}{\bot}$! This is not the case in our setting, as the
\Code{true} branch contributes all the desired outputs. Formally, this
property is checked by the following assumption of the coverage typing
rule for conditionals:
\begin{align*}
  &\Code{n}{:}\nuut{int}{\top},
  \Code{b}{:}\nuut{bool}{\nu{\iff}\Code{n} \;\Code{mod}\; 2 = 0}
  \\&\qquad\vdash \nuut{int}{\textcolor{DeepGreen}{(\Code{b} \land \nu = \Code{n})} \lor
  \textcolor{red}{(\neg \Code{b} \land \bot)}}
  <: \nuut{int}{\nu\; \Code{mod}\; 2 == 0}
\end{align*}\noindent
The $\textcolor{DeepGreen}{\Code{b} \land \nu = n}$ and
$\textcolor{red}{\neg \Code{b} \land \bot}$ subformulas correspond to
the types of the \Code{true} and \Code{false} branches\footnote{As is
  standard in dependent type systems, the types of both branches have
  been refined to reflect the path conditions under which they will be
  executed.}, respectively. Taking the disjunction of these two
formulas describes the set of values that can be produced by either
branch\footnote{This is similar to how the derived rule of choice
  in IL uses disjunction to reason about both branches of a nondeterministic
  choice statement.}; this subtyping relationship guarantees this
type is at least as large as the type expected by the entire
conditional. To check that this subtyping relationship holds, our type checker
generates the following formula:
\begin{align}
    \forall \nu,
    (\nu\; \Code{mod}\; 2 = 0 ) \implies
    (\exists \Code{n}, \top \land \exists \Code{b}, \Code{b}{\iff} \Code{n}\; \Code{mod}\; 2 = 0 \land \textcolor{DeepGreen}{(\Code{b} \land \nu = \Code{n})} \lor \textcolor{red}{(\neg \Code{b} \land \bot)})
  \label{frm:subtype+ex}
\end{align}\noindent
This formula aligns with the intuitive meaning of
(\ref{frm:if+type}): in our type system, coverage types of variables
in the typing context tell us what values they must (at least)
produce. When checking whether a particular subtyping or typing
relationship holds, we are free to choose \emph{any} instantiation of
the variables that entails the desired property. Accordingly, in
(\ref{frm:subtype+ex}), the variables $n$ and $b$ are
\emph{existentially} quantified to indicate there exists an execution
path that instantiates these local variables in a way that produces
the output $\nu$, instead of being universally quantified as they
would be in a standard refinement type system.

\paragraph{Function types}
To type functions, most refinement type systems add a restricted form
of the \emph{dependent function types} found in full-spectrum
dependent type systems. Such types allow the qualifiers in the result
type of a function to refer to its parameters, enabling the expression
of rich safety conditions governing the arguments that may be supplied
to the function. To see how this capability might be useful in our
setting, consider the test generator \Code{bst\_gen} from the
introduction. The complete version of this function produces all BSTs
whose elements fall between the range specified by its two parameters,
\Code{lo} and \Code{hi}.  For the bounds \Code{0} and \Code{3}, the
application \Code{bst\_gen\ 0\ 3} can be typed as:
$\nuut{int}{\I{bst}(\nu)\, \land\, \forall u, \mem(\nu, u)\impl 0 < u
  < 3}$.  Using the standard typing rule for functions, the only way
to encode this relationship in the type of \Code{bst\_gen} is:
\begin{align*}
    \urt{\Int}{\nu = 0}\sarr \urt{\Int}{\nu = 3}\sarr \nuut{int\ tree}{\I{bst}(\nu)\, \land\, \forall u, \mem(\nu, u)\impl 0 < u < 3}
\end{align*}\noindent
Of course, this specification fails to account for the behaviors of
\Code{bst\_gen} when supplied with different bounds: for example, the
application \Code{bst\_gen\ 2\ 7} will fail to typecheck against this
type.

Since the desired coverage property of \Code{bst\_gen} fundamentally
depends on the kinds of inputs given to it, our type system includes
dependent products of the form:
\begin{align*}
    &\Code{lo}{:}\ort{\Int}{\top}\sarr \Code{hi}{:}\ort{\Int}{\Code{lo} \leq \nu}\sarr \\&\qquad \nuut{int\ tree}{\I{bst}(\nu) \land \forall u, \mem(\nu, u)\impl \Code{lo} < u < \Code{hi}}
\end{align*}\noindent
We use the notation
$\textcolor{magenta}{\textbf{\{}}...\textcolor{magenta}{\textbf{\}}}$
to emphasize that the argument types of a dependent arrow have a
similar purpose and interpretation as in standard refinement type
systems. Thus, the above type can be read as ``if the inputs \Code{lo}
and \Code{hi} are \emph{any} number such that \Code{lo} $\leq$
\Code{hi}, then the output \emph{must} cover all possible BSTs whose
elements are between \Code{lo} and \Code{hi}''. Using this type for
\Code{bst\_gen} allows our system to seamlessly type-check both
(\Code{bst\_gen\ 0\ 3}) and (\Code{bst\_gen\ 2\ 7}). Our typing algorithm
will furthermore flag the call (\Code{bst\_gen\ 3\ 1}) as being
ill-typed, since the function's type dictates that the generator's
second argument (\Code{1}) \emph{may} only be greater than or equal to
its first (\Code{3}).

\begin{figure}[t!]
  \begin{minted}[fontsize = \normalsize]{ocaml}
let bst_gen_low_bound (low: int) =
    let (high: int) = int_gen () in
    bst_gen low high
\end{minted}
  \caption{This function generates a BST with a supplied lower
    bound, \Code{low}.}
\label{fig:application}
\end{figure}

\paragraph{Function Application} Since the type of a function parameter is interpreted as a normal
(overapproximate, ``may'') refinement type, while arguments in an application
may be typed using (underapproximate, ``must'') coverage types, we need to be
able to bridge the gap between may and must types when typing function
applications.  Intuitively, our type system does so by ensuring that
the set of values in the coverage type of the argument has a nonempty
overlap with the set of possible values expected by the function.  We
establish this connection by using the fact that the typing context
captures the control flow paths that may and must exist when the
function is called.  To illustrate this intuition concretely, consider
the function \Code{bst\_gen\_low\_bound} shown in
\autoref{fig:application}. This function generates all non-empty BSTs
whose elements are integers with the lower bound given by its
parameter. The judgment we need to check is of the form:
\begin{flalign*}
& \Code{bst\_gen} : \Code{lo}{:}\ort{\Int}{\top}\sarr
     \Code{hi}{:}\ort{\Int}{\Code{lo} \leq \nu}\sarr \nuut{int\ tree}{...},\\
&\Code{low}{:}\nuot{int}{\top},
 ~\Code{high}{:}\nuut{int}{\top} \\
 & \vdash \Code{bst\_gen\ low\ high} : \ldots
\end{flalign*}\noindent
Note that the type for \Code{low} is a normal refinement type
that specifies a safety condition for function \Code{bst\_gen\_low\_bound}, namely that
\Code{low} \emph{may} be any number.  In contrast,  the type
for \Code{high} is a coverage type, representing the result of
\Code{int\_gen()} that indicates that it \emph{must} (i.e., guaranteed
to) be any possible integer.  However, the signature for
\Code{bst\_gen} demands that parameter \Code{hi} only be supplied
values greater than its first argument (\Code{lo}); we incorporate
this requirement by \emph{strengthening} \Code{high}'s
type (via a subsumption rule) to reflect this additional constraint
when typing the body of the let expression in which \Code{high} is
bound.  This strengthening, which is tantamount to a more refined
underapproximation, allows us to typecheck the application
(\Code{bst\_gen low\ high}) in the
following context:
\begin{flalign*}
& \Code{bst\_gen}: \Code{lo}{:}\ort{\Int}{\top}\sarr
    \Code{hi}{:}\ort{\Int}{\Code{lo} \leq \nu}\sarr \nuut{int\ tree}{...},\\
&\Code{low}{:}\nuot{int}{\top},
\,\Code{high}{:}\nuut{int}{\Code{low} \leq \nu} \\
 & \vdash \Code{bst\_gen\ low\ high} : \ldots
\end{flalign*}\noindent
The coverage type associated with \Code{high} guarantees
that \Code{int\_gen()} must produce values greater than
\Code{low} (along with possibly other values).  To ensure that the
result type of the call reflects the underapproximate (coverage)
dependences that exist between \Code{low} and \Code{high}, we
introduce existential quantifiers in the type's qualifier:
\begin{align*}
&..., \Code{low}{:}\nuot{int}{\top} \\
&\vdash
\nuut{int\ tree}{\I{bst}(\nu)\, \land\, \exists \Code{high}, \Code{low} \leq \Code{high} \land \forall u, \mem(\nu, u)\impl \Code{low} < u < \Code{high}}
\end{align*}\noindent
This type properly captures the behavior of the generator: it is
guaranteed to generate all BSTs characterized by a lower bound given
\Code{low} such that there exists an upper bound \Code{high} where \Code{low} $\leq$ \Code{high} and in
which every element in the tree is contained within these bounds.

\paragraph{Summary} Coverage types invert many of the expected
relationships that are found in a normal refinement type system. Here,
qualifiers provide an \emph{underapproximation} of the values that an
expression may evaluate to, in contrast to the typically provided
\emph{overapproximation}.  This, in turn, causes the subtyping
relation to invert the standard relationship entailed by logical
implication between type qualifiers.  Our coverage analysis also
considers the \emph{disjunction} of the coverage guarantees provided
by the branches of control-flow constructs, instead of their
\emph{conjunction}. Finally, when applying a function with a dependent
arrow type to a coverage type, we check semantic inclusion between the
overapproximate and underapproximate constraints provided by the two
types, and manifest the paths that witness the elements guaranteed to
be produced by the coverage type through existentially-quantified
variables in the application's result type.

\section{Language}
\label{sec:lang}

\begin{figure}[t!]
{
    \begin{alignat*}{2}
    \text{\textbf{Variables }}& \quad &\quad& x, y, z, f, u, ... \\
    \text{\textbf{Data constructors }}& \quad &d ::= \quad & () ~|~ \Code{true} ~|~ \Code{false} ~|~ \Code{O} ~|~ \Code{S} ~|~ \Code{Cons} ~|~ \Code{Nil} ~|~ \Code{Leaf} ~|~ \Code{Node} \\
    \text{\textbf{Constants }}& \quad &c ::= \quad & \mathbb{B} ~|~ \mathbb{N} ~|~ \mathbb{Z} ~|~ \ldots ~|~ d\ \overline{c}\\
    \text{\textbf{Operators }}& \quad & \I{op} ::= \quad &d ~|~ {+} ~|~ {==} ~|~ {<} ~|~ \Code{mod} ~|~ \randomnat ~|~  \Code{int\_gen} ~|~ ...\\
    \text{\textbf{Values }}& \quad  & v ::= \quad & c ~|~ \I{op} ~|~ x ~|~ \zlam{x}{\tau}{e} ~|~ \zfix{f}{\tau}{x}{\tau}{e}  \\
    \text{\textbf{Terms}}& \quad & e ::=\quad & v ~|~ \exn
    ~|~ \zlet{x}{e}{e}
    ~|~ \zlet{x}{op\ \overline{v}}{e}\\
    &\quad&\quad& |~ \zlet{x}{v\ v}{e} ~|~ \match{v} \overline{d\ \overline{y} \to e}   \\
    \text{\textbf{Base Types}}& \quad & b  ::= \quad &  \Code{unit} ~|~ \Code{bool} ~|~ \Code{nat} ~|~ \Code{int} ~|~ \Code{float} ~|~ \Code{char} ~|~ \Code{string} \\
    &\quad&\quad& |~ b \times b ~|~ \{\overline{x{:}b}\} ~|~ b\ \Code{option} ~|~ b\ \Code{list} ~|~ b\ \Code{tree} ~|~ \ldots   \\
    \text{\textbf{Basic Types}}& \quad & t  ::= \quad & b ~|~ t\, \sarr\,t \\
    \text{\textbf{Uninterpreted Functions }}& \quad & \I{uf} ::= \quad & \I{emp} ~|~ \I{hd} ~|~ \I{mem} ~|~ ...\\
    \text{\textbf{Literals }}& \quad & l ::= \quad & c ~|~ x \\
    \text{\textbf{Propositions}}& \quad & \phi ::= \quad &  l ~|~ \bot ~|~ \top  ~|~ \I{op}(\overline{l}) ~|~ \I{uf}(\overline{x}) \\
    &\quad&\quad& |~ \neg \phi ~|~ \phi \land \phi ~|~ \phi \lor \phi ~|~ \phi \impl \phi ~|~  \forall u{:}b.\; \phi ~|~ \exists u{:}b.\; \phi \\
    \text{\textbf{Refinement Types}}& \quad &\tau  ::= \quad &  \nuut{\textit{b}}{\phi} ~|~ \nuot{\textit{b}}{\phi} ~|~  x{:}\tau\sarr\tau \\
    \text{\textbf{Type Contexts}}& \quad &\Gamma ::= \quad & \emptyset ~|~ \Gamma, x{:}\tau
  \end{alignat*}
}
    \caption{\langname{} syntax.}\label{fig:syntax}
\end{figure}

\paragraph{Terms} In order to formalize our typed-based verification approach of input
test generators, we introduce a core calculus for test generators,
\langname{}.  The language, whose syntax is summarized in
\autoref{fig:syntax}, is a call-by-value lambda-calculus with
pattern-matching, inductive datatypes, and well-founded (i.e.,
terminating) recursive functions whose argument must be structurally
decreasing in all recursive calls made in the function's body.  The
syntax of \langname{} is expressed in monadic normal-form
(MNF)~\cite{JO94}, a variant of A-Normal Form (ANF)~\cite{FSDF93} that
allows nested let-bindings. \cbnewadding{Terms in a standard lambda-calculus can be expressed as syntax sugar in ANF, e.g., an application ($e_1\;e_2$) can be defined as:
\begin{align*}
    e_1\;e_2 \doteq \zlet{f}{e_1}{\zlet{x}{e_2}{\zlet{y}{f\;x}{y}}}
\end{align*}\noindent
To enable type-checking, the parameters of lambda function (i.e. $\zlam{x}{\tau}{e}$) and fixpoint functions (i.e. $\zfix{f}{\tau}{x}{\tau}{e}$) are also notated with coverage types.}
The language additionally allows
faulty programs to be expressed using the error term \Code{err}.  As
discussed in~\autoref{sec:motivation}, this term is important in our
investigation because coverage types capture an expression's
reachability properties, and we need to ensure the guarantees offered
by such types are robust even in the presence of stuck computations
induced by statements like \Code{err}.  The language is also equipped
with primitive operators to generate natural numbers, integers, etc.
($\randomnat\; (), \Code{int\_gen}(), \ldots$) that can be used to
express various kinds of non-deterministic behavior relevant to test
input generation. As an example, the $\oplus$ choice operator used
in~\autoref{fig:gen-ex} can be defined as:
\begin{align*}
    e_1 \oplus e_2 \doteq &\ \zlet{n}{\Code{int\_gen}\; ()\ \Code{mod}\ 2}{\match{n}0 \to e_1 ~|~\_ \to e_2}
\end{align*}\noindent
Note that the primitive generators of \langname{} are
completely agnostic to the specific sampling strategy they employ, as
long as they ensure every value in their range has a nonzero
likelihood of being generated.
\langname{} has a completely standard
small-step operational semantics.




\paragraph{Types}Like other refinement type systems~\cite{JV21},
\langname{} supports three classes of types: \emph{base types},
\emph{basic types}, and \emph{refinement types}.  Base types ($b$)
include primitive types such as $\Code{unit}$, $\Code{bool}$,
$\Code{nat}$, etc., \cbnewadding{product types ($b \times b$), record
  types $\{\overline{x{:}b}\}$,} and inductive datatypes (e.g.,
\Code{int list}, \Code{bool tree}, \Code{int list list}, etc.). Basic
types ($t$) extend base types with function types.  Refinement types
($\S{\tau}$) qualify base types with both underapproximate and
overapproximate propositions, expressed as predicates defined in
first-order logic (FOL).  Function parameters can also be qualified
with overapproximate refinements that specify when it is safe to apply
this function. In contrast, the return type of a function can only be
qualified using an underapproximate refinement, reflecting the
coverage property of the function's result and thus characterizing the
values the function is guaranteed to produce.  \cbnewadding{The type
  context is defined normally, i.e., as a sequence of variable-type
  bindings consisting of overapproximate refinement types,
  underapproximate coverage types, and arrow (function) types.}


\paragraph{Qualifiers}
To express rich shape properties
over inductive datatypes, we allow propositions to reference uninterpreted predicates, as it is straightforward to generate verification
conditions using these uninterpreted predicates that can be handled by
an off-the-shelf theorem prover like Z3~\cite{de2008z3}. As we
describe in Section~\ref{sec:algorithm}, our typechecking algorithm
imposes additional constraints on the form propositions can take, in
order to ensure that its validity is decidable. In particular, we
require that uninterpreted predicates in our language are stratified and that the Z3 queries generated by our typechecker to check
refinement validity are always over effectively propositional (EPR)
sentences (i.e., prenex-quantified formulae of the form
$\exists^*\forall^*\varphi$ where $\varphi$ is quantifier-free).

\section{Type System}
\label{sec:rules}

\cbnewadding{ Similar to other contemporary refinement type
  systems~\cite{JV21}, our type system assumes all terms are
  well-typed in a basic (aka non-refined) type system\footnote{The
    basic typing rules, proofs of theorems, and the details of our
    evaluation are provided in \releasesource{}.}, i.e.,
  $\emptyset \basicvdash e : t$. For example, a term $e$ that has
  refinement type $\tau$ has basic type $\eraserf{\tau}$ where
  $\eraserf{...}$ is a type erasure function, as shown in
  \autoref{fig:aux-rules}, which erases all qualifiers from refinement
  types and type contexts.  }

Despite superficial similarities to other contemporary type
systems, the typing rules of \langname{} differ in significant ways
from those of its peers, due to the fundamental semantic distinction
that arises when viewing types as an underapproximation and not
overapproximation of program behavior.

\begin{figure}[!t]
{\small
{\normalsize\begin{flalign*}
 &\text{\textbf{Type Erasure}} & \fbox{$\eraserf{\tau} \quad \eraserf{\Gamma}$}
\end{flalign*}}
\vspace{-1em}
\begin{alignat*}{5}
    \eraserf{\ort{b}{\phi}} &\doteq b \quad&
    \eraserf{\urt{b}{\phi}} &\doteq b \quad&
    \eraserf{x{:}t \sarr \tau} &\doteq \eraserf{t}\sarr\eraserf{\tau} \quad&
    \eraserf{\emptyset} &\doteq \emptyset \quad&
    \eraserf{x{:}\tau, \Gamma} &\doteq x{:}\eraserf{\tau}, \eraserf{\Gamma}
\end{alignat*}
}
\vspace{-1em}
{\small
{\normalsize
\begin{flalign*}
 &\text{\textbf{Well-Formedness }} &
 \fbox{$\wfvdash \Gamma$ \quad $\Gamma \wfvdash \tau$}
\end{flalign*}
}
\\ \
\mprooftr{40mm}
{$\wfvdash \Gamma$ \quad $\eraserf{\Gamma}, \vnu{:}b \basicvdash \phi : \Bool$}
{WfBase}
{$\Gamma \wfvdash \urt{b}{\phi}$}
\quad
\mprooftr{40mm}
{$\Gamma, x{:}\ort{b}{\phi} \wfvdash \tau$}
{WfRes}
{$\Gamma \wfvdash x{:}\ort{b}{\phi}\sarr\tau$}
\\ \ \\ \ \\
\mprooftr{40mm}
{$\Gamma, x{:}(y{:}\tau_y\sarr\tau) \wfvdash \tau$}
{WfRes}
{$\Gamma \wfvdash x{:}(y{:}\tau_y\sarr\tau)\sarr\tau$}
\\ \ \\ \ \\
\mprooftr{10mm}
{}
{WfCtxEmp}
{$\wfvdash \emptyset$}
\quad
\mprooftr{40mm}
{$\wfvdash \Gamma$ \quad $\Gamma \wfvdash \urt{b}{\phi}$}
{WfCtxOBase}
{$\wfvdash \Gamma, x{:}\ort{b}{\phi}$}
\\ \ \\ \ \\
\mprooftr{40mm}
{$\wfvdash \Gamma$ \quad $\Gamma \wfvdash \urt{b}{\phi}$ \quad
$\exn \not\in \denot{\urt{b}{\phi}}_{\Gamma}$
}
{WfCtxUBase}
{$\wfvdash \Gamma, x{:}\urt{b}{\phi}$}
\quad
\mprooftr{40mm}
{$\wfvdash \Gamma$ \quad $\Gamma \wfvdash y{:}\tau_y\sarr\tau$}
{WfCtxArr}
{$\wfvdash \Gamma, x{:}(y{:}\tau_y\sarr\tau)$}
}
{\small
{\normalsize
\begin{flalign*}
 &\text{\textbf{Subtyping }} & \fbox{$\Gamma \vdash \tau_1 <: \tau_2$}
\end{flalign*}
}
\\ \
\begin{prooftree}
\hypo{
\parbox{40mm}{\center
$\denot{\urt{b}{\phi_1}}_{\Gamma} \subseteq \denot{\urt{b}{\phi_2} }_{\Gamma}$
}}
\infer1[\textsc{SubUBase}]{
\Gamma \vdash \urt{b}{\phi_1} <: \urt{b}{\phi_2}
}
\end{prooftree}
\quad
\begin{prooftree}
\hypo{
\parbox{37mm}{\center
$\denot{\rawnuot{b}{\phi_1}}_{\Gamma} \subseteq \denot{\rawnuot{b}{\phi_2}}_{\Gamma}$
}}
\infer1[\textsc{SubOBase}]{
\Gamma \vdash \rawnuot{b}{\phi_1} <: \rawnuot{b}{\phi_2}
}
\end{prooftree}
 \\ \ \\ \ \\
\begin{prooftree}
\hypo{
\parbox{47mm}{\center
$\Gamma \vdash \tau_{21} <: \tau_{11}$ \quad
$\Gamma, x{:}\tau_{21} \vdash \tau_{12} <: \tau_{22}$
}
}
\infer1[\textsc{SubArr}]{
\Gamma \vdash x{:}\tau_{11}\sarr\tau_{12} <: x{:}\tau_{21}\sarr\tau_{22}
}
\end{prooftree}
{\normalsize
\begin{flalign*}
 &\text{\textbf{Disjunction }} & \fbox{$\Gamma \vdash \tau_1 \lor \tau_2 = \tau_3$}
\end{flalign*}
}
\begin{prooftree}
\hypo{
\denot{\tau_1}_{\Gamma} \cap \denot{\tau_2}_{\Gamma} = \denot{\tau_3}_{\Gamma}}
\infer1[\textsc{Disjunction}]{
\Gamma \vdash \tau_1 \lor \tau_2 = \tau_3
}
\end{prooftree}
}
\caption{Auxiliary typing relations}\label{fig:aux-rules}
\end{figure}

Our type system depends on three auxiliary relations shown in
\autoref{fig:aux-rules}.  \cbnewadding{ The first group defines
  well-formedness conditions on type contexts ($\wfvdash \Gamma$) as
  well as refinement types under a particular type context
  ($\Gamma \wfvdash \tau$).}  A type $\tau$ that is well-formed under
a well-formed type context $\Gamma$ needs to meet three criteria:
\cbnewadding{ (1) the qualifier in $\tau$ needs to be well-typed under
  the erasure of the current typing context in the basic type system
  (e.g., requirement
  $\eraserf{\Gamma}, \vnu{:}b \basicvdash \phi : \Bool$ in
  \textsc{WfBase}), (2) overapproximate types may only appear in the
  domain of a function type (\textsc{WfArg}); and, (3)
  underapproximate coverage types may only appear in the range of a
  function type (only underapproximate and function type are allowed
  as parameter type in \textsc{WfArg} and\textsc{WfRes} respectively).
  A type context is well-formed when all types in a type context are
  well-formed under the context derived from the former's bindings,
  moreover, the denotations\footnote{The definition of a type's
    denotation is given in \autoref{sec:denotation}.} of coverage base
  types ($\urt{b}{\phi}$) should not include $\exn$
  (\textsc{WfCtxUBase}). } To understand the motivation for this
constraint, observe that a type context in our setting provides a
witness to feasible execution paths in the form of bindings to local
variables.  \cbnewadding{Accordingly, the type context
  $x{:}\nuut{nat}{\bot}$ or
  $x{:}\nuot{nat}{\nu > 0}, y{:}\nuut{nat}{x = 0 \land \nu = 2}$ are
  not well-formed,} as neither context corresponds to a valid manifest
execution path. On the other hand, a well-formed type is allowed to
include an error term in its denotation, e.g., type $\nuut{nat}{\bot}$
is well-formed under type context $x{:}\nuot{nat}{\nu > 0}$ as it
always corresponds to a valid underapproximation.

Our second set of judgments defines a largely standard subtyping
relation based on the underlying denotation of the types being
related.  In general, an overapproximate type and an underapproximate
type are inhabited by incomparable sets of terms, even when they share
a qualifier. A non-singleton underapproximate type
can only be assigned to nondeterministic terms, while the
corresponding overapproximate type can be given to many possible
deterministic terms. On the other hand, the unconstrained generator
for a base type, e.g., \Code{int\_gen()} or \Code{tree\_gen()}, can be
checked against an underapproximate type qualified with any predicate,
but it can only be assigned the overapproximate type qualified with
$\top$. As a consequence, our subtyping relation does not relate
overapproximate and underapproximate types, and our typing rules
tightly control when one can be treated as another.

The disjunction rule (\textsc{Disjunction}), which was informally
introduced in \autoref{sec:motivation}, merges the coverage types
found along distinct control paths.  Intuitively, the type
$\nuut{nat}{\nu = 1 \lor \nu = 2}$ is the disjunction of the types
$\nuut{nat}{\nu = 1}$ and $\nuut{nat}{\nu = 2}$. Notice that only an
inhabitant of both $\nuut{nat}{\nu = 1}$ and $\nuut{nat}{\nu = 2}$
should be included in their disjunction: e.g., the term $1\oplus2$ is
one such inhabitant. Thus, we formally define this relation as the
\emph{intersection} of the denotations of two types.

\begin{figure}[!ht]
  {\small
{\normalsize
\begin{flalign*}
 &\text{\textbf{Typing }} & \fbox{$\Gamma \vdash e : \tau$ \quad $\Gamma \vdash \I{op} : \tau$ \quad $\Gamma \vdash d : \tau$}
\end{flalign*}
}
\\ \
\begin{prooftree}
\hypo{\Gamma \wfvdash \urt{b}{\bot}}
\infer1[\textsc{TErr}]{
\Gamma  \vdash \exn : \urt{b}{\bot}
}
\end{prooftree}
\quad
\begin{prooftree}
\hypo{\Gamma \wfvdash \S{Ty}(c)}
\infer1[\textsc{TConst}]{
\Gamma \vdash c : \S{Ty}(c)
}
\end{prooftree}
\quad
\begin{prooftree}
\hypo{\Gamma \wfvdash \S{Ty}(d)}
\infer1[\textsc{TDt}]{
\Gamma \vdash d : \S{Ty}(d)
}
\end{prooftree}
\quad
\begin{prooftree}
\hypo{\Gamma \wfvdash \S{Ty}(\I{op})}
\infer1[\textsc{TOp}]{
\Gamma \vdash \I{op} : \S{Ty}(\I{op})
}
\end{prooftree}
\\ \ \\ \ \\
\begin{prooftree}
\hypo{\Gamma \wfvdash \urt{b}{\nu = x}}
\infer1[\textsc{TVarBase}]{
\Gamma \vdash x : \urt{b}{\nu = x}
}
\end{prooftree}
\quad
\begin{prooftree}
\hypo{\Gamma(x) = (a{:}\tau_a\sarr\tau_b) \quad \Gamma \wfvdash a{:}\tau_a\sarr\tau_b}
\infer1[\textsc{TVarFun}]{
\Gamma \vdash x : (a{:}\tau_a\sarr\tau_b)
}
\end{prooftree}
\\ \ \\ \ \\
\begin{prooftree}
\hypo{
\Gamma, x{:}\tau_x \vdash e : \tau \quad
\Gamma \wfvdash x{:}\tau_x\sarr\tau
}
\infer1[\textsc{TFun}]{
\Gamma \vdash \zlam{x}{\tau_x}{e} : (x{:}\tau_x\sarr\tau)
}
\end{prooftree}
\\ \ \\
\begin{prooftree}
\hypo{
\parbox{100mm}{\center
$\Gamma, x{:}\rawnuot{b}{\phi}, f{:}(x{:}\rawnuot{b}{\nu{~\prec~}x ~\land~ \phi}\sarr\tau)\vdash e :\tau$ \quad $\Gamma \vdash x{:}\rawnuot{b}{\phi}\sarr\tau$
}}
\infer1[\textsc{TFix}]{
\Gamma \vdash \zfix{f}{(x{:}\rawnuot{b}{\phi}\sarr\tau)}{x}{\rawnuot{b}{\phi}}{e} \;:\; x{:}\rawnuot{b}{\phi}\sarr\tau
}
\end{prooftree}
\\ \ \\ \ \\
\begin{prooftree}
\hypo{
\parbox{27mm}{\center
$\emptyset \vdash \tau <: \tau'$\quad
$\emptyset \vdash e : \tau$ \quad $\Gamma \wfvdash \tau'$
}}
\infer1[\textsc{TSub}]{
\Gamma \vdash e : \tau' }
\end{prooftree}
\quad
\begin{prooftree}
\hypo{
\parbox{33mm}{\center
$\Gamma \vdash \tau' <: \tau$ \quad
$\Gamma \vdash \tau <: \tau'$ \\
$\Gamma \vdash e : \tau$ \quad
$\Gamma \wfvdash \tau'$
}}
\infer1[\textsc{TEq}]{
\Gamma \vdash e : \tau' }
\end{prooftree}
\\ \ \\ \ \\
\begin{prooftree}
\hypo{
\parbox{40mm}{\center
$\Gamma \vdash e : \tau_1$\quad
$\Gamma \vdash e : \tau_2$\quad
$\Gamma \vdash \tau_1 \lor \tau_2 = \tau$ \quad
$\Gamma \wfvdash \tau$
}}
\infer1[\textsc{TMerge}]{
\Gamma \vdash e : \tau }
\end{prooftree}
\quad
\begin{prooftree}
\hypo{
\parbox{43mm}{\center
$\Gamma \vdash e_x : \tau_x$\quad
$\Gamma, x{:}\tau_x \vdash e : \tau$\quad
$\Gamma \wfvdash \tau$
}
}
\infer1[\textsc{TLetE}]{
\Gamma \vdash \zlet{x}{e_x}{e} : \tau
}
\end{prooftree}
\\ \ \\ \ \\
\begin{prooftree}
\hypo{
\parbox{43mm}{\center
$\Gamma \vdash \I{op} : \overline{a_i{:}\ort{b_i}{\phi_i}}\sarr\tau_x$\quad
$\forall i, \Gamma \vdash v_i : \urt{b_i}{[\phi_i}$\quad
$\Gamma, x{:}\tau_x[\overline{a_i \mapsto v_i}] \vdash e : \tau$ \quad
$\Gamma \wfvdash \tau$
}
}
\infer1[\textsc{TAppOp}]{
\Gamma \vdash \zlet{x}{\I{op}\ \overline{v_i}}{e} : \tau
}
\end{prooftree}
\quad
\begin{prooftree}
\hypo{
\parbox{43mm}{\center
$\Gamma \vdash v_1 : (\tau_1\sarr\tau_2)\sarr\tau_x$\quad
$\Gamma \vdash v_2 : \tau_1\sarr\tau_2$\quad
$\Gamma, x{:}\tau_x \vdash e : \tau$ \quad
$\Gamma \wfvdash \tau$
}
}
\infer1[\textsc{TAppFun}]{
\Gamma \vdash \zlet{x}{v_1\ v_2}{e} : \tau
}
\end{prooftree}
\\ \ \\ \ \\
\begin{prooftree}
\hypo{
\parbox{43mm}{\center
$\Gamma \vdash v_1 : a{:}\ort{b}{\phi}\sarr\tau_x$\quad
$\Gamma \vdash v_2 : \urt{b}{\phi}$\quad
$\Gamma, x{:}\tau_x[a \mapsto v_2] \vdash e : \tau$ \quad
$\Gamma \wfvdash \tau$
}
}
\infer1[\textsc{TApp}]{
\Gamma \vdash \zlet{x}{v_1\ v_2}{e} : \tau
}
\end{prooftree}
\quad
\begin{prooftree}
\hypo{
\parbox{55mm}{\center
$\Gamma \vdash v : \tau_v$\quad
$\Gamma \wfvdash \tau$\quad
$\Gamma, \overline{{y}{:}\tau_y} \vdash d_i(\overline{{y}}) : \tau_v$\quad
$\Gamma, \overline{{y}{:}\tau_y} \vdash e_i : \tau$
}}
\infer1[\textsc{TMatch}]{
\Gamma \vdash (\match{v} \overline{d_i\ \overline{y} \to e_i}) : \tau}
\end{prooftree}
}
    \caption{Typing rules}\label{fig:type-rules}
\end{figure}

\cbnewadding{ The rules of our type system are defined in
  \autoref{fig:type-rules}, and include typing judgments for terms
  ($\Gamma \vdash e : \tau$), primitive operators
  ($\Gamma \vdash \I{op} : \tau$), and data constructors
  ($\Gamma \vdash d : \tau$). All our typing rules assume that terms
  are well-typed according to the basic type system; the basic type
  system is completely standard and we omit it here. } The rules
collectively maintain the invariant that terms can only be assigned a
well-formed type. \cbnewadding{The rules for constants
  (\textsc{TConst}), operators (\textsc{TOp}), and data constructors
  (\textsc{TDt}) are straightforward.} It relies on an auxiliary
function, $\S{Ty}$, to assign types to the primitives of
\langname{}. \autoref{tab:prim-types} presents some examples of the
typings provided by $\S{Ty}$. We use uninterpreted predicates in the types of
constructors: the types for list constructors, for example, use
$\I{emp}$, $\I{hd}$ and $\I{tl}$, to precisely capture that \Code{[]}
constructs an empty list, and that $(\Code{Cons}\, x\, y)$ builds a
list containing $x$ as its head and $y$ as its tail.

\begin{table}[t]
\renewcommand{\arraystretch}{0.9}

\caption{Example typings for \langname{} primitives}

\small
\begin{tabular}{l|l}
  \toprule
  Constants&
  $\S{Ty}(\Code{true}) = \nuut{bool}{\nu}$ \quad $\S{Ty}(\Code{8}) = \nuut{nat}{\nu = 8}$...\\
  \midrule
  Data & $\S{Ty}(\Code{[]}) = \urt{\List[b]}{\emp(\nu)}$\\
  Constructors&
  $\S{Ty}(\Code{{Cons}}) = x{:}\ort{b}{\top} \sarr y{:}\ort{\List[b]}{\top}\sarr \urt{\List[b]}{\hd(\nu, x) \land \I{tl}(\nu, y)}$...\\
  \midrule
  Operators&
  $\S{Ty}(\randomnat) = \nuut{unit}{\top}\sarr\nuut{int}{0 \leq \vnu}$ \\
  &$\S{Ty}(+) = x{:}\ort{\Int}{\top}\sarr y{:}\ort{\Int}{\top}\sarr \nuut{int}{\nu = x+ y}$...\\
  \bottomrule
\end{tabular}
\label{tab:prim-types}

\end{table}

The typing rules for function abstraction (\textsc{TFun}) and error
(\textsc{TErr}) are similarly straightforward.  \cbnewadding{In a
  $\lambda$-abstraction, the type of the bound variable ($\tau_x$) is
  explicitly given.}  The error term can be assigned an arbitrary
bottom coverage base type.  The variable rule (\textsc{TVarBase})
establishes that the variable $x$ in the type context with a base type
can also be typed with the tautological qualifier $\nu = x$ (the
rule's well-formedness assumption ensures that $x$ is not free). This
judgment allows us to, for example, type the function
$\lambda x:\mathit{nat}.x$ with the type
$x{:}\nuot{nat}{\top}\sarr\nuut{nat}{\nu = x}$, indicating that the
return value is guaranteed to be exactly equal to the input $x$.
Observe that the type of $x$ under the type context
$x{:}\nuot{nat}{\top}$ (generated by the function rule \textsc{TFun})
is \emph{not} $\nuut{nat}{\top}$. We cannot simply duplicate the
qualifier for $x$ from the type context here, as this is only sound
when types characterize an overapproximation of program behavior. As
an example, $\nuot{nat}{\top}$ is a subtype of $\nuot{nat}{\nu = x}$
under the type context $x{:}\nuot{nat}{\top}$; in our underapproximate
coverage type system, in contrast, $\nuut{nat}{\top}$ is not a subtype
of $\nuut{nat}{\nu = x}$ under the type context $x{:}\nuot{nat}{\top}$
.


In a call-by-value language, a variable must refer to the same value
throughout its scope; this property extends to the guarantees provided
by its coverage type.  Consider the following example:
\begin{minted}[fontsize = \small, escapeinside=&&]{ocaml}
  &$\not\vdash$& let x = int_gen () in let y = 3 / x in x : &$\nuut{int}{\vnu = 0}$&
\end{minted}
Although $x$ may be bound any to any integer, this program will never
evaluate to $0$, as the evaluation of the second \Code{let} expression
will raise a division-by-zero exception when $x$ is $0$.  Our type
system enforces the invariant that the coverage type of a variable is
fixed when it is bound, ensuring that the variable provides consistent
coverage guarantees whenever it is subsequently used.  In the example
above, to ensure safe execution, $\Code{int\_gen\;()}$ must be
assigned a coverage type that excludes $0$, i.e.,
$\nuut{int}{\vnu \neq 0}$.  To achieve this, our subsumption rule
(\textsc{TSub}) is restricted to closed terms, permitting us to
strengthen the coverage guarantee only when a closed generator like
$\Code{int\_gen\;()}$ is introduced; and disallowing any coverage
changes of varaible $x$ at the places where it is used.

The typing rule for application (\textsc{TApp}) requires both its
underapproximate argument type and the overapproximate parameter type
to have the same qualifier, and furthermore requires that the type of
the body ($\tau$) is well-formed under the original type context
$\Gamma$, enforcing that $x$ (the result of the application) does not appear
free in $\tau$.  When argument and parameter qualifiers are not
identical, a subsumption rule is typically used to bring the two types
into alignment.  Recall the following example from
\autoref{sec:motivation}, suitably modified to conform to \langname's
syntax:
\begin{minted}[fontsize = \small, escapeinside=&&]{ocaml}
  &$\Code{bst\_gen}: \Code{lo}{:}\ort{\Int}{\top}\sarr \Code{hi}{:}\ort{\Int}{\Code{lo \leq \nu}}\sarr \nuut{int\ tree}{...}, \Code{low}: \nuot{int}{\top}\vdash$&
        let (g: unit -> int) = int_gen in let (x: unit) = () in
        let (high: int) = g x in let (y: int tree) = bst_gen low high in y
\end{minted}
\noindent Here, the type of \Code{high},
$\nuut{int}{\top}$ is stronger than the type expected
for the second parameter of \Code{bst\_gen},
$\nuut{int}{\Code{lo \leq \nu}}$. The subsumption rule
(\textsc{TSub}), that would normally allow us to strengthen the type
of \Code{high} to align with the required parameter type, is
applicable to only closed terms, which \Code{high} is not. For the
same reason, we cannot use \textsc{TSub} to strengthen the type of
\Code{high} when it is bound to \Code{g x}. Thankfully, we can
strengthen \Code{g} when it is bound to \Code{int\_gen}: according to
\autoref{tab:prim-types}, the operator $\Code{int\_gen}$ has type
$\ort{\Unit}{\top}\sarr \nuut{int}{\top}$ and is also closed, and can
thus be strengthened via \textsc{TSub}, allowing us to type the call
to \Code{bst\_gen} under the following, stronger type context:
\begin{minted}[fontsize = \small, escapeinside=&&]{ocaml}
&$\Code{bst\_gen}: \Code{lo}{:}\ort{\Int}{\top}\sarr\Code{hi}{:}\ort{\Int}{\Code{lo \leq \nu}}\sarr\nuut{int\ tree}{...}, \Code{low}: \nuot{int}{\top},$&
&$\Code{g}: \urt{\Unit}{\top}\nuut{int}{\textcolor{red}{\Code{low} \leq \nu}}, \Code{x} : \nuut{unit}{\top}, \Code{high}: \nuut{int}{\textcolor{red}{\Code{low} \leq \nu}}\vdash$&
       let (y: int tree) = bst_gen low high in y
\end{minted}
The subsumption rule allows us to use \Code{int\_gen} in a context that
requires \emph{fewer} guarantees than \Code{int\_gen} actually
provides, namely those values of $\Code{high}$ required by the
signature of $\Code{bst\_gen}$.  Intuitively, since our notion of
coverage types records feasible executions in the type context in the
form of existentials that serve as witnesses to an underapproximation,
the strengthening provided by the subsumption rule establishes an
invariant that all bindings introduced into a type context only
characterize valid behaviors in a program execution. When coupled with
\textsc{TMerge}, this allows us to \emph{split} a typing derivation
into multiple plausible strengthenings when a variable is introduced
into the typing context and then \emph{combine} the resulting types to
reason about multiple feasible paths. 

Now, using \textsc{TApp} to type $\Code{bst\_gen\ low\ high}$, and \textsc{TVarBase} to type
the body of the \Code{let} gives us:
\begin{minted}[fontsize = \small, escapeinside=&&]{ocaml}
&$\Code{bst\_gen}: \Code{lo}{:}\ort{\Int}{\top}\sarr\Code{hi}{:}\ort{\Int}{\Code{lo \leq \nu}}\sarr\nuut{int\ tree}{...},$&
&$\Code{low}: \nuot{int}{\top}, \Code{high}:\nuut{int}{\textcolor{red}{\Code{low} \leq \nu}},$&
&$\Code{y}: \nuut{int\ tree}{\I{bst}(\nu) \land \forall u, \mem(\nu, u)\impl \Code{lo} < u <  \Code{hi}}[\Code{lo} \mapsto \Code{low}][\Code{hi} \mapsto \Code{high}]$&
 &$\vdash \Code{y}:\nuut{int\ tree}{\nu = \Code{y}}$&
\end{minted}
Observe that \textsc{TVarBase} types the body as:
$\nuut{int\ tree}{\nu = \Code{y}}$, which is not closed.  To construct a
well-formed term, we need a formula equivalent to this type that
accounts for the type of \Code{y} in the current type context.  The
\textsc{TEq} rule allows us to interchange formulae that are
equivalent under a given type context to ensure the well-formedness of
the types constructed. Unlike \textsc{TSub}, it simply changes the
form of a type's qualifiers, \emph{without altering} the scope of
feasible behaviors under the current context.  In this example, such an
equivalent closed type, given the binding for \Code{y} in the type
context under which the expression is being type-checked, would be:
\begin{minted}[fontsize = \small, escapeinside=&&]{ocaml}
  let (y: int tree) = bst_gen low high in y :
    &$\nuut{int\ tree}{\exists \Code{y}, (\I{bst}(\Code{y}) \land \forall u, \mem(\Code{y}, u)\impl \Code{low} < u <  \Code{high}) \land \nu = \Code{y}}$&
\end{minted}

\noindent With these pieces in hand, we can see that the typing rule
for \Code{match} is a straightforward adaptation of the components we
have already seen, where the type of matched variable $v$ is assumed
to have been strengthened by the rule \textsc{TSub} to fit the type
required to take the $i$th branch
$\Gamma, \overline{{y}{:}\tau_y} \vdash d_i(\overline{{y}}) :
\tau_v$.  We can also safely assume the type of the branch $\tau_i$ is
closed under original type context $\Gamma$, relying on \textsc{TEq}
to meet this requirement. While \textsc{TMatch} only allows for a
single branch to be typechecked, applying \textsc{TMerge} allows us to
reason about the coverage provided by multiple branches, which have
all been typed according to this rule.

\cbnewadding{As in \textsc{TFun}, the typing rule for recursive
  functions (\textsc{TFix}) requires the types of the self-reference
  to $f$ and the function parameter $x$ to be explicitly given. }
Since types in our language serve as witnesses to feasible executions,
the result type of any recursive procedure must characterize the set
of values the procedure can plausibly return. Thus, the \textsc{TFix}
rule forces its first argument to always decrease according to some
well-founded relation $\prec$. To see why we impose this restriction,
consider the function $\S{loop}$:
\begin{minted}[fontsize = \small,xleftmargin = 10pt, escapeinside=&&]{ocaml}
let rec loop (n: nat) = loop n
\end{minted}
\noindent Without our termination check, this function can be assigned
the type $\nuot{nat}{\top}\sarr\nuut{nat}{\nu = 3}$, despite the fact
that this function never returns \Code{3}--- or any value at all!  The
body of this expression can be type-checked under the following type
context (via \textsc{TFix} and \textsc{TFun}):
\begin{minted}[fontsize = \small, escapeinside=&&]{ocaml}
  &$\Code{n}{:}\nuot{nat}{\top},\, \Code{loop}{:}(\Code{n}{:}\ort{\Nat}{\top}\sarr\nuut{nat}{\nu = 3}) \vdash$& &$\Code{loop}\ \Code{n}$& : &$\nuut{nat}{\nu = 3}$&
\end{minted}
This judgment reflects an infinitely looping execution,
however. Indeed, the same reasoning allows us to type this function
with any result type. Constraining \Code{loop}'s argument type to be
decreasing according to $\prec$ yields the following typing obligation:
\begin{minted}[fontsize = \small, escapeinside=&&]{ocaml}
  &$\Code{n}{:}\nuot{nat}{\top},\, \Code{loop}{:}(\Code{n}{:}\nuot{nat}{\nu \prec \Code{n}}\sarr\nuut{nat}{\nu = 3}) \vdash\ \Code{n}$& : &$\nuut{nat}{\nu = \Code{n}}$&
\end{minted}
where the qualifiers $\nu \prec \Code{n}$ and $\nu = \Code{n}$
conflict, raising a type error, and preventing \Code{loop} from being
recursively applied to \Code{n}.

\section{Type Soundness}
\label{sec:denotation}

\paragraph{Type Denotations} Assuming a standard typing judgement for
basic types, $\emptyset \basicvdash e : t$,
a type denotation for a type $\tau$, $\denot{\tau}$, is a set of
closed expressions:
\begin{align*}
    &\llbracket \rawnuot{b}{\phi} \rrbracket &&\doteq \{ v ~|~ \emptyset \basicvdash v : b \land \phi[\nu\mapsto v] \}\\
    &\llbracket \urt{b}{\phi} \rrbracket &&\doteq \{ e ~|~ \emptyset \basicvdash e : b \land \forall v{:}b, \phi[\nu\mapsto v] \impl e \hookrightarrow^* v \}\\
    &\llbracket x{:}\tau_x\sarr\tau \rrbracket &&\doteq \{ f~|~ \emptyset \basicvdash f : \eraserf{\tau_x\sarr\tau} \land
    \forall v_x \in \llbracket \tau_x \rrbracket \impl f\ v_x \in  \llbracket \tau[x\mapsto v_x ] \rrbracket \}
\end{align*}
In the case of an overapproximate refinement type, $\rawnuot{b}{\phi}$,
the denotation is simply the set of all values of type $b$
whose elements satisfy the type's refinement predicate ($\phi$), when
substituted for all free occurrences of $\nu$ in $\phi$.\footnote{The
  denotation of an overapproximate refinement type is more generally
  $\{ e{:}b ~|~ \emptyset \vdash e: b \land \forall
  v{:}b, e \hookrightarrow^* v \impl \phi[x\mapsto v] \}$.
  However, because such types are only used for function parameters,
  and our language syntax only admits values as arguments, our
  denotation uses the simpler form.}  Dually, the denotation of an
underapproximate coverage type is the set of expressions that evaluate
to $v$ whenever $\phi[\nu \mapsto v]$ holds, where $\phi$ is the
type's refinement predicate.  Thus, every expression in such a
denotation serves as a witness to a feasible, type-correct,
execution.  The denotation for a function type is defined in terms of
the denotations of the function's argument and result in the usual
way, ensuring that our type denotation is a logical predicate.

\paragraph{Type Denotations under a Type Context} The denotation of a
refinement type $\tau$ under a type context $\Gamma$ (written
$\denot{\tau}_{\Gamma}$) is:\footnote{In the last case, since $\hat{e}_x$ may
  nondetermistically reduce to multiple values, we employ
  intersection (not union), similar to the \textsc{Disjunction}
  rule.}$^,$\footnote{When reasoning about a subset relation between the denotations of two types under a type context
  $\denot{\urt{b}{\phi_1}}_{\Gamma} \subseteq \denot{\urt{b}{\phi_2} }_{\Gamma} $ we require that the denotations
  be computed using the same $\Gamma$; details are provided in the \releasesource{}.}
{\small
\begin{alignat*}{2}
    \denot{\tau}_{\emptyset} &\doteq \denot{\tau}
    \\ \denot{\tau}_{x{:}\tau_x, \Gamma} &\doteq \{ e ~|~
    \forall v_x \in \denot{\tau_x}. \zlet{x}{v_x}{e} \in
    \denot{\tau [x\mapsto v_x]}_{\Gamma [x\mapsto v_x]} \} \hspace{1.8cm} \text{ if $\tau
      \equiv \rawnuot{b}{\phi}$} \\
      \denot{\tau}_{x{:}\tau_x,
      \Gamma} &\doteq \{ e ~|~ \exists \hat{e}_x \in
    \denot{\tau_x}. \forall e_x \in \denot{\tau_x}.
    \\ &\hspace{1.8cm} \zlet{x}{e_x}{e} \in \bigcap_{\hat{e}_x \hookrightarrow^*
      v_x} \denot{\tau
    [x\mapsto v_x]}_{\Gamma [x\mapsto v_x]} \}
    \hspace{1.7cm} \text{ otherwise}
\end{alignat*}}\noindent
The denotation of an overapproximate refinement type under a type
context is mostly unsurprising, other than our presentation choice to
use a let-binding, rather than substitution, to construct the expressions
included in the denotations.

For a coverage type, however, the definition precisely captures our
notion of a reachability witness by explicitly constructing an
execution path as a sequence of \Code{let}-bindings that justifies the
inhabitant of the target type $\tau$.  Using \Code{let}-bindings
forces expressions in the denotation to make consistent choices when
evaluated. The existential introduced in the definition captures the
notion of an underapproximation, while the use of set intersection
allows us to reason about non-determinism introduced by primitive
generators like \Code{nat\_gen\;()}.

\begin{example} The term $x + 1$ is included in the denotation of the
  type $\nuut{nat}{\nu = x + 1 \lor \nu = x + x}$ under the type
  context $x:\nuut{nat}{\nu = 1}$.  This is justified by picking $1$
  for $\hat{e}_x$, which yields a set intersection that is equivalent
  to $\denot{\nuut{nat}{\nu = 2}}$. Observe that \emph{any}
  expression in $\denot{\nuut{nat}{\nu = 1}}$, e.g. $0\oplus1$
  and $1\oplus2$, yields an expression, $\zlet{x}{0\oplus1}{x+1}$ or
  $\zlet{x}{1\oplus2}{x+1}$, included in this intersection.
\end{example}

\begin{example}
  On the other hand, the term $x$ is not a member of the denotation of the type $\nuut{nat}{\nu = x + 1 \lor \nu = x + x}$. To see why, let us pick \Code{nat\_gen()} for
  $\hat{e}_x$. This yields a set intersection that is equivalent to
  $\denot{\nuut{nat}{\top}}$. While specific choices for $e_x$,
  e.g., \Code{nat\_gen()}, are included in this denotation, it does
  not work for all terms $e_x \in \denot{\nuut{nat}{\nu = 1}}$.
  As one example, $0\oplus1\oplus2$ is an element of this set, but
  $\zlet{x}{0\oplus1\oplus2}{x}$ is clearly not a member of
  $\denot{\nuut{nat}{\top}}$. Suppose instead that we picked a
  more restrictive expression for $\hat{e}_x$, like the literal $1$
  from the previous example. Here, it is easy to choose
  $e_x \in \denot{\nuut{nat}{\nu = 1}}$ (e.g., the literal $1$) such that
  $\zlet{x}{e_x}{x} \not\in \denot{\nuut{nat}{\nu = 2}}$.
\end{example}

Our main soundness result establishes the correctness of type-checking
in the presence of coverage types with respect to a type's denotation:

\begin{theorem}\label{theorem:type-sound}[Type Soundness] For all type contexts
  $\Gamma$, terms $e$ and coverage types $\tau$,
  $\Gamma \vdash e : \tau \implies e \in
  \denot{\tau}_{\Gamma}$.
\end{theorem}
\noindent It immediately follows that a closed input generator \Code{e} with
coverage type $\urt{b}{\phi}$ must produce every value satisfying
$\phi$, as desired.

\section{Typing Algorithm} 
\label{sec:algorithm}

The declarative typing rules are highly nondeterministic, relying on a
combination of the \textsc{TMerge} and \textsc{TSub} rules to both
explore and combine the executions needed to establish the desired
coverage properties. In addition, each of the auxillary typing
relations depend on logical properties of the semantic interpretation
of types. Any effective type checking algorithm based on these rules
must address both of these issues. Our solution to the first problem
is to implement a bidirectional type
checker~\cite{BidirectionalTyping} whose type synthesis phase
characterizes a set of feasible paths and whose type checking phase
ensures those paths produce the desired results. Our solution to the
second is to encode the logical properties into a \emph{decidable}
fragment of first order logic that can be effectively discharged by an
SMT solver.

\subsection{Bidirectional Typing Algorithm}

\begin{figure}[h]
{\small
{\normalsize
\begin{flalign*}
 &\text{\textbf{Type Synthesis }} &
 \fbox{$\Gamma \vdash e \typeinfer \tau$}&&
  &\text{\textbf{Type Check }} &
 \fbox{$\Gamma \vdash e \typecheck \tau$}
\end{flalign*}
}
\\ \
\begin{prooftree}
\hypo{\Gamma \wfvdash \S{Ty}(c)}
\infer1[\textsc{SynConst}]{
\Gamma \vdash c \typeinfer \S{Ty}(c)
}
\end{prooftree}
\quad
\begin{prooftree}
\hypo{\Gamma \wfvdash \urt{b}{\bot}}
\infer1[\textsc{SynErr}]{
\Gamma  \vdash \exn \typeinfer \urt{b}{\bot}
}
\end{prooftree}
\quad
\begin{prooftree}
\hypo{\Gamma \wfvdash \urt{b}{\nu = x}}
\infer1[\textsc{SynVarBase}]{
\Gamma \vdash x \typeinfer \urt{b}{\nu = x}
}
\end{prooftree}
\\ \ \\ \ \\
\quad
\begin{prooftree}
\hypo{\Gamma(x) = a{:}\tau_a\sarr\tau_b \quad \Gamma \wfvdash a{:}\tau_a\sarr\tau_b}
\infer1[\textsc{SynVarFun}]{
\Gamma \vdash x \typeinfer  a{:}\tau_a\sarr\tau_b
}
\end{prooftree}
\quad
\mprooftr{45mm}{
    $\Gamma, x{:}\ort{b}{\top} \vdash e \typeinfer \tau$ \quad
    $\Gamma \wfvdash x{:}\ort{b}{\top}\sarr\tau$
}{ChkFun} {
    $\Gamma \vdash \lambda x{:}b.e \typeinfer x{:}\ort{b}{\top}\sarr\tau$
}
\\ \ \\
\begin{prooftree}
\hypo{
\parbox{42mm}{\center
  $\Gamma \vdash v_1 \typeinfer \tau_2\sarr\tau_x$\quad $\eraserf{\tau_2}\notin b$\quad
  $\Gamma \vdash v_2 \typecheck \tau_2$\quad
  $\Gamma' = x{:}\tau_x$\quad
$\Gamma, \Gamma' \vdash e \typeinfer \tau$ \quad
$\tau' = \existsop(\Gamma', \tau)$\quad
$\Gamma \wfvdash \tau'$
}
}
\infer1[\textsc{SynAppFun}]{
\Gamma \vdash \zlet{x}{v_1\ v_2}{e} \typeinfer \tau'
}
\end{prooftree}
\quad
\begin{prooftree}
\hypo{
\parbox{45mm}{\center
  $\Gamma \vdash v_1 \typeinfer a{:}\ort{b}{\phi}\sarr\tau_x$\quad
  $\Gamma' = a{:}\urt{b}{\nu = v_2 \land \phi}, x{:}\tau_x$\quad
  $\Gamma, \Gamma' \vdash e \typeinfer \tau$ \quad
$\tau' = \existsop(\Gamma', \tau)$\quad
$\Gamma \wfvdash \tau'$
}
}
\infer1[\textsc{SynAppBase}]{
\Gamma \vdash \zlet{x}{v_1\ v_2}{e} \typeinfer \tau'
}
\end{prooftree}
\\ \ \\
\begin{prooftree}
\hypo{
\parbox{43mm}{\center
$\S{Ty}(\I{op}) \typeinfer \overline{a_i{:}\ort{b_i}{\phi_i}}\sarr\tau_x$\quad
$\Gamma' = \overline{a_i{:}\urt{b_i}{\nu = v_i \land \phi_i}}, x{:}\tau_x$\quad
$\Gamma, \Gamma' \vdash e \typeinfer \tau$ \quad
$\tau' = \existsop(\Gamma', \tau)$\quad
$\Gamma \wfvdash \tau'$
}
}
\infer1[\textsc{SynAppOp}]{
\Gamma \vdash \zlet{x}{\I{op}\ \overline{v_i}}{e} \typeinfer \tau'
}
\end{prooftree}
\quad
\begin{prooftree}
\hypo{
\parbox{43mm}{\center
$\Gamma \vdash e_x \typeinfer \tau_x$\quad
$\Gamma' = x{:}\tau_x$\quad
$\Gamma, \Gamma' \vdash e \typeinfer \tau$\quad
$\tau' = \existsop(\Gamma', \tau)$\quad
$\Gamma \wfvdash \tau'$
}
}
\infer1[\textsc{SynLetE}]{
\Gamma \vdash \zlet{x}{e_x}{e} \typeinfer \tau
}
\end{prooftree}
\\ \ \\
\begin{prooftree}
\hypo{
\parbox{115mm}{\center
$\forall i, \S{Ty}(d_i) =  \overline{{y}{:}\ort{b_y}{\theta_y}}\sarr\urt{b}{\psi_i}$ \quad $\Gamma_i' = \overline{{y}{:}\nuut{b_y}{\theta_y}}, a{:}\urt{b}{\nu = v_a \land \psi_i}$\quad
$\Gamma,  \Gamma_i'\vdash e_i \typeinfer \tau_i$ \quad $\tau_i' = \existsop(\Gamma_i', \tau_i)$ \quad $\Gamma \wfvdash \disjop(\overline{\tau_i'}) $
}}
\infer1[\textsc{SynMatch}]{
\Gamma \vdash \match{v_a} \overline{d_i\ \overline{y} \to e_i} \typeinfer \disjop(\overline{\tau_i'})}
\end{prooftree}
\\ \ \\ \ \\
\begin{prooftree}
\hypo{
\emptyset \vdash e \typeinfer \tau \quad
\Gamma \vdash \tau <: \tau' \quad
\Gamma \wfvdash \tau'
}
\infer1[\textsc{ChkSub}]{
\Gamma \vdash e \typecheck \tau'
}
\end{prooftree}
\quad
\begin{prooftree}
\hypo{
\Gamma, x{:}\tau_x \vdash e \typecheck \tau \quad
\Gamma \wfvdash x{:}\tau_x\sarr\tau
}
\infer1[\textsc{ChkFun}]{
\Gamma \vdash \lambda x{:}\eraserf{\tau_x}.e \typecheck (x{:}\tau_x\sarr\tau)
}
\end{prooftree}
\\ \ \\
\begin{prooftree}
\hypo{
\parbox{115mm}{\center
$\forall i, \S{Ty}(d_i) =  \overline{{y}{:}\ort{b_y}{\theta_y}}\sarr\urt{b}{\psi_i}$ \quad $\Gamma_i' = \overline{{y}{:}\nuut{b_y}{\theta_y}}, a{:}\urt{b}{\nu = v_a \land \psi_i}$\quad
$\Gamma,  \Gamma_i'\vdash e_i \typeinfer \tau_i$ \quad $\tau_i' = \existsop(\Gamma_i', \tau_i)$ \quad $\Gamma \vdash \disjop(\overline{\tau_i'}) <: \tau' \quad \Gamma \wfvdash \tau'$
}}
\infer1[\textsc{ChkMatch}]{
\Gamma \vdash \match{v_a} \overline{d_i\ \overline{y} \to e_i} \typecheck \tau'}
\end{prooftree}
\\ \ \\
\begin{prooftree}
\hypo{
\parbox{120mm}{\center
$\Gamma \vdash \lambda x{:}b.\lambda f{:}(b\sarr\eraserf{\tau}).e \typecheck (x{:}\ort{b}{\phi}\sarr f{:}(x{:}\ort{b}{\nu{\prec}x \land \phi}\sarr\tau)\sarr\tau)$ \quad $\Gamma \wfvdash x{:}\ort{b}{\phi}\sarr\tau$
}}
\infer1[\textsc{ChkFix}]{
\Gamma \vdash \S{fix}f{:}(b\sarr\eraserf{\tau}).\lambda x{:}b.e \typecheck (x{:}\ort{b}{\phi}\sarr\tau)
}
\end{prooftree}
}
\caption{Bidirectional Typing Rules}\label{fig:bi-type-rules}
\end{figure}

As is standard in bidirectional type systems, our typing algorithm
consists of a type synthesis judgement
($\Gamma \vdash e \typeinfer \tau$) and a type checking judgment
($\Gamma \vdash e \typecheck \tau$) as presented in \autoref{fig:bi-type-rules}.
\cbnewadding{Most of rules are similar to the declarative typing rules, e.g., the type synthesis rule \textsc{SynVarBase} has the same premise as the declarative typing rule \textsc{TVarBase} shown in \autoref{fig:type-rules}. We use one type checking rule \textsc{ChkMatch} and one synthesis rule \textsc{SynAppBase} as examples to illustrate how we convert declarative typing rules into a typing algorithm. }

\paragraph{Typing \Code{match}}
As we saw in \autoref{sec:rules}, applying the declarative typing rule
for \Code{match} expressions typically requires first using several
other rules to get things into the right form: \textsc{TMerge} is
required to analyze and combine the types of each branch,
\textsc{TSub} is used to equip each branch with the right typing
context, and \textsc{TEq} is used to remove any local or pattern
variables from the type of a branch. Our bidirectional type system
combines all of these into the single \textsc{ChkMatch} rule shown in
\autoref{fig:bi-type-rules}. At a high level, this rule synthesizes a
type for all the branches and then ensures that, in combination, they
cover the desired type.

Similarly to other refinement type systems, when synthesizing the type
for the branch for constructor $d_i$ , we use a ghost variable
$a{:}\urt{b}{\nu = v_a \land \psi_i}$ to ensure that the types of any
pattern variables $\overline{y}$ are consistent with the parameters of
$d_i$. This strategy allows us to avoid having to apply \textsc{TSub}
to focus on a particular branch: instead, we simply infer a type for
\emph{each} branch, and then combine them using our disjunction
operation. In order for the inferred type of a branch to make sense,
we need to remove any occurrences of pattern variables or the ghost
variable $a$. To do, we use the $\existsop$ function, which
intuitively allows us to embed information from the typing context
into a type.
  This function takes as input a typing context
$\Gamma$ and type $\tau$ and produces an equivalent type in which pattern and ghost variables do not appear
free.  Finally, \textsc{ChkMatch} uses $\disjop$ to ensure that the
combination of the types of all the branches cover the required type
$\Gamma \vdash \disjop(\overline{\tau_i'}) <: \tau$.

\begin{example}
  \label{ex:typing+ex1}
  Consider how we might check that the body of the generator for
  even numbers introduced in \autoref{sec:motivation} has the
  expected type $\nuut{int}{\nu ~\Code{mod}~ 2 = 0}$:\footnote{We have
    replaced the \Code{if} from the original example with a
    \Code{match} expression, to be consistent with the syntax of
    \langname{}.}
  \begin{minted}[fontsize = \normalsize,xleftmargin=10pt, escapeinside=&&]{ocaml}
 &$\Code{int\_gen} {:} \ort{\Unit}{\top}\sarr\nuut{int}{\top} \vdash$&
    let (n: int) = int_gen() in let (b: bool) = n mod 2 == 0 in
    match b with true -> err | false -> n &$\typecheck \nuut{int}{\nu\ \Code{mod}\ 2 = 0}$&
\end{minted}
Our typing algorithm first adds the local variable $\Code{n}$ and
$\Code{b}$ to the type context, and then checks the pattern-matching
expression against the given type:
\begin{minted}[fontsize = \normalsize,xleftmargin=10pt, escapeinside=&&]{ocaml}
 &$\Code{int\_gen}{:}\ort{\Unit}{\top}\sarr\nuut{int}{\top}, \Code{n}{:}\nuut{int}{\top}, \Code{b}{:}\nuut{bool}{\nu {\iff} \Code{n}\ \Code{mod}\ 2 = 0} \vdash$&
    match b with true -> err | false -> n &$\typecheck \nuut{int}{\nu\ \Code{mod}\ 2 = 0}$&
\end{minted}
The \textsc{ChkMatch} rule first synthesizes types for the two
branches separately.  Inferring a type of the first branch using the
existing type context:
\begin{minted}[fontsize = \normalsize,xleftmargin=10pt, escapeinside=&&]{ocaml}
&$..., \Code{b}{:}\nuut{bool}{\nu {\iff} \Code{n}\ \Code{mod}\ 2 = 0}, \Code{b'}{:}\nuut{bool}{\nu = \Code{b} \land \nu} \vdash$& err &$\typeinfer \nuut{int}{\bot}$&
\end{minted}
adds a ghost variable $\Code{b'}$ to reflect the fact that $\Code{n\;mod\;2}$ must be equal to $0$ in this branch. By next applying the
\textsc{TErr} rule, our algorithm infers the type $\nuut{int}{\bot}$
for this branch. The rule next uses $\existsop$ to manifest
$\Code{b'}$ in the inferred type, encoding the path constraints under
which this type holds (i.e. $b$ is true).
\begin{minted}[fontsize = \normalsize,xleftmargin=10pt, escapeinside=&&]{ocaml}
&$...,\Code{b}{:}\nuut{bool}{\nu {\iff} \Code{n}\ \Code{mod}\ 2 = 0}, \Code{b'}{:}\nuut{bool}{\nu = \Code{b} \land \nu} \vdash$&
    err &$\typeinfer \nuut{int}{\exists \Code{b'}, \Code{b'} = \Code{b} \land \Code{b'} \land \bot}$&
\end{minted}
Thus, the synthesized type for the first branch is
$\nuut{int}{\Code{b} \land \bot}$ modulo some trivial simplification. The type of
the second branch provides a better demonstration of why $\existsop$
is needed:
\begin{minted}[fontsize = \normalsize,xleftmargin=10pt, escapeinside=&&]{ocaml}
&$...,\Code{b}{:}\nuut{bool}{\nu {\iff} \Code{n}\ \Code{mod}\ 2 = 0}, \Code{b'}{:}\nuut{bool}{\nu = \Code{b} \land \neg \nu} \vdash$&
    n &$\typeinfer \nuut{int}{\nu = \Code{n}}$&
\end{minted}
After applying $\existsop$, the inferred type is $\nuut{int}{\exists
  \Code{b'}, \Code{b'} = \Code{b} \land \neg \Code{b'} \land \nu = \Code{n}}$; after simplification, this
becomes $\nuut{int}{\neg \Code{b} \land \nu = \Code{n}}$.  The disjunction of these
two types:
\begin{minted}[fontsize = \normalsize,xleftmargin=10pt, escapeinside=&&]{ocaml}
 &$\disjop(\nuut{int}{\Code{b} \land \bot}, \nuut{int}{\neg \Code{b} \land \nu = \Code{n}}) = \nuut{int}{(\Code{b} \land \bot) \lor (\neg \Code{b} \land \nu = \Code{n})}$&
\end{minted}
 results in exactly the type shown in \autoref{sec:motivation}.
 This can be then successfully checked against the target type $\nuut{nat}{\Code{\nu}\ \Code{mod}\ 2 = 0}$.
\end{example}


\paragraph{Application} Our type synthesis rules for function
application adopt a strategy similar to \textsc{ChkMatch}'s, trying to
infer the strongest type possible for an expression that uses the
result of a function application. The rule for a function whose
parameter is an overapproximate refinement type (\textsc{SynAppBase})
is most interesting, since it has to bridge the gap with an argument
that has an underapproximate coverage type.  When typing $e$, the
expression that uses the result of the function call, the rule
augments the typing context with a ghost variable $a$. This variable
records that the coverage type of the argument must overlap with the
type expected by the function (both must satisfy the refinement
predicate $\phi$): if this intersection is empty, i.e., the type of
$a$ is equivalent to $\bot$, we will fail to infer a type for $e$, as
no type will be well-formed in this context. As with
\textsc{ChkMatch}, \textsc{SynAppBase} uses $\existsop$ to ensure that
it does not infer a type that depends on $a$.




\subsection{Auxiliary Typing Functions}


\begin{algorithm}[ht!]
  \Params{A type context $\Gamma$, and two base coverage types $\urt{b}{\phi_1}$ and $\urt{b}{\phi_2}$. } %
  \Output{A FOL formula that can be automated checked by SMT solver.  }
      \Procedure{$\subquery(\Gamma, \urt{b}{\phi_1}, \urt{b}{\phi_2}) := $}{
      \Match{$\Gamma$}{
        \Case{$\emptyset$}{
          \Return{$\forall \nu{:}b, \phi_2 \implies \phi_1$}\;
        }
        \Case{$\Gamma, x{:}(a{:}\tau_a\sarr\tau)$}{
          $\phi \leftarrow \subquery((\Gamma, \urt{b}{\phi_1}, \urt{b}{\phi_2})$\;
          \Return{$\phi$}\;
        }
        \Case{$\Gamma, x{:}\ort{b_x}{\phi_x}$}{
          $\tau_1 \leftarrow \urt{b}{\forall x{:}b_x, \phi_x[\nu\mapsto x] \impl \phi_1}$\;
          $\tau_2 \leftarrow \urt{b}{\forall x{:}b_x, \phi_x[\nu\mapsto x] \impl \phi_2}$\;
          \Return{$\subquery(\Gamma, \urt{b}{\tau_1}, \urt{b}{\tau_2})$}\;
        }
        \Case{$\Gamma, x{:}\urt{b_x}{\phi_x}$}{
          $\tau_1 \leftarrow \urt{b}{\exists x{:}b_x, \phi_x[\nu\mapsto x] \land \phi_1}$\;
          $\tau_2 \leftarrow \urt{b}{\exists x{:}b_x, \phi_x[\nu\mapsto x] \land \phi_2}$\;
          \Return{$\subquery(\Gamma, \urt{b}{\tau_1}, \urt{b}{\tau_2})$}\;
         }
        }
      }
      \caption{Subtyping Query Encoding}
      \label{algo:subtyping}
  \end{algorithm}

The auxillary $\disjop$ and $\existsop$ operations are a straightforward syntactic transformations; their
full definitions can be found in the \techreport{}.
More interesting is how
we algorithmically check well-formedness and subtyping. Our type
checking algorithm translates both obligations into logical formulae
that can be discharged by a SMT solver. Both obligations are encoded
by the $\subquery$ subroutine shown in \autoref{algo:subtyping}.
$\subquery(\Gamma, \urt{b}{\phi_1}, \urt{b}{\phi_2})$ encodes the
bindings in $\Gamma$ in the typing context from right to left, before
checking whether $\phi_1$ implies $\phi_2$.  Variables with function
types, on the other hand, are omitted entirely, as qualifiers cannot
have function variables in FOL. Variables with an overapproximate
(underapproximate) type are translated as a universally (existential)
quantified variable, and are encoded into the refinement of both
coverage types.

\begin{example} Consider the subtyping obligation generated by
  Example~\ref{ex:typing+ex1} above:
\begin{minted}[fontsize = \small,xleftmargin=10pt, escapeinside=&&]{ocaml}
  &$\Code{int\_gen}{:}\ort{\Unit}{\top}\sarr\nuut{int}{\top},  \Code{n}{:}\nuut{int}{\top}, \Code{b}{:}\nuut{bool}{\nu {\iff} \Code{n}\ \Code{mod}\ 2 = 0} \vdash$&
  &$\nuut{int}{(b \land \bot) \lor (\neg b \land \nu = n)} <: \nuut{int}{\nu \geq 0}$&
\end{minted}
\noindent
This obligation is encoded by the following call to $\subquery$
{\small
\begin{align*}
  &\;\subquery( \Code{int\_gen}{:}\ort{\Unit}{\top}\sarr\nuut{int}{\top},
              \Code{n}{:}\nuut{int}{\top}, \Code{b}{:}\nuut{bool}{\nu
              {\iff} \Code{n}\ \Code{mod}\ 2 = 0})\\
            &\qquad\qquad\quad\nuut{int}{(\Code{b} \land \bot) \lor (\neg \Code{b} \land \nu = \Code{n})},
              \nuut{int}{\nu \geq 0}) \\
  \equiv &\;\subquery( \Code{int\_gen}{:}\ort{\Unit}{\top}\sarr\nuut{int}{\top},
              \Code{n}{:}\nuut{int}{\top}, \\
            &\qquad\qquad\quad\nuut{int}{\exists \Code{b}, \Code{b}{\iff} \Code{n}\ \Code{mod}\ 2 = 0 \land (\Code{b} \land \bot) \lor (\neg \Code{b} \land \nu = \Code{n})},\\
            &\qquad\qquad\quad\nuut{int}{\exists \Code{b}, \Code{b}{\iff}\Code{n}\ \Code{mod}\ 2 = 0\land \nu \geq 0}) \\
  \equiv &\;\subquery( \Code{int\_gen}{:}\ort{\Unit}{\top}\sarr\nuut{int}{\top}, \\
            &\qquad\qquad\quad\nuut{int}{\exists \Code{n}, \top \land \exists \Code{b}, \Code{b}{\iff}\Code{n}\ \Code{mod}\ 2 = 0 \land (\Code{b} \land \bot) \lor (\neg \Code{b} \land \nu = \Code{n})},\\
            &\qquad\qquad\quad\nuut{int}{\exists \Code{n}, \top \land \exists \Code{b}, \Code{b}{\iff} \Code{n}\ \Code{mod}\ 2 = 0
              \land \nu \geq 0}) \\
  \equiv &\;\subquery(\emptyset, \nuut{int}{\exists \Code{n}, \top \land \exists \Code{b}, \Code{b} {\iff}\Code{n}\ \Code{mod}\ 2 = 0 \land (\Code{b} \land \bot) \lor (\neg \Code{b} \land \nu = \Code{n})},\\
            &\qquad\qquad\quad\quad \nuut{int}{\exists \Code{n}, \top \land \exists \Code{b}, \Code{b}{\iff} \Code{n}\ \Code{mod}\ 2 = 0
              \land \nu \geq 0}) \\
    \equiv &\; \forall \nu, \exists \Code{n}, \top \land \exists \Code{b}, \Code{b} {\iff} \Code{n}\ \Code{mod}\ 2 = 0 \land \nu \geq 0)\implies\\
            &\quad\;\;\;\; \exists \Code{n}, \top \land \exists \Code{b}, \Code{b}{\iff} \Code{n}\ \Code{mod}\ 2 = 0 \land (\Code{b} \land \bot) \lor (\neg \Code{b} \land \nu = \Code{n})
\end{align*}
}\noindent
This is equivalent to formula (\ref{frm:subtype+ex}) from \autoref{sec:motivation}:
\begin{minted}[fontsize = \small,xleftmargin=10pt, escapeinside=&&]{ocaml}
 &$\forall \nu, (\nu \geq 0)\impl (\exists \Code{n}, \exists \Code{b}, \Code{b} {\iff} \Code{n}\ \Code{mod}\ 2 = 0 \land (\Code{b} \land \bot) \lor (\neg \Code{b} \land \nu = \Code{n}))$&
\end{minted}
\end{example}

Using $\subquery$, it is straightforward to discharge well-formedness
and subtyping obligations using the rules shown in
\autoref{fig:aux-rules-algo}. In the case of \textsc{WfBase}, for
example, observe
that the error term $\exn$ is always an inhabitant of the type
$\urt{b}{\bot}$ for arbitrary base type $b$. Thus, to check the last
assumption of \textsc{WfBase}, it suffices to iteratively check if any
coverage types in the type context are a supertype of their associated
bottom type.

\begin{figure}
  \centering
{\small
\begin{prooftree}
\hypo{
  \not\models \subquery(\Gamma,\urt{b}{\bot}, \urt{b}{\phi})
}
\infer1{
  \exn \not\in \denotation{\urt{b}{\phi}}_{\Gamma}
}
\end{prooftree}
\quad
\begin{prooftree}
\hypo{
  \models \subquery(\Gamma,\urt{b}{\phi_1}, \urt{b}{\phi_2})
}
\infer1{
\Gamma \vdash \urt{b}{\phi_1} <: \urt{b}{\phi_2}
}
\end{prooftree}
\\ \ \\ \ \\ \
\begin{prooftree}
\hypo{
  \models \subquery(\Gamma,\urt{b}{\phi_2}, \urt{b}{\phi_1})
}
\infer1{
\Gamma \vdash \ort{b}{\phi_1} <: \ort{b}{\phi_2}
}
\end{prooftree}
}

\caption{Auxiliary Algorithmic Typing Functions}\label{fig:aux-rules-algo}
\end{figure}

Discharging subtyping obligations is slightly more involved, as we
need to ensure that the formulas sent to the SMT solver are
decidable. Observe that in order to produce effectively decidable
formulas, the encoding strategy realized by $\subquery$ always
generates a formula of the form
$\forall\overline{x}. \exists\overline{y}. \phi$, i.e. it does not
allow for arbitrary quantifier alternations. To ensure that this is
sound strategy, we restrict all overapproximate refinement types in a
type context to not have any free variables that have a coverage
type. This constraint allows us to safely lift all universal
quantifiers to the top level, thus avoiding arbitrary quantifier
alternations.

As an example of a scenario disallowed by this restriction, consider
the following type checking judgment:
\begin{align*}
x{:}\nuut{nat}{\nu > 0} \vdash \lambda \Code{y}:nat. \,  \Code{x + y} \typecheck y{:}\nuot{nat}{\nu > x + 1}{\;\shortrightarrow\;}\nuut{nat}{\phi}
\end{align*}
This judgment produces the following subtyping check:
\begin{align*}
x{:}\nuut{nat}{\nu > 0}, y {:} \nuot{nat}{\nu > x + 1} \vdash \nuut{nat}{\nu = x + y} <: \nuut{nat}{\phi}
\end{align*}
where the normal refinement type $\nuot{nat}{\nu > x + 1}$ in the type context has free variable $x$ that has coverage type.
Evaluating this judgment entails solving the formula:
\begin{align*}
\forall \nu, (\exists x, x > 0 \land (\forall y, y > x + 1 \impl \phi)) \implies (\exists x, x > 0 \land (\forall y, y > x + 1 \impl \nu = x + y))
\end{align*}
which is not decidable due to the quantifier alternation
$\forall\nu\exists x\forall y$.

\begin{theorem}\label{theorem:algo-sound}[Soundness of Algorithmic Typing] For all type context
  $\Gamma$, term $e$ and coverage type $\tau$,
  $\Gamma \vdash e \typecheck \tau {\implies} \Gamma \vdash e
  : \tau$
\end{theorem}

\begin{theorem}\label{theorem:algo-complete}[Completeness of Algorithmic Typing] Assume an oracle
  for all formulas produced by the $\subquery$ subroutine.  Then for any type context
  $\Gamma$, term $e$ and coverage type $\tau$, $\Gamma \vdash e: \tau {\implies} \Gamma \vdash e \typecheck \tau$.
  \end{theorem}

\section{Polymorphic Coverage Type}\label{sec:poly}


\begin{figure}
    \centering
    \begin{minted}[xleftmargin=5pt, numbersep=4pt, linenos = true, fontsize = \small, escapeinside=??]{OCaml}
    val union: (unit -> 'a) -> (unit -> 'a) -> (unit -> 'a)
    let union g1 g2 = fun () -> (g1 () ?$\oplus$? g2 ())

    val pos_gen: unit -> int (* positive number generator *)
    val neg_gen: unit -> int (* negative number generator *)
    
    let bool_gen = union (fun () -> true) (fun () -> false)
    let nat_gen = union pos_gen (fun () -> 0)
    let int_gen = union neg_gen nat_gen
    \end{minted}
    \caption{A $\Code{union}$ combinator example.}
    \label{fig:poly-example}
\end{figure}

Most mainstream PBT frameworks provide a set of combinators that enable users to compositionally build test generators. For example, the \Code{QCheck} framework provides a $\Code{union}$ combinator shown in \autoref{fig:poly-example} that produces a new generator which covers the combined test cases of two test input generators.
Unlike $\oplus$ which takes two base values as arguments, the $\Code{union}$ combinator combines two input generators, defined as function values. As shown in \autoref{fig:poly-example}, the $\Code{union}$ combinator is polymorphic and cannot be assigned a (principal) type in the monomorphic coverage type system we have discussed so far. As a consequence, we need to extend this type system in order to, e.g., type the simple generator on lines 7-9. 




To support this flexibility, we extend this type system to align with how OCaml implements polymorphism. Thus, a polymorphic term is assigned a \emph{principal} type (i.e., a most general type) that includes universally quantified type variables like $\tvar{a}$.

\begin{example}\label{ex:oplus} In this extended type system, the non-deterministic choice operator $\oplus$ is now assigned the following polymorphic function coverage type:
\begin{align*}
    \tau_\oplus \doteq x{:}\ort{\tvar{a}}{\top} \sarr y{:}\ort{\tvar{a}}{\top} \sarr \urt{\tvar{a}}{\vnu = x \lor \vnu = y}
\end{align*}\noindent
The type $\tau_\oplus$ imposes no constraint on its two parameters $x$ and $y$, and guarantees the return value ($\vnu$) must cover both (i.e., $\vnu = x \lor \vnu = y$).
\end{example}

While type polymorphism is sufficient to capture the basic behavior of generator combinators, it falls short in a type system, where we aim to specify and provide coverage guarantees. Consider lines 8-9 in \autoref{fig:poly-example}: the $\Code{union}$ combinator combines a generator that produces only positive integers ($\Code{pos\_gen}$) with the constant generator that always returns $0$ ($\Code{\mykw{fun}\ ()\ \to\ 0}$), to build a natural number generator ($\Code{nat\_gen}$). It is then used again to create an integer generator ($\Code{int\_gen}$) from a negative number generator ($\Code{neg\_gen}$) and the aforementioned natural number generator. Thus, $\Code{union}$ takes test generators with \emph{arbitrary} coverage guarantees, and build generator whose coverage guarantees combines those guarantees.

Our goal is to assign the $\Code{union}$ combinator a principal coverage type $\tau_{\Code{union}}$ that reflects that it builds a generator that returns a superset of the values produced by its arguments. This guarantee is similar to that of the non-deterministic choice operator $\oplus$ shown in Example~\ref{ex:oplus}; but because the $\Code{union}$ combinator takes functions as parameters, the analog of the qualifier ``$\vnu = x \lor \vnu = y$'' for the $\Code{union}$ combinator cannot be expressed in our current qualifier language. One potential solution is to add union and intersection types (e.g., $\tau \sqcap \tau$ and $\tau \sqcup \tau$) into our type system, but this strategy will not work for more complicated combinators like $\Code{bind}$, $\Code{frequency}$, and $\Code{fix}$ shown in \autoref{fig:qcheck-example}.

To address this, we draw inspiration from other variants of refinement type systems (\citet{ParamRelationalType} and \citet{AbstractRefinementType}) and introduce \emph{qualifier polymorphism} to allow type to be parameterized over \emph{predicate variables}.
With this extension, the principal type of $\tau_{\Code{union}}$ becomes: 
\begin{align*}
    \tau_{\Code{union}} \doteq \forall P_1\, P_2 : \tvar{a} \sarr \Bool.\, &(\Unit \sarr \urt{\tvar{a}}{P_1(\vnu)}) \sarr (\Unit \sarr \urt{\tvar{a}}{P_2(\vnu)}) \sarr \\
    &(\Unit \sarr \urt{\tvar{a}}{P_1(\vnu) \lor P_2(\vnu)})
\end{align*}\noindent
Here, $P_1$ and $P_2$ are universally quantified predicate variable of type $\tvar{a} \sarr \Bool$. This type states that ``for all possible qualifiers $P_1(\vnu)$ and $P_2(\vnu)$, if the two input generators cover $P_1(\vnu)$ and $P_2(\vnu)$ respectively, then the generator that results from combining them with a union operation covers $P_1(\vnu) \lor P_2(\vnu)$''.


Much like type variables, predicate variable must appear at the front of the type and be instantiated with concrete qualifiers during function application. For instance, when applying $\Code{union}$ combinator on line $8$ in \autoref{fig:poly-example}, we instantiate the predicate variables according to the following assignment:
\begin{align*}
    P_1 \mapsto \lambda \vnu.0 < \vnu \quad \text{and} \quad P_2 \mapsto \lambda \vnu.\vnu = 0
\end{align*}\noindent
This allows us to assign $\Code{union}$ combinator the following coverage type:
\begin{align*}
    &\tau_{\Code{union}_1} \doteq (\Unit\sarr\urt{\Int}{0 < \vnu}) \sarr (\Unit\sarr\urt{\Int}{\vnu = 0}) \sarr (\Unit\sarr\urt{\Int}{0 \leq \vnu }) 
\end{align*}\noindent
The rest of program can be checked using this type, which align the types of 
 $\Code{neg\_gen}$ and $\Code{\mykw{fun}\ ()\ \to\ 0}$. 
Our extended bidirectional typing rules can automatically infer predicate variable assignments from the qualifier in the corresponding arguments using a simple form of syntactic unification, which we discuss below.

\subsection{Extended Type Syntax}

\begin{figure}[t!]
    \begin{alignat*}{2}
    \text{\textbf{Type Variable}}& \quad &  \quad &  \textcolor{DeepGreen}{\tvar{a}, \tvar{b}, \tvar{c}, ...} \\
    \text{\textbf{Base Types}}& \quad & b  ::= \quad &  \textcolor{DeepGreen}{\tvar{a}} ~|~ \Code{unit} ~|~ \Code{bool} ~|~ \Code{nat} ~|~ \Code{int} ~|~  b\ \Code{list} ~|~ b\ \Code{tree} ~|~ \ldots \\
    \text{\textbf{Basic Types}}& \quad & t  ::= \quad &  b ~|~ t\sarr t ~|~ \textcolor{DeepGreen}{\tvar{a}.t} \\
    \text{\textbf{Predicate Variable}}& \quad &\quad &  \textcolor{DeepGreen}{P, ...} \\
    \text{\textbf{Predicates}}& \quad & \I{pred} ::=  \quad & \textcolor{DeepGreen}{P ~|~ \overline{\lambda x}.\phi} \\
    \text{\textbf{Qualifier}}& \quad & \phi ::= \quad &  l ~|~ \bot ~|~ \top  ~|~ \I{op}(\overline{l}) ~|~ \I{uf}(\overline{x}) ~|~ \neg \phi ~|~ \phi \land \phi ~|~ \phi \lor \phi ~|~ \phi \impl \phi \\
    &\quad&\quad& |~  \forall u{:}b.\; \phi ~|~ \exists u{:}b.\; \phi ~|~ \textcolor{DeepGreen}{ \I{pred}(\overline{l})} \\
    \text{\textbf{Refinement Types}}& \quad &\tau  ::= \quad & \urt{b}{\phi} ~|~ \ort{b}{\phi} ~|~ x{:}\tau\sarr\tau ~|~ \textcolor{DeepGreen}{\tvar{a}.\tau ~|~ \forall P{:}(\overline{b}\sarr \Bool).\tau} \\
    \text{\textbf{Type context}}& \quad &\Gamma  ::= \quad & \emptyset ~|~ \Gamma, x{:}\tau ~|~ \textcolor{DeepGreen}{\Gamma, P{:}t ~|~ \Gamma, \tvar{a}}
  \end{alignat*}
    \caption{Extended \langname{} syntax. }\label{fig:extended-syntax}
\end{figure}

The syntax of our extended type system is presented in \autoref{fig:extended-syntax}, with new additions highlighted in green.

\paragraph{Basic Types}  
Following the conventions of standard polymorphic type systems~\cite{TAPL}, the basic type system of \langname{} supports both type variables ($\tvar{a}$) and type quantification ($\tvar{a}.t$). The type variable $\tvar{a}$ in our type system can be used in refinement types $\urt{\tvar{a}}{\phi}$, and thus can only be instantiated with base type ($b$). As in OCaml, type and predicate quantifiers are omitted when the context is clear, e.g., $\tvar{a}.\forall P{:}\tvar{a}\sarr\Bool.\Unit\sarr\urt{\tvar{a}}{P(\vnu)}$ is written as $\Unit\sarr\urt{\tvar{a}}{P(\vnu)}$.

\paragraph{Qualifiers and Refinement Types}
In addition to supporting type quantification (e.g., $\tvar{a}.\tau$), coverage refinement type also allow quantified \emph{predicate variables} $P$.
As is standard, \emph{predicates} are functions (e.g., $\overline{\lambda x}.\phi$) whose parameters must be base types arguments and which return boolean values ($\overline{b} \sarr \Bool$).
Fully applied predicate variables may appear in qualifiers, like $P(\overline{l})$, analogous to primitive operators ($\I{op}$). Predicate application $(\overline{\lambda x}.\phi)(\overline{l})$ is simplified as $\phi\overline{[x\mapsto l]}$. Notably, the extension of the qualifier language yields  both universal (for type-checking polymorphic functions) and existential (for type-checking polymorphic function application) quantified type variables and predicate variables. Our subtyping rules instantiate these variables such that derived verification conditions only contain universal quantified variables. Type and predicate variables are encoded as \emph{uninterpreted sorts} and \emph{uninterpreted predicates} in the verification conditions generated by our type system, ensuring that type-checking remains  within the decidable fragment of first order logic. 


\paragraph{Type context and type erasure function}
To support these extensions, the type context $\Gamma$ is extended to include type variables ($\Gamma, \tvar{a}$) as well as predicate variables ($\Gamma, P{:}t$). The updated type erasure function $\eraserf{\tau}$ preserves all quantified type variables and discards all quantified predicate variables.

\subsection{Extended Typing Rules}

\begin{figure}[!t]
{\small
{\normalsize
\begin{flalign*}
&\text{\textbf{Well-Formedness }} &
 \fbox{$\Gamma \wfvdash t$ \quad $\Gamma \wfvdash \tau$ \quad $\wfvdash \Gamma$}
\end{flalign*}
}
\begin{prooftree}
\hypo{
\parbox{40mm}{\centering
$\mathbf{FreeTypeVar}(t) \subseteq \Gamma $}}
\infer1[\textsc{WfBasicType}]{
  \Gamma \wfvdash t
}
\end{prooftree}
\quad
\mprooftr{40mm}
{$\wfvdash \Gamma$ \quad $\eraserf{\Gamma}, \vnu{:}b \basicvdash \phi : \Bool$ \quad $\textcolor{DeepGreen}{\Gamma \wfvdash b}$}
{WfBase}
{$\Gamma \wfvdash \urt{b}{\phi}$}
\\ \ \\ \ 
\begin{prooftree}
\hypo{
\parbox{40mm}{\centering
$\Gamma, \overline{\tvar{a}} \wfvdash \tau$\quad 
\text{no type quantification in $\tau$}}}
\infer1[\textsc{WfPolyType}]{
  \Gamma \wfvdash \overline{\tvar{a}}.\tau
}
\end{prooftree}
\quad
\begin{prooftree}
\hypo{
\parbox{40mm}{\centering
$\Gamma, \overline{P{:}\overline{b}\sarr\Bool} \wfvdash \tau$\quad 
\text{no predicate quantification in $\tau$}}}
\infer1[\textsc{WfPolyPred}]{
\Gamma \wfvdash \overline{\forall P{:}\overline{b}\sarr\Bool}.\tau
}
\end{prooftree}
\\ \
{\normalsize
\begin{flalign*}
 &\text{\textbf{Subtyping }} & \fbox{$\Gamma \vdash \tau_1 <: \tau_2$}
\end{flalign*}
}
\begin{prooftree}
\hypo{\Gamma \wfvdash \tau[\tvar{a} \mapsto b]}
\infer1[\textsc{SubPolyType}]{
\Gamma \vdash \tvar{a}.\tau <: \tau[\tvar{a} \mapsto b]
}
\end{prooftree}
\quad
\begin{prooftree}
\hypo{
\Gamma, \overline{x{:}b} \wfvdash \phi
}
\infer1[\textsc{SubPolyPred}]{
\Gamma \vdash \forall P: \overline{b}\sarr\Bool.\tau <: \tau[P\mapsto\overline{\lambda x}.\phi]
}
\end{prooftree}
{\normalsize
\begin{flalign*}
 &\text{\textbf{Typing }} & \fbox{$\Gamma \vdash e : \tau$}
\end{flalign*}
}
\\ \
\mprooftr{38mm}{
$\Gamma, \tvar{a} \vdash e : \tau$ \quad $\Gamma \wfvdash \tau$}
{TPolyType}{
  $\Gamma \vdash e : \tvar{a}.\tau$
}
\quad
\mprooftr{48mm}{
$\Gamma, P{:}\overline{b}\sarr\Bool \vdash e : \tau$ \quad $\Gamma \wfvdash \tau$}
{TPolyPred}{
  $\Gamma \vdash e : \forall P{:}\overline{b}\sarr\Bool.\tau$
}
}
\caption{Extended typing relations}\label{fig:extended-typing}
\end{figure}

The introduction of polymorphism induces the three new sets of typing rules shown in \autoref{fig:extended-typing}. The additional well-formedness \textsc{WfPolyType} and \textsc{WfPolyPred} rules ensure that refinement types are in prenex normal form, i.e., all quantifiers appear at the front of types. The subtyping extensions are also straightforward: the \textsc{SubPolyType} rule permits instantiating type variables with any valid base type, while the \textsc{SubPolyPred} rule allows a predicate variable $P$ to be instantiated with any well-formed predicate $\overline{\lambda x}.\phi$. Finally, we have two new declarative typing rules (\textsc{TPolyType} and \textsc{TPolyPred}) that move type and predicates quantifiers from the type into the type context.

\paragraph{Type Denotation and Soundness}  
The extended denotational semantics for types is:
\begin{align*}
    \denot{\tvar{a}.\tau} &\doteq \bigcup_{b} \denot{\tau[\tvar{a} \mapsto b]} \quad \text{where $b$ contains no type variables} \\
    \denot{\forall P{:}(\overline{b}\sarr\Bool).\tau} &\doteq \bigcup_{\phi} \denot{\tau[P \mapsto \overline{\lambda x}.\phi]} \quad \text{where } \overline{x{:}b} \basicvdash \phi : \Bool
\end{align*}\noindent
The fundamental theorem and overall type soundness remain unchanged.

\subsection{Extended Typing Algorithm}

\begin{figure}[h]
{\small
{\normalsize
\begin{flalign*}
 &\text{\textbf{Type Synthesis }} &
 \fbox{$\Gamma \vdash e \typeinfer \tau$}&&
  &\text{\textbf{Type Check }} &
 \fbox{$\Gamma \vdash e \typecheck \tau$}
\end{flalign*}
}
\\ \
\mprooftr{35mm}{
$\Gamma, \tvar{a} \typecheck e : \tau$ \quad $\Gamma \wfvdash \tau$}
{ChkPolyType}{
  $\Gamma \vdash e \typecheck \tvar{a}.\tau$
}
\quad
\mprooftr{45mm}{
$\Gamma, p{:}\overline{b}\sarr\Bool \typecheck e : \tau$ \quad $\Gamma \wfvdash \tau$}
{ChkPolyPred}{
  $\Gamma \vdash e \typecheck \forall p{:}\overline{b}\sarr\Bool.\tau$
}
\\ \
\begin{prooftree}
\hypo{
\parbox{110mm}{\center
  \adjustbox{margin=2pt {\fboxsep},bgcolor=LightGrey}{
  $\Gamma \vdash v_1 \typeinfer \tau_1$\quad 
  $\Gamma \vdash v_2 \typeinfer \tau_2$\quad 
  $\eraserf{\tau_2} \not\in b$\quad
  $\instop(\Gamma, \tau_1, \tau_2) = a{:}\tau_2'\sarr\tau_x$\quad
  $\Gamma \vdash \tau_2 <: \tau_2'$}\\
  $\Gamma' = x{:}\tau_x$\quad
$\Gamma, \Gamma' \vdash e \typeinfer \tau$ \quad
$\tau' = \existsop(\Gamma', \tau)$\quad
$\Gamma \wfvdash \tau'$
}
}
\infer1[\textsc{SynAppFun}]{
\Gamma \vdash \zlet{x}{v_1\ v_2}{e} \typeinfer \tau'
}
\end{prooftree}
\\ \
\begin{prooftree}
\hypo{
\parbox{110mm}{\center
  \adjustbox{margin=2pt {\fboxsep},bgcolor=LightGrey}{
  $\Gamma \vdash v_1 \typeinfer \tau_1$\quad
  $\Gamma \vdash v_2 \typeinfer \urt{b}{\phi_2}$\quad
  $\instop(\Gamma, \tau_1, \urt{b}{\phi_2}) = a{:}\ort{b}{\phi}\sarr\tau_x$}\quad
  $\Gamma' = a{:}\urt{b}{\nu = v_2 \land \phi}, x{:}\tau_x$\quad
  $\Gamma, \Gamma' \vdash e \typeinfer \tau$ \quad
$\tau' = \existsop(\Gamma', \tau)$\quad 
$\Gamma \wfvdash \tau'$
}
}
\infer1[\textsc{SynAppBase}]{
\Gamma \vdash \zlet{x}{v_1\ v_2}{e} \typeinfer \tau'
}
\end{prooftree}
}
\caption{Extended bidirectional Typing Rules}\label{fig:extended-bi-type-rules}
\end{figure}

The extended bidirectional typing algorithm is shown in \autoref{fig:extended-bi-type-rules}. We introduce two new rules \textsc{ChkPolyType} and \textsc{ChkPolyPred} to type-check polymorphic terms, which are similar to \textsc{TPolyType} and \textsc{TPolyPred} from \autoref{fig:extended-typing}. More challenging is, function application, which in the declarative type system relies on the auxiliary \textsc{SubPolyType} and \textsc{SubPolyPred} subtyping rules to instantiate both type variables and predicate variables in polymorphic refinement types.
We modify the previous typing algorithm for function application (i.e., \textsc{SynAppFun} and \textsc{SynAppBase}) as shown in \autoref{fig:extended-bi-type-rules}, where modified parts are colored in gray. These two rules first infer a function type ($v_1$) and type for the corresponding argument ($v_2$), then invoke the instantiation subroutine $\instop$, which instantiates the polymorphic function type with respect to the argument type (e.g., instantiates $\tau_1$ on $\tau_2$ in the rule \textsc{SynAppFun}). Since we perform type synthesis instead of type checking for the function $v_1$ in \textsc{SynAppFun}, we additionally inline the rule \textsc{ChkSub} to ensure  the argument type $\tau_2$ is a subtype of the instantiated parameter type $\tau_2'$.
The output of the $\instop$ subroutine also ensures that the resulting parameter type (e.g., $\ort{b}{\phi}$ in \textsc{SynAppBase}) contains no free predicate variables.
This guarantees that the remaining typing rules can be directly copied from the original typing algorithm.

\begin{algorithm}[t!]
\Params{A type context $\Gamma$, a function type $\forall\overline{P{:}t}.x{:}\tau_\Code{param}\sarr\tau_\Code{ret}$, and an argument type $\tau_2$}
\Output{the instantiated function type}
    \Procedure{$\instop(\Gamma, \forall\overline{P{:}t}.x{:}\tau_\Code{param}\sarr\tau_\Code{ret}, \tau_2) := $}{
                $\overline{P{:}t} \leftarrow \overline{P{:}t} \setminus \mathbf{FreePredVar}(\tau_\Code{param})$\;
                $\sigma \leftarrow \Code{UnifyType}(\Gamma, \emptyset, [(\tau_\Code{param},\tau_2)])$\;
                $x{:}\tau_\Code{param}\sarr\tau_\Code{ret} \leftarrow \sigma(x{:}\tau_\Code{param}\sarr\tau_\Code{ret})$\;
                \Return{$x{:}\tau_\Code{param}\sarr(\forall\overline{P{:}t}.\tau_\Code{ret})$}\;
    }
    \caption{Predicate Variable Instantiation}
    \label{algo:instantiate}
\end{algorithm}

As shown in \autoref{algo:instantiate}, the $\instop$ subroutine assumes that type qualifiers (e.g., $\tvar{a}.\tau$) have been eliminated, and divides the input function type $\tau_1$ into three parts: quantified predicates $\overline{P{:}t}$, parameter type $\tau_\Code{param}$, and the return type $\tau_\Code{ret}$. Newly quantified predicates are updated as shown on line $2$ to guarantee that all predicate variables appearing in $\tau_\Code{param}$ (i.e., $\mathbf{FreePredVar}(\tau_\Code{param})$) are instantiated. The algorithm then uses the unification subroutine to obtain a predicate assignment $\sigma$ that ensures  predicate variables within the parameter type $\tau_\Code{param}$ are consistent with the argument type $\tau_2$ (line $3$). After substituting predicate variables with the corresponding assignments (line $4$), $\instop$ subroutine then returns an instantiated function type (line $5$) whose return type is properly quantified.

\begin{algorithm}[t!]
\Params{A type context $\Gamma$, an initial predicate variable assignment $\sigma$, and a list of type alignment constraints $l$}
\Output{A predicate variable assignment that satifies all input constraints.}
    \Procedure{$\Code{UnifyType}(\Gamma, \sigma, l) := $}{
    \Match{$l$}{
        \lCase{$[]$}{ \Return{$\sigma$} }
        \Case{$(\ort{b}{\phi_1}, \ort{b}{\phi_2}) :: l$ or $(\urt{b}{\phi_1}, \urt{b}{\phi_2}) :: l$}{
            \If{$\mathbf{FreePredVar}(\phi_1) \setminus \mathbf{Dom}(\Gamma) \not= \emptyset$}{
                $\sigma_q = \Code{UnifyQualifier}(\Gamma, \emptyset, [\phi_1, \phi_2])$\;
                $\sigma = \sigma_q(\sigma)$\;
                $l = \sigma_q(l)$\;
                \Return{$\Code{UnifyType}(\Gamma, \sigma\cup\sigma_q, l)$}\;
            }\Else{
                \Return{$\Code{UnifyType}(\Gamma, \sigma, l)$}\;
            }
        }
        \Case{$(x{:}\tau_{11}\sarr\tau_{12}, x{:}\tau_{21}\sarr\tau_{22}) :: l$}{
            \Return{$\Code{UnifyType}(\sigma, (\tau_{11}, \tau_{21}) :: (\tau_{21}, \tau_{22}) :: l)$}\;
        }
        \Case{$(\tau_1, \tau_2) :: l$}{
            \textbf{Raise Failure}\;
          }
        }
    }

\Params{A type context $\Gamma$, an initial predicate variable assignment $\sigma$, and a list of qualifier alignment constraints $l$}
\Output{A predicate variable assignment that satifies all input constraints.}
    \Procedure{$\Code{UnifyQualifier}(\Gamma, \sigma, l) := $}{
    \Match{$l$}{
        \lCase{$[]$}{
            \Return{$\sigma$}
        }
        \Case{$(P(\overline{x}), \phi) :: l$}{
            $\sigma_q = \{P \mapsto \overline{\lambda x}.\phi\}$\;
            $\sigma = \sigma_q(\sigma)$\;
            $l = \sigma_q(l)$\;
            \Return{$\Code{UnifyQualifier}(\Gamma, \sigma\cup\sigma_q, l)$}\;
        }
        \Case{$(\phi_1 \lor \phi_2, \phi_1' \lor \phi_2') :: l$}{
            \Return{$\Code{UnifyQualifier}(\Gamma, \sigma, (\phi_1, \phi_1') :: (\phi_2, \phi_2') :: l)$}\;
        }
        \Case{$(\phi_1 \land \phi_2, \phi_1' \land \phi_2') :: l$}{
            \Return{$\Code{UnifyQualifier}(\Gamma, \sigma, (\phi_1, \phi_1') :: (\phi_2, \phi_2') :: l)$}\;
        }
        \Case{$(\phi_1 \impl \phi_2, \phi_1' \impl \phi_2') :: l$}{
            \Return{$\Code{UnifyQualifier}(\Gamma, \sigma, (\phi_1, \phi_1') :: (\phi_2, \phi_2') :: l)$}\;
        }
        \Case{$(\neg\phi, \neg\phi') :: l$}{
            \Return{$\Code{UnifyQualifier}(\Gamma, \sigma, (\phi, \phi') :: l)$}\;
        }
        \Case{$(\forall x.\phi, \forall x.\phi') :: l$}{
            \Return{$\Code{UnifyQualifier}(\Gamma, \sigma, (\phi, \phi') :: l)$}\;
        }
        \Case{$(\exists x.\phi, \exists x.\phi') :: l$}{
            \Return{$\Code{UnifyQualifier}(\Gamma, \sigma, (\phi, \phi') :: l)$}\;
        }
        \Case{$(\phi, \phi') :: l$}{
            \textbf{Raise Failure}\;
        }
        }
    }
    \caption{Predicate Unification}
    \label{algo:unification}
\end{algorithm}

The unification algorithm is shown in \autoref{algo:unification} and consists of two subroutines $\Code{UnifyType}$ and $\Code{UnifyQualifier}$ that collectively unify refinement type and qualifiers in similar manner to the standard type unification algorithm~\cite{TAPL}. The $\Code{UnifyType}$ subroutine takes an initial predicate assignment solution ($\sigma$) and a list of type constraints ($l$). When encountering a base refinement type containing a free predicate variable that is not included in the type context (line $5$), the algorithm invokes the qualifier unification subroutine (line $6$) to get a predicate assignment $\sigma_q$. It then unions the predicate assignment and continues to process the remaining constraints after substitution (line $6$ and $7$). A function type is  unified piecewise (line $12-13$); otherwise, a mismatched case leads to a unification error (line $15$). The $\Code{UnifyQualifier}$ subroutine works similarly but recursively processes over the syntactic structure of the qualifier instead.

\begin{example} As shown on line $5$ in \autoref{fig:poly-example}, type-checking the application $\Code{union\;neg\_gen\;nat\_gen}$ requires instantiating the quantified types predicates of $\tau_\Code{union}$ based on its negative and natural number generators arguments. 
\begin{align*}
    &\Code{union}:(\Unit\sarr\urt{\Int}{P_1(\vnu)}) \sarr
    \\&\hspace{1.16cm}(\Unit\sarr\urt{\Int}{P_2(\vnu)}) \sarr (\Unit\sarr\urt{\Int}{P_1(\vnu) \lor P_2(\vnu)}),
    \\&\Code{neg\_gen}:\Unit\sarr\urt{\Int}{\vnu < 0}, \Code{nat\_gen}:\Unit\sarr\urt{\Int}{0 \leq \vnu} \vdash
    \\&\quad \Code{(union\;neg\_gen)\;nat\_gen} \typeinfer ?
\end{align*}\noindent
We type check the inner application $\Code{(union\;neg\_gen)}$ first. The rule \textsc{SynAppFun} uses the $\instop$ subroutine to unify the parameter type $\Unit\sarr\urt{\Int}{P_1(\vnu)}$ against $\Unit\sarr\urt{\Int}{\vnu < 0}$. The predicate assignment after unification is $P_1\mapsto \lambda \vnu.\vnu < 0$, and the synthesized type of the application $\Code{(union\;neg\_gen)}$ is
\begin{align*}
    &(\Unit\sarr\urt{\Int}{P_2(\vnu)}) \sarr (\Unit\sarr\urt{\Int}{\vnu < 0 \lor P_2(\vnu)})
\end{align*}\noindent
Now we can type check the second application $\Code{(union\ neg\_gen)\ nat\_gen}$ in a similar way, where the $\instop$ subroutine unifies the parameter type $\Unit\sarr\urt{\Int}{P_2(\vnu)}$ against $\Unit\sarr\urt{\Int}{0 \leq \vnu}$. The predicate assignment after unification is $P_2 \mapsto \lambda \vnu.0\leq \vnu$, and the coverage type of the final application is $\Unit\sarr\urt{\Int}{\vnu < 0 \lor 0 \leq \vnu}$, which covers all integers.
\end{example}

Note that our instantiation algorithm is not complete, as it relies on the assumption that the syntactic structures of two types under unification are aligned. For example, the $\instop$ subroutine cannot unify $\urt{\Int}{P_1(\vnu) \lor P_2(\vnu)}$ with $\urt{\Int}{\vnu \geq 0}$, although one possible solution is assigning $P_1$ to $\lambda \vnu.\vnu > 0$ and $P_2$ to $\lambda \vnu.\vnu = 0$.


\begin{theorem}\label{theorem:extended-algo-sound}[Soundness of Extended Algorithmic Typing] For all type contexts
  $\Gamma$, term $e$ and coverage type $\tau$,
  $\Gamma \covervdash e \typecheck \tau {\implies} \Gamma \covervdash e
  : \tau$
\end{theorem}

\begin{theorem}\label{theorem:extended-algo-complete}[Completeness of Extended Algorithmic Typing] Assuming an oracle for all formulas produced by the $\subquery$ subroutine, and assuming $\instop$ subroutine always succeeds, then for any type context
  $\Gamma$, term $e$ and coverage type $\tau$, $\Gamma \covervdash e: \tau {\implies} \Gamma \covervdash e \typecheck \tau$.
  \end{theorem}

\section{The Coverage Monad and Monadic Generator Combinators}
\label{sec:monad}

As discussed in \autoref{sec:intro}, real-world PBT frameworks like
\Code{QCheck} define the type of input generators as a monad and
provide combinators for compositionally defining input generators. In
order to align with this practice, this section defines a
\emph{coverage monad} that provides coverage guarantees about input
generators defined in this style. This monad and the monadic generator
combinators described in this section can all be desugared into the
core language from the previous section.




\subsection{Coverage Monad}

\begin{figure}
    \centering
    \begin{minted}[xleftmargin=5pt, numbersep=4pt, linenos = true, fontsize = \normalsize, escapeinside=??]{OCaml}
    module Gen = struct
        (* RS.t is the internal state of PRNG *)
        type 'a t = RS.t -> 'a
        let return (x: 'a) = fun (st: RS.t) -> x
        let bind (g: 'a t) (f: 'a -> 'b t) =
           fun (st: RS.t) -> let (x: 'a) = gen st in f x s
    end
    \end{minted}
    \caption{The definition of the test generator monad in
      \Code{QCheck}.}
    \label{fig:coverage-monad-imp}
\end{figure}

In real-world programming languages, computational nondeterminism is
typically introduced via a primitive pseudorandom number generator
(PRNG) that produces unpredictable sequences of numbers based on an
random seed. This facility is used by PBT frameworks like the
generator library $\Code{Gen}$ of \Code{QCheck} to parameterize the
behavior of generators by a PRNG state, allowing test executions to be
replayed by reusing the same initial seed.
\autoref{fig:coverage-monad-imp} shows the definition of $\Code{Gen}$
in \Code{QCheck}. The type of a generator $\tvar{a}\TM$ is a function
type $\Code{RS.t} \to \tvar{a}$ (line $3$), which takes a random seed
and returns generated test value of type $\tvar{a}$. The
$\Code{return}$ operator returns a constant value, regardless of the
input state (line $4$); while the $\Code{bind}$ operator passes the
PRNG state through both the generator and the bind function (lines
$5$–$6$), respectively.


\begin{example} In \Code{QCheck}, a generator for positive integers
  can be compositionally defined by using the monadic \ocamlinline{bind}
  combinator to get a random value from the natural number generator
  \ocamlinline{nat_gen} and replacing it with 1 when it is a non-positive
  integer:
  \begin{minted}[xleftmargin=5pt, numbersep=4pt, linenos = true,
    fontsize = \normalsize, escapeinside=??]{OCaml}
    val nat_gen: int Gen.t
    let pos_gen: int Gen.t =
        bind nat_gen (fun (n: int) ->
            if 0 < n then return n else return 1)
    \end{minted}
\end{example}





\begin{figure}[h]
    \centering
\begin{align*}
  & \Murt{b}{\phi} \doteq \ort{\Code{RS.t}}{\top}\sarr\urt{b}{\phi} \\
  & \bindC{}~\Murt{b_1}{\phi_1}~
  \;(x{:}\ort{b_1}{\phi_2}\sarr\Murt{b_2}{\phi_3}) \doteq \\
  & \quad\quad \Murt{b_2}{\exists x{:}\;b_1. (\phi_1\land \phi_2)[\vnu\mapsto x] \land \phi_3}
  \\\\&\Code{return} : x{:}\ort{\tvar{a}}{\top} \sarr \Murt{\tvar{a}}{\vnu = x}
  \\&\Code{bind} :
  \Murt{\tvar{a}}{P_1(\vnu)} \sarr (x{:}\ort{\tvar{a}}{P_2(\vnu)}\sarr\Murt{\tvar{b}}{P_3(x, \vnu)}) \sarr
  \\&\quad\quad \bindC{}~\Murt{\tvar{a}}{P_1(\vnu)}~ (x{:}\ort{\tvar{a}}{P_2(\vnu)}\sarr\Murt{\tvar{b}}{P_3(x, \vnu)})
  \\&\quad\quad \adjustbox{margin=2pt
  {\fboxsep},bgcolor=LightGrey}{$\equiv \Murt{\tvar{b}}{\exists x. P_1(x)\land P_2(x)\land P_3(x, \vnu)}$ \quad \text{(after unfolding $\bindC{}$)}}
\end{align*}
\caption{The coverage monad and the signatures of its operators.}
\label{fig:coverage-monad-type}
\end{figure}


While \ocamlinline{nat_gen} and \ocamlinline{pos_gen} produce
different sets of values, they both have the same basic type:
\ocamlinline{int Gen.t}. To better capture the coverage guarantees of
these sorts of monadic generators, we define the coverage monad
$\Murt{b}{\phi}$, shown in \autoref{fig:coverage-monad-type}. As
expected, the shape of this monad augments the type parameter of
\ocamlinline{Gen.t} with a qualifier that captures its coverage
properties. The definition of $\Murt{b}{\phi}$ has two consequences:
first, since only non-function values can be qualified, the coverage
monad can only range over base types. Second, the coverage guarantees
of a monadic generator cannot reference the random seed. This latter
restriction makes intuitive sense, as a monadic test generator of type
$\M{\urt{b}{\phi}}$ should be guaranteed to produce values satisfying
predicate $\phi$ regardless of the random seed. We can use the
coverage monad to capture the different coverages of
\ocamlinline{nat_gen} and \ocamlinline{pos_gen}, as
$\M{\urt{\Int}{0 < \vnu}}$ and $\M{\urt{\Int}{0 \leq \vnu}}$,
respectively.

\autoref{fig:coverage-monad-type} also gives the signatures of the
operators of the coverage monad; their definitions are the same as in
\autoref{fig:coverage-monad-imp}. The signature of $\Code{return}$
reflects that lifts an arbitrary value into a generator that always
produces that value (i.e., $\vnu = x$). The signature of the
\ocamlinline{bind} operator uses a special notation, $\bindC{}$, for
its codomain; this alias will be helpful in the upcoming discussion of
generator combinators, which are built using \ocamlinline{bind}.
Intuitively, the qualifier in this type introduces an existentially
quantified variable $x$ that represents the intermediate value by the
first argument of \ocamlinline{bind}. This value is required to
satisfy both the coverage guarantee of the test generator ($\phi_1$)
as well as the safety constraint associated with the bind function
($\phi_2$).

To understand the definition of $\bindC{}$, consider how the
definition of \ocamlinline{bind} in \autoref{fig:coverage-monad-imp}
can be typed using the polymorphic type system of the previous
section. After unfolding $\bindC{}$, the initial typing goal is:
\begin{align*}
  &\Code{g}{:}\Murt{\tvar{a}}{P_1(\vnu)}, \Code{f}{:}(x{:}\ort{\tvar{a}}{P_2(\vnu)}\sarr\Murt{\tvar{b}}{P_3(x, \vnu)}),
  \\&\quad \vdash \Code{\zlam{st}{RS.t}{\zlet{x}{g\;st}{f\;x\;st}}} : \urt{\tvar{b}}{\exists x. P_1(x)\land P_2(x)\land P_3(x, \vnu)}
\end{align*}
After moving the PRNG state and the value produce by the first generator into the context, the typing goal is:
\begin{align*}
  &\Code{g}{:}\Murt{\tvar{a}}{P_1(\vnu)}, \Code{f}{:}(x{:}\ort{\tvar{a}}{P_2(\vnu)}\sarr\Murt{\tvar{b}}{P_3(x, \vnu)}), \Code{s}{:}\Code{RS.t}, \Code{x}{:}\urt{\tvar{a}}{P_1(\vnu)}
  \\&\quad \vdash \Code{f\;x\;s} : \urt{\exists x. \tvar{b}}{P_1(x) \land P_2(x) \land P_3(\Code{x}, \vnu)}
\end{align*}
Similarly to how \ocamlinline{bst_gen} was typed in
\autoref{sec:rules}, we can use \textsc{TEq} to remove the existential
quantifier in the target type, and then use \textsc{TSub} to bring
\ocamlinline{x} into alignment with the argument type of
\ocamlinline{f} to yield the expected type.

\begin{figure}[t!]
    \centering
    \begin{minted}[xleftmargin=5pt, numbersep=4pt, linenos = true, fontsize = \normalsize, escapeinside=!!]{OCaml}
    val power: !$m{:}\ort{\Int}{0 \leq v}\sarr x{:}\ort{\Int}{0 \leq v} \sarr \urt{\Int}{\vnu = m^x}$!
    let power_gen = bind (int_bound 10: !$\urt{\Int}{\vnu \leq 10}$!)
         (fun (x: !$\ort{\Int}{0 \leq v}$!) -> return (power 2 x))
    \end{minted}
    \caption{A generator for powers of $2$ less than $2^{10}$.}
    \label{fig:bind-example}
\end{figure}

\begin{example}
  The coverage guarantees of generators built from monadic combinators
  are also derived compositionally.As an example,
  consider the definition of a generator for powers of $2$ from $2^0$
  to $2^{10}$ shown in \autoref{fig:bind-example}. After randomly
  generating an integer using \ocamlinline{int_bound 10},
  \ocamlinline{power_gen} raises 2 to this power and returns the
  result. Note that the coverage guarantees of \ocamlinline{power} are
  only provided for arguments greater than $0$, however, a fact that
  will be reflected in the final type of \ocamlinline{power_gen}.
  Importantly, the instantiations of three predicate variables in the
  type of $\bindC$ come directly from the qualifiers of its arguments:
  \begin{align*}
    P_1\mapsto \underbrace{\lambda \vnu.\vnu \leq 10}_\Code{int\_bound\;10}
    \quad\quad P_2\mapsto \underbrace{\lambda \vnu.0 \leq
    \vnu}_\Code{fun\; x: \ort{\Int}{0 \leq v}\;->}
    \quad\quad P_3 \mapsto \underbrace{\lambda x.\lambda \vnu.\vnu = 2^x}_\Code{power\;2\;x}
  \end{align*}
  \noindent Allowing us to derive the desired coverage type for
  \ocamlinline{power_gen}:
  \begin{align*}
    \Code{power~:~\ldots}~\vdash \Code{power\_gen}~:~\Murt{\Int}{\exists x. 0 \leq x \land x \leq 10 \land \vnu = 2^x}
  \end{align*}
\end{example}

\begin{figure}[h!]
    \centering
    \begin{minted}[xleftmargin=5pt, numbersep=4pt, linenos = true, fontsize = \small, escapeinside=!!]{OCaml}
    let biased_nat_gen (st: RS.t) =
        union (int_range 0 10) (return Int.max_int)

    !$\vdash$! biased_nat_gen: !$\M{\ort{\Int}{0 \leq \vnu \leq \Code{Int.max\_int}}}$!
    \end{minted}
    \caption{Example of safety monad.}
    \label{fig:over-under}
\end{figure}





\paragraph{Over- and under-approximate monads}
Monadic combinators also lend themselves well to an over-approximate
analysis via traditional refinement types. In the context of test
input generators, such specifications are useful in contexts where
developers want to prioritize certain \emph{safe} values that always
meet the preconditions of the system under test, e.g., when probing
boundary cases. As an example of this situation, consider the
following generator:
\begin{minted}[xleftmargin=5pt, numbersep=4pt, linenos = true, fontsize = \small, escapeinside=!!]{OCaml}
  let biased_nat_gen (st: RS.t) =
  union (int_range 0 10) (return Int.max_int)

  !$\vdash$! biased_nat_gen: !$\M{\ort{\Int}{0 \leq \vnu \leq \Code{Int.max\_int}}}$!
\end{minted}
\noindent $\Code{biased\_nat\_gen}$ does not enumerate all natural
numbers; instead, it samples from a biased distribution that
emphasizes small numbers (e.g., $[0, 10]$) and the maximum integer
value (\Code{Int.max\_int}), the latter of which covers the critical
edge case for overflow testing. To cover this scenario, we can
similarly define a dual \emph{safety} monad for generators
$\M{\ort{b}{\phi}}$. Here, $\Code{biased\_nat\_gen}$ can be assigned
the safety monad type
$\M{\ort{\Int}{0 \leq \vnu \leq \Code{Int.max\_int}}}$, ensuring that
all sampled values fall within this range. We further discuss
real-world instances of such usage in \autoref{sec:evaluation}.


Using monad combinators to define generators enables the seamless use
of both safety and coverage monads to provide richer specifications
that capture their ability to produce every value that meets a
precondition and only such values. Consider a generator $g_1$ for
lists of integers which has the following type: %
\begin{align*} %
  &\vdash \Code{g_1} : \bindC{}~ \M{\ort{\Int}{0 \leq \vnu}}~
    (n{:}\Int\sarr\M{\urt{\List[\Int]]}{\I{len}(\vnu, n) \land \I{sorted}(\vnu)}})
\end{align*} %
\noindent This type specifies that $\Code{g_1}$ produces all possible
sorted lists of integers with length $n$, where $n$ is drawn from a generator
assigned a safety monad type (e.g., $\Code{biased\_nat\_gen}$) which is biased toward values believed to be
important by developers. Conversely, consider a generator $\Code{g_2}$
with the following type: %
\begin{align*}
  &\vdash \Code{g_2} : \bindC{}~ \M{\urt{\Int}{0 \leq \vnu}}~ (n{:}\Int \sarr\M{\ort{\List[\Int]}{\I{len}(\vnu, n) \land \I{sorted}(\vnu)}})
\end{align*}\noindent
Here, $\Code{g_2}$ produces all possible non-negative list lengths,
while imposing no specific coverage obligations over the list contents
of a generated sorted list of a given length. This form is useful when the
target property under test is sensitive to shape of the datatype
(i.e., list length) but not necessarily to the elements it contains.
Developers are then allowed to specialize the generated list of length $n$ as a fixed choice (e.g, $[1; 2; \dots; n]$), rather than generating all possible sorted lists, which would require more effort.
In both of these examples, the combined use of safety and coverage
monads provides developers more flexibility to specify generators that
are neither totally over-approximated nor totally under-approximated.

\subsection{Monadic Combinator Examples}

This section provides examples of the coverage guarantees our system
can provide for several monadic combinators typically found in PBT
frameworks like \Code{QCheck}. We show the (monadic) coverage type of
each example combinator first using the coverage monad, and then a
version in which $\returnC{}$ and $\bindC{}$ are unfolded.

\begin{figure}[h!]
  \begin{mdframed}
    \centering
    \begin{align*}
    \Code{option}:\;&\Murt{\tvar{a}}{P(\vnu)}\sarr
    \\& \bindC{}~ \Murt{\tvar{a}}{P(\vnu)}~ \;(x{:}\tvar{a}\sarr \returnC{}\;\urt{\Option[\tvar{a}]}{\vnu = \None \lor \vnu = \Some[x]})
    \\&\adjustbox{margin=2pt
      {\fboxsep},bgcolor=LightGrey}{$\equiv\Murt{\Option[\tvar{a}]}{\vnu =
      \None \lor \exists x.P(x) \land \vnu = \Some[x]}$ \quad
      \text{(after unfolding)}}
    \\\Code{pair}:\;& \Murt{\tvar{a}}{P_1(\vnu)}\sarr \Murt{\tvar{b}}{P_2(\vnu)}\sarr
      \\& \bindC{}~ \Murt{\tvar{a}}{P_1(\vnu)} ~ \\
      & \quad (x{:}\tvar{a}\sarr
      \bindC{}~ \Murt{\tvar{b}}{P_2(\vnu)} ~(y{:}\tvar{b}\sarr \returnC{}\;\urt{\tvar{a} \times \tvar{b}}{\vnu = (x, y)}))
    \\&\adjustbox{margin=2pt {\fboxsep},bgcolor=LightGrey}{$\equiv
      \Murt{\tvar{a} \times \tvar{b}}{P_1(\fst(\vnu)) \land
      P_2(\snd(\vnu))}$ \quad \text{(after unfolding)}}
    \end{align*}
  \end{mdframed}
\end{figure}
The $\Code{option}$ combinator lifts a generator into a version for
option types that covers both the $\Some$ and $\Code{None}$ cases
($\vnu = \None \lor \vnu = \Some[x]$). Similarly, the $\Code{pair}$ combinator returns a pair
containing the results of two input test generators ($\vnu = (x, y)$). 

\begin{figure}[h!]
  \begin{mdframed}
    \centering
    \vspace{-.5cm}
    \begin{align*}
      \Code{list\_repeat}:\;& s{:}\ort{\Int}{0 \leq \vnu}\sarr g{:}(\Murt{\tvar{a}}{P(\vnu)}) \sarr
      \\& \Murt{\List[\tvar{a}]}{\I{len}(\vnu, s) \land \forall x. \mem(\vnu, x) \impl P(x)}
    \end{align*}
  \end{mdframed}
\end{figure}

The $\Code{list\_repeat}$ combinator uses an input generator ($g$) to
construct a list of $s$ elements. Its coverage type guarantees that it
produces all lists with of the stipulated length $\I{len}(\vnu, s)$
that contain elements covered by the input generators (i.e.,
$\forall x. \mem(\vnu, x) \impl P_2(x)$). As a simple example, the
following generator produces all lists with three positive numbers:
\begin{minted}[xleftmargin=5pt, numbersep=4pt, linenos = true, fontsize = \small, escapeinside=!!]{OCaml}
  let pos_list_gen = list_repeat 3 pos_gen
\end{minted}

\begin{figure}[h!]
  \begin{mdframed}
    \centering
    \vspace{-.5cm}
    \begin{align*}
    \Code{oneof}:\;& s{:}\ort{\Int}{0 < \vnu} \sarr (i{:}\ort{\Int}{P_1(\vnu)}\sarr\Murt{\tvar{a}}{P_2(i,\vnu)})\sarr
    \\& \bindC{}\;(\returnC{}\;\urt{\Int}{0 \leq \vnu < s})\; (i{:}\ort{\Int}{P_1(\vnu)}\sarr\Murt{\tvar{a}}{P_2(i,\vnu)})
    \\&\adjustbox{margin=2pt
      {\fboxsep},bgcolor=LightGrey}{$\equiv \Murt{\tvar{a}}{\exists i. 0 \leq
      i < s \land P_1(i) \land P_2(i,\vnu)}$ \quad \text{(after unfolding)}}
    \\\Code{frequency}:\;& s{:}\ort{\Int}{0 < \vnu} \sarr (i{:}\ort{\Int}{P_1(\vnu)}\sarr\Murt{\Int\times \tvar{a}}{P_2(i,\vnu)})\sarr
    \\&\bindC{}\;(\bindC{}\;(\returnC{}\;\urt{\Nat}{0 \leq \vnu <
      s})\; \\
      &\hspace{1.5cm}  (i{:}\ort{\Int}{P_1(\vnu)}\sarr\Murt{\Int\times \tvar{a}}{P_2(i,\vnu)}))
    \\&\hspace{1.0cm} (p{:}\Nat\times \tvar{a} \sarr \returnC{}\;\urt{\tvar{a}}{\fst(p) > 0 \land \snd(p) = \vnu})
    \\&\adjustbox{margin=2pt
      {\fboxsep},bgcolor=LightGrey}{$\equiv \Murt{\tvar{a}}{\exists i. 0 \leq
      i < s \land P_1(i) \land \exists w. w > 0 \land P_2(i,(w, \vnu))
      }$ \quad \text{(after unfolding)}}
    \end{align*}
    \end{mdframed}
\end{figure}

The $\Code{oneof}$ combinator is a more powerful version of the
$\Code{union}$ combinator that combines an arbitrary number of test
generators, instead of only two. This combinator takes two arguments:
the total number of input generators ($s$) and an indexed function
($f$) where $f\;i$ represents the $i^\text{th}$ input
generator.\footnote{The $\Code{oneof}$ and
  $\Code{frequency}$ combinators in QCheck use lists instead of
  a function to hold its input generators. Our formulation of both
  combinators allows them to straightforwardly combine generators with
  different coverage guarantees, without having to deal with
  heterogeneous lists. } As captured by its type signature, the number
of input generators is constrained by predicate variable $P_1$ and the
coverage guarantee of each input generator is constrained by $P_2$.
The coverage type of \Code{union} guarantees that it builds a
generator that covers everything that the input generators do (i.e.,
$\exists\ i. 0 \leq i < s$). As an example, the following (redundant)
generator for integers uses $\Code{oneof}$ to combine $3$ generators:
a negative number generator ($\Code{neg\_gen}$), a constant generator
that only produce $0$ ($\Code{return\;0}$), and a positive number
generator ($\Code{pos\_gen}$).
\begin{minted}[xleftmargin=5pt, numbersep=4pt, linenos = true, fontsize = \small, escapeinside=!!]{OCaml}
    let int_gen = oneof 3 (function
                           | 0 -> neg_gen
                           | 1 -> return 0
                           | _ -> pos_gen)
\end{minted}
The $\Code{frequency}$ combinator is similar to $\Code{oneof}$, but it
additionally allows the developer to provide a weight for each input
generator that governs their sampling distribution. The type of this
combinator requires that each of these weights are positive ($w > 0$).

\begin{figure}
  \begin{mdframed}
    \centering
    \vspace{-.85cm}
    \begin{align*}
      \\\Code{fix}:\;& (m{:}\Nat\sarr (n{:}\ort{\Nat}{\vnu < m}\sarr\Murt{\tvar{a}}{P(n, \vnu)}) \sarr \Murt{\tvar{a}}{P(m, \vnu)}) \sarr
    \\& (m{:}\Nat\sarr\Murt{\tvar{a}}{P(m, \vnu)})
    \end{align*}
  \end{mdframed}
\end{figure}

The $\Code{fix}$ combinator constructs a test generator that produces
larger test inputs from a function that generators smaller test
inputs; the first parameter of this function is a natural number that
limits the number of recursive calls. Unsurprisingly, this combinator
has the same premises as \textsc{TFix} in \autoref{fig:type-rules}.

\begin{figure}
    \begin{minted}[xleftmargin=5pt, numbersep=4pt, linenos = true, fontsize = \small, escapeinside=!!]{OCaml}
    type 'a tree = Leaf of 'a | Node of 'a tree * 'a tree
    let tree_gen : int -> int tree Gen.t = fix
      (fun n self -> match n with
        | 0 -> return err
        | 1 -> bind nat_gen (fun x -> return (Leaf x))
        | n ->
          frequency 2
            (function
            | 0 -> (1, bind nat_gen (fun x -> return (Leaf x)))
            | _ -> (2, bind (self n/2) (fun tr1 ->
                         bind (self n/2) (fun tr2 ->
                           return (Node (tr1, tr2))))))
      )
    \end{minted}
    \caption{Defining a tree generator using $\Code{frequency}$ and
      $\Code{fix}$.}
    \label{fig:frequency-fix}
\end{figure}

\begin{figure}[h!]
    \centering
    \includegraphics[width=0.35\linewidth]{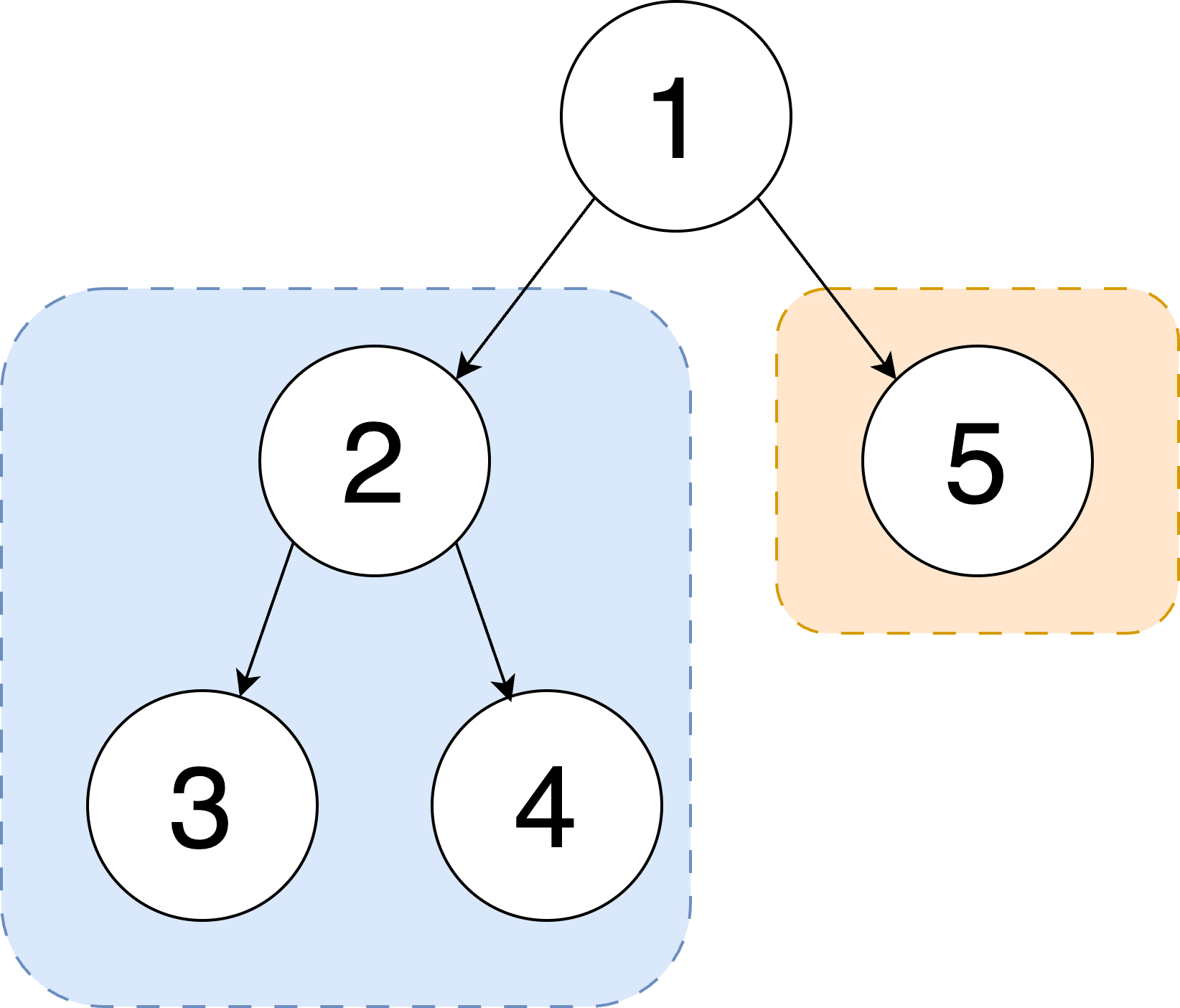}
    \caption{A 5-node tree that cannot be produced by $\Code{tree\_gen\; 5}$.}
    \label{fig:tree-ex}
\end{figure}

\begin{example}
  We demonstrate how these combinators work by reimplementing the tree
  generator example from \autoref{sec:intro} using the $\Code{fix}$
  and $\Code{frequency}$ combinators. The new implementation is given
  in \autoref{fig:frequency-fix}. Since every \ocamlinline{tree}
  contains at least one element, the base case returns an error.
  Otherwise, if the target number of elements is $1$ (line $5$),
  we generate a leaf node by $\Code{bind}$ing the
  natural number generator with the \Code{Leaf} constructor. In the
  other cases, we use the $\Code{frequency}$ combinator to combine two
  test generators with weight $1$ and $2$ respectively (line $9$ -
  $12$). The first generator is a singleton tree generator,
  while the second generator calls $\Code{self}$ recursively to
  generate left and right subtrees containing half the nodes. Note
  that each subtree has at most half the target number of nodes, to
  ensure that the final tree contains the expected number of elements.

    Although this tree generator accepts a bound $n$ as input, it cannot be assigned the coverage type $n{:}\Nat\sarr\Murt{\Tree[\Nat]}{\I{numElem}(\vnu) \leq n}$, which would require it to generate all trees containing at most $n$ elements. This is because the implementation divides the input bound evenly during recursive calls (lines $10$ and $11$), making it unable to construct trees that are highly unbalanced. For instance, with input $5$, the generator cannot produce the tree illustrated in \autoref{fig:tree-ex}, where the left subtree (highlighted in blue) contains $3$ elements which is exceeding the divided bound $5/2$. Instead, we can assign it the coverage type $n{:}\Nat\sarr\Murt{\Tree[\Nat]}{2^{\I{depth}(\vnu)} \leq n}$, which captures the idea that all trees of depth $m$ can be covered when the input bound is at least $2^m$. 
    According to this type, any tree can be generated by this tree generator given a sufficiently large input bound.
    This alternative interpretation of the input bound as ``depth'' is coverage-complete but less intuitive than the ``number of nodes'' interpretation, highlighting the intricate nature involved in underapproximate-style reasoning.

\end{example}


\section{Implementation and Evaluation}
\label{sec:evaluation}

\paragraph{Implementation}
We have implemented a coverage type checker, called \name{}, based on
the above approach. \name{} targets functional, non-concurrent OCaml programs that rely on
libraries to manipulate algebraic data types; it consists of
approximately 11K lines of OCaml and uses Z3~\cite{de2008z3} as
its backend solver.

\name{} takes as input an Ocaml program representing a test input
generator and a user-supplied coverage type for that generator.  After
basic type-checking and translation into MNF, \name{} applies
bidirectional type inference and checking to validate that the
program satisfies the requirements specified by the type.  Our
implementation provides built-in coverage types for a number of OCaml
primitives, including constants, various arithmetic operators, and
data constructors for a range of datatypes.  Refinements defined in
coverage types can also use predefined (polymorphic) uninterpreted predicates
that capture non-trivial datatype shape properties.  For example, the
uninterpreted predicate $\mem(\Code{l},\Code{u})$ indicates the element $\Code{u}{:}\tvar{a}$
is contained in the data type instance $\Code{l}{:}\List[\tvar{a}]$; the uninterpreted predicate $\I{len}(\Code{l}, 3)$ indicates the list $\Code{l}$ has length $3$,
or the tree $\Code{l}$ has depth $3$. The semantics of these uninterpreted predicates are defined as a set of FOL-encoded lemmas and axioms to
facilitate automated verification; e.g., the lemma $\I{len}(\Code{l}, 0)
\impl \forall \Code{u}, \neg \mem(\Code{l},\Code{u})$ indicates that the empty datatype
instance contains no element.

\cbnewadding{
In order to improve  efficiency, verification conditions generated by \name{} are simplified before being solved by the underlying SMT solver. For example, the $\existsop$ function generates queries of the form $\exists x. x = v \land \phi$, which can be simplified to $\phi[x\mapsto v]$. This transformation can significantly reduce both the number of quantifiers and the overall size of queries.
}

\begin{table}[]
\renewcommand{\arraystretch}{0.8}
\caption{Experimental results on hand-written
  generators. Existing benchmarks are annotated with their source:
  QuickChick~\cite{quickchick} (*),
  QuickCheck~\cite{claessen2011quickcheck} ($^\circ$), and bespoke
  test input generators from \cite{quickcheck-coverage-guided}
  ($^\star$) and \cite{ZDDJ21} ($^\diamond$).  }
\footnotesize
\begin{tabular}{ccccc|ccc}
\toprule
 & \#Branch & Rec & \#LVar & \#UP & \#Query & (max. \#$\forall$,\#$\exists$) & total(avg. time) (ms)\\
\midrule
\textsf{SizedList}* & $3$ & $\checkmark$ & $9$ & $3$ & $8$ & $(2, 5)$ & $10.45(1.31)$ \\
\textsf{SortedList}* & $3$ & $\checkmark$ & $13$ & $4$ & $11$ & $(4, 7)$ & $31.21(2.84)$ \\
\textsf{UniqueList}$^\diamond$ & $3$ & $\checkmark$ & $10$ & $5$ & $9$ & $(2, 5)$ & $12.28(1.36)$ \\
\midrule
\textsf{SizedTree}* & $3$ & $\checkmark$ & $12$ & $4$ & $11$ & $(2, 7)$ & $15.00(1.36)$ \\
\textsf{CompleteTree}$^\star$ & $2$ & $\checkmark$ & $9$ & $5$ & $8$ & $(2, 5)$ & $12.01(1.50)$ \\
\textsf{RedBlackTree}* & $6$ & $\checkmark$ & $44$ & $7$ & $41$ & $(5, 20)$ & $83.28(2.03)$ \\
\textsf{SizedBST}$^\star$ & $4$ & $\checkmark$ & $28$ & $6$ & $25$ & $(6, 12)$ & $42.69(1.71)$ \\
\midrule
\textsf{BatchedQueue}$^\diamond$ & $1$ &   & $5$ & $1$ & $5$ & $(2, 4)$ & $6.58(1.65)$ \\
\textsf{BankersQueue}$^\diamond$ & $1$ &   & $5$ & $1$ & $5$ & $(2, 4)$ & $7.53(1.51)$ \\
\midrule
\textsf{Stream}$^\diamond$ & $3$ & $\checkmark$ & $10$ & $4$ & $9$ & $(2, 6)$ & $11.84(1.32)$ \\
\midrule
\textsf{SizedHeap}$^\circ$ & $4$ & $\checkmark$ & $17$ & $5$ & $15$ & $(5, 9)$ & $23.10(1.54)$ \\
\textsf{LeftistHeap}$^\diamond$ & $2$ & $\checkmark$ & $13$ & $5$ & $12$ & $(2, 7)$ & $47.06(3.92)$ \\
\midrule
\textsf{SizedSet}$^\circ$ & $3$ & $\checkmark$ & $26$ & $6$ & $23$ & $(5, 11)$ & $45.17(1.96)$ \\
\textsf{UnbalanceSet}$^\diamond$ & $4$ & $\checkmark$ & $35$ & $6$ & $31$ & $(6, 14)$ & $67.22(2.17)$ \\
\bottomrule
\end{tabular}
\label{tab:evaluation}
\end{table}

\subsection{Completeness of Hand-Written Generators}
We have evaluated \name{} on a corpus of hand-written, non-trivial test input
generators drawn from a variety of sources (see
Table~\ref{tab:evaluation}). These benchmarks provide test input
generators over a diverse range of datatypes, including various kinds
of lists, trees, queues, streams, heaps, and sets.  For each datatype
implementation, \name{} type checks the provided implementation
against its supplied coverage type to verify that the generator is
able to generate all possible datatype instances consistent with this
type.  Our uninterpreted predicates allow us capture non-trivial structural properties. For example, to
verify a red-black tree generator, we use the predicate
$\I{black\_height}(\nu, n)$ to indicate that all branches of the tree
$\nu$ have exactly $n$ black nodes, the predicate
$\I{no\_red\_red}(\nu)$ to indicate $\nu$ contains no red node with
red children, and the predicate $\I{root\_color}(\nu, b)$ to indicate
the root of the tree $\nu$ has the red (black) color when the boolean
value $b$ is true (false).\footnote{These uninterpreted predicates can be
  found in the implementation of the red-black tree generator given
  in~\cite{quickchick}. }

Given this rich set of predicates, it is straightforward to express
interesting coverage types.  For example, given size $\Code{s}$ and
lower bound $\Code{lo}$, we can express the property that a sorted
list generator $\Code{sorted\_list\_gen}$ \emph{must} generate all
possible sorted lists with the length $\Code{s}$ and in which all
elements are greater than or equal to $\Code{lo}$, as the following
type:
{\small
\begin{align*}
    \Code{s}{:}\ort{\Int}{\nu \leq 0}\sarr\Code{lo}{:}\ort{\Int}{\top}\nuut{int\ list}{\I{len}(\nu, \Code{s}) \land \I{sorted}(\nu) \land \forall u, \mem(\nu, u) \impl \Code{lo} \leq u }
\end{align*}
}\noindent
Notice that this type is remarkably similar to a normal refinement type:
{\small
\begin{align*}
    \Code{s}{:}\ort{\Int}{\nu \leq 0}\sarr\Code{lo}{:}\ort{\Int}{\top}\nuot{int\ list}{\I{len}(\nu, \Code{s}) \land \I{sorted}(\nu) \land \forall u, \mem(\nu, u) \impl \Code{lo} \leq u }
\end{align*}
}\noindent
albeit with the return type marked as a coverage type to capture
our desired must-property.

The first group of columns in \autoref{tab:evaluation} describes the
salient features of our benchmarks.  Each benchmark exhibits
non-trivial control-flow, containing anywhere from 1 to 6 nested
branches; the majority of our benchmarks are also recursive (column
Rec).  The number of local (i.e., let-bound) variables (column
\#LVars) is a proxy for path lengths that must be encoded within
the types inferred by our type-checker; column \#UP indicates the
number of uninterpreted predicates found in the benchmark's type
specification.

The second group of columns presents type checking results. The column
\#Query indicates the number of SMT queries that are triggered during
type checking. The column \#$(\forall,\exists)$ indicates the maximum
number of universal and existential quantifiers in these queries,
respectively.  The $\exists$ column is a direct reflection of
control-flow (path) complexity --- complex generators with deeply nested
match-expressions like \textsf{RedBlackTree} result in queries with
over 20 existential quantifiers.  These numbers broadly track with the
values in columns \#Branch and \#LVar. Despite the complexity of
some of these queries, as evidenced by the number of their
quantifiers, overall verification time (average verification time per query, resp.), reported in the last column, is
quite reasonable,
\cbnewadding{
with times ranging from 6.58 to 83.28 milliseconds, i.e., all benchmarks finishing in less than 0.1 second.}

\subsection{Case Study: Well-Typed STLC Terms}
\label{sec:stlc}

\begin{figure}[t!]
  \begin{minted}[fontsize = \small,xleftmargin = 0pt,escapeinside=<>]{ocaml}
type ty = Ty_nat | Ty_arr of ty * ty

type term =
  | Const of int
  | Var of int
  | Abs of ty * term
  | App of term * term

type tyctx = ty list
  \end{minted}
\caption{ Datatypes from the STLC case study.}
\label{fig:stlc-lang}
\end{figure}
We have also applied \name{} to a more substantial example:
  a generator for well-typed simply typed lambda calculus (STLC) terms
  adapted from \citet{LG+17,PCRH+11}. Such a generator can be used to
  test that the typing relation guarantees the expected runtime
  behaviors of programs, e.g. progress and preservation. In addition
  to the complexity of the coverage property itself (well-typedness),
  this case study features multiple inductive datatypes (for types,
  terms, and typing contexts, as shown in \autoref{fig:stlc-lang}),
  and 13 auxiliary functions. The coverage type of
  $\Code{gen\_term\_size}$, the top-level generator, stipulates that
  it can generate all terms of a desired type, up to a user-provided
  size bound:
\begin{align*}
  \Code{gen\_term\_size}:
  \underbrace{\Code{n}{:}\nuot{int}{0 \leq \nu}}_\texttt{maximum term
  size} {\shortrightarrow}
  &\underbrace{\Code{t}{:}\nuot{\Code{ty}}{\top}}_\texttt{type of term} {\shortrightarrow}
    \underbrace{\Gamma{:}\nuot{\Code{tyctx}}{\top}}_\texttt{typing context} {\shortrightarrow}
  \\
  &\underbrace{\nuut{\Code{term}}{\I{has\_ty}(\Gamma,\nu,\Code{t})
    \land \I{max\_app\_num}(\nu,\Code{n})}}_\texttt{result is well-typed and has at most $\Code{n}$ applications}
\end{align*}\noindent
The results of using \name{} to verify that
  $\Code{gen\_term\_size}$ meets the above specification are shown in
  Table~\ref{tab:stlc}. The table also reports the results for the
  most interesting auxiliary functions used by the
  function.
 The last
  column shows that \name{} is able to verify these functions
  within a reasonable time, ranging from \cbnewadding{$0.01$ to $4.74$}
  seconds. Although more complex functions (as indicated by the
  column labeled \#Branch) require more time to verify, total verification time is
  nonetheless reasonable: \cbnewadding{less than $5$ seconds} in
  total. Taken together, these results highlight the compositionality
  of \name{}'s type-based approach: each of the $13$ auxiliary
  functions used by $\Code{gen\_term\_size}$ is individually
  type-checked against its signature; these signatures are then used to
  verify any procedures that call the function.

Interestingly, the function $\Code{type\_eq}$ has a longer
  average query time than most of other functions, despite having fewer
  local variables and uninterpreted predicates.  This function implements a
  deterministic equality test, returning $\Code{true}$ when two types
  are the same and $\Code{false}$ otherwise. Thus, the coverage type
  of this function degenerates into a singleton type for each of the
  branches, resulting in stricter queries to the SMT solver that take
  longer to find a valid witness.

\begin{table}[]
\renewcommand{\arraystretch}{0.8}
\caption{
Experimental results from the STLC case study. Each function is implemented as a wrapper around a subsidiary function that takes an additional strictly decreasing argument to ensure termination (the original QuickChick implementations uses Coq/Rocq’s Program command for this purpose). These subsidiary functions are responsible for the bulk of the computation, so we report the results for those functions here. }
\footnotesize
\begin{tabular}{ccccc|ccc}
\toprule
 & \#Branch & Rec & \#LVar & \#UP & \#Query & (max. \#$\forall$,\#$\exists$) & total (avg. time) (ms)\\
\midrule

\textsf{type\_eq} & $5$ & $\checkmark$ & $6$ & $4$ & $4$ & $(3, 12)$ & $57.15(14.29)$ \\
\textsf{gen\_type} & $2$ & $\checkmark$ & $9$ & $3$ & $8$ & $(2, 5)$ & $9.47(1.18)$ \\
\textsf{vars\_with\_type} & $4$ & $\checkmark$ & $7$ & $5$ & $5$ & $(3, 6)$ & $8.38(1.68)$ \\
\textsf{gen\_term\_no\_app} & $3$ & $\checkmark$ & $8$ & $12$ & $6$ & $(3, 7)$ & $13.88(2.31)$ \\
\textsf{gen\_term\_size} & $4$ & $\checkmark$ & $36$ & $11$ & $32$ & $(5, 16)$ & $4735.37(147.98)$ \\
\bottomrule
\end{tabular}
\label{tab:stlc}
\end{table}

\paragraph{Discussion} To handle the complexity of this
  benchmark, \name{} requires $16$ uninterpreted predicates and $24$ axioms,
  the large majority of which correspond to helper definitions and lemmas
  from the original development. The predicates that encode typing and
  the bounds on the number of applications in a term ($\I{has\_ty}$
  and $\I{max\_num\_app}$, resp.) come directly from the QuickCheck
  version, for example.  The following axiom encodes the semantic
  relationship of these predicates
\begin{align*}
    &\forall \Gamma{:}\Code{tyctx}.\;\forall\Code{t{:}ty}.\; \forall\Code{e{:}term}.\;
    \\&\quad \I{has\_ty}(\Gamma,\Code{e},\Code{t}) \iff \exists \Code{n}{:}\Nat.\;\I{max\_num\_app}(\Code{e},\Code{n}) \land \I{has\_ty}(\Gamma, \Code{e},\Code{t})
\end{align*}\noindent
and is analogous to the helper lemma
$\Code{has\_ty\_max\_tau\_correct}$ in the original Coq/Rocq
development. In addition, some predicates and axioms are independent
of this particular case study: the typing context is implemented as a
list of STLC types, and thus we were able to reuse generic predicates
and axioms about polymorphic lists.

\subsection{Completeness of Synthesized Generators}
\label{sec:synthesis}

\begin{table}[h!]
\renewcommand{\arraystretch}{0.8}
    \caption{ Quantifying the space of safe and complete test input
      generators constructed by an automated program synthesis
      tool.}
    \centering
    \footnotesize
    \begin{tabular}{cccc}
    \toprule
    Benchmark & \#Total & \#Complete \\
    \midrule
    \textsf{UniqueList}  & $284$ & $10$\\
    \textsf{SizedList}  & $126$ & $28$\\
    \textsf{SortedList}  & $30$ & $8$\\
    \textsf{SizedTree}  & $103$ & $2$\\
    \textsf{SizedBST}  & $229$ & $54$\\
    \textsf{RedBlackTree}  & $234$ & $2$\\
    \bottomrule
    \end{tabular}
    \label{tab:evaluation-synthesis}
\end{table}
An underlying hypothesis motivating our work is that writing sound and
\emph{complete} test input generators can be subtle and tricky, as
demonstrated by our motivating example (Figure~\ref{fig:gen-ex}).  To
justify this hypothesis, we repurposed an existing deductive
component-based program synthesizer~\cite{MJ22} to automatically
synthesize correct (albeit possibly incomplete) generators that
satisfy a specification given as an overapproximate refinement type;
these generators are then fed to \name{} to validate their
completeness.  We provided the synthesizer with a datatype definition
and a set of specifications describing constraints on that datatype
the synthesized generator should use, along with a library of
functions, including primitive generators such as \Code{nat\_gen},
available to the synthesizer for construction.  A refinement
type-guided enumeration is performed to find all correct programs
consistent with the specification.  Since the space of these programs
is potentially quite large (possibly infinite), we constrain the
synthesizer to only generate programs with bounded function call
depths; in our experiments, this bound was set to three.  The
generator outputs all programs that are safe with respect to the
specification. \autoref{tab:evaluation-synthesis} shows results of
this experiment for five of the benchmarks given in
Table~\ref{tab:evaluation}; results for the other benchmarks are
similar. We report the total number of synthesized generators
(\#Total) constructed and the number of those that are {\it correct
  and complete} as verified by \name{} (\#Complete).  The table
confirms our hypothesis that the space of complete generators with
respect to the supplied coverage type is significantly smaller than
the space of safe generators, as defined by an
overapproximate refinement type specification.

More concretely, Figure~\ref{fig:synth_sizedlist} shows three
synthesized generators that satisfy the following specification of a
list generator that is meant to construct \emph{all}
lists no longer than some provided bound:
 \begin{align*}
    \Code{size}{:}\ort{\Int}{\nu \leq 0}\sarr\nuut{int\ list}{\forall u, \I{len}(\nu, u)  \impl (0 \leq u \land u \leq \Code{size}) }\\[-15pt]
\end{align*}

\begin{figure}
\begin{subfigure}[b]{0.95\textwidth}
\begin{minted}[fontsize = \small, xleftmargin=2pt,numbers=left]{ocaml}
let rec sized_list_gen
    (size : int) : (int list) =
  if (size == 0) then []
  else
    if bool_gen () then sized_list_gen (size - 1)
    else (int_gen ()) :: (sized_list_gen (size - 1))
\end{minted}
\vspace{-0.3cm}
\caption{A sound and complete generator.}
\label{fig:synth_sizedlist_a}
\end{subfigure}
\vspace{0.2cm}
\\
\begin{subfigure}[b]{0.95\textwidth}
\begin{minted}[fontsize = \small,xleftmargin=2pt]{ocaml}
let rec sized_list_gen
    (size : int) : (int list) =
  if (size == 0) then []
  else (int_gen ()) :: (sized_list_gen (size - 1))
    \end{minted}
\vspace{-0.3cm}
\caption{A sound but incomplete generator.}
\label{fig:synth_sizedlist_b}
\end{subfigure}
\vspace{0.2cm}
\\
\begin{subfigure}[b]{0.95\textwidth}
\begin{minted}[fontsize = \small,xleftmargin=2pt]{ocaml}
let rec sized_list_gen
    (size : int) : (int list) =
  if (size == 0) then []
  else
    if bool_gen () then sized_list_gen (size - 1)
    else size :: (sized_list_gen (size - 1))
        \end{minted}
\vspace{-0.3cm}
\caption{Another sound but incomplete generator.}
\label{fig:synth_sizedlist_c}
    \end{subfigure}
    \caption{Three example generators that generate size-bounded lists.}
    \label{fig:synth_sizedlist}
\end{figure}\noindent
\autoref{fig:synth_sizedlist_b} is incomplete because it never
generates an empty list when the size parameter \Code{size} is greater
than 0.  On the other hand, while \autoref{fig:synth_sizedlist_c} does
generate empty lists, the else branch of its second conditional has a
fixed first element and will therefore never generate lists with
distinct elements.  The complete generator shown in
\autoref{fig:synth_sizedlist_a} incorporates a control-flow path (line
5) that can non-deterministically choose to make a recursive call to
\Code{sized\_list\_gen} with a smaller size, thereby allowing it to
generate lists of variable size up to the \Code{size} bound, including
the empty list; another conditional branch uses \Code{int\_gen()} to
generate a new randomly selected list element, thereby allowing the
implementation to generate lists containing distinct elements. We
again emphasize that \name{} was able to verify the correct generator
and discard the two incorrect generators \emph{automatically}, without
any user involvement.

\subsection{Case Study: Checking Coverage in Practice}

\begin{table}[h!]
\renewcommand{\arraystretch}{0.8}
\caption{
Experimental results that use monadic combinators from the \Code{Gen} library of the \Code{QCheck} framework. The columns are the same as in \autoref{tab:evaluation};  we additionally count the number of type variables ($\#\tvar{a}$) and predicate variables ($\#P$) for these polymorphic combinators.  A number of combinator types do not require uninterpreted predicates since they do not make use of datatypes; \textsf{list\_repeat} (\Code{list}), \textsf{option} (\Code{option}), \textsf{pair} (\Code{pair}), and \textsf{frequency} (\Code{pair}), do however, employ such predicates.}
\footnotesize
\bgroup
\setlength\tabcolsep{5pt}
\begin{tabular}{ccccccc|ccc}
\toprule
 & \#Branch & Rec & \#LVar & \#$\tvar{a}$ & \#P &\#UP & \#Query & (max. \#$\forall$,\#$\exists$) & total (avg. time) (ms)\\
\midrule
\textsf{return} & $1$ &   & $2$ & $1$ & $0$ & $0$ & $1$ & $(0, 0)$ & $0.14(0.14)$ \\
\textsf{bind} & $1$ &   & $5$ & $2$ & $2$ & $0$ & $3$ & $(2, 1)$ & $28.74(9.58)$ \\
\textsf{fmap} & $1$ &   & $4$ & $2$ & $2$ & $0$ & $2$ & $(2, 1)$ & $2.23(1.11)$ \\
\textsf{pair} & $1$ &   & $5$ & $2$ & $2$ & $2$ & $3$ & $(1, 2)$ & $2.61(0.87)$ \\
\textsf{option} & $2$ &   & $7$ & $1$ & $1$ & $3$ & $14$ & $(2, 3)$ & $5.91(0.42)$ \\
\textsf{union} & $2$ &   & $4$ & $1$ & $2$ & $0$ & $12$ & $(1, 1)$ & $2.48(0.21)$ \\
\textsf{oneof} & $1$ &   & $4$ & $1$ & $1$ & $0$ & $3$ & $(2, 1)$ & $4.35(1.45)$ \\
\textsf{frequency} & $1$ &   & $5$ & $1$ & $1$ & $2$ & $3$ & $(3, 2)$ & $3.04(1.01)$ \\
\textsf{list\_repeat} & $2$ & $\checkmark$ & $12$ & $1$ & $1$ & $4$ & $12$ & $(3, 5)$ & $11.48(0.96)$ \\
\textsf{fix} & $1$ &   & $5$ & $1$ & $1$ & $0$ & $15$ & $(2, 0)$ & $11.23(0.75)$ \\
\bottomrule
\end{tabular}
\egroup
\label{tab:qcheck}
\end{table}


In this section, we focus on test input generators from real-world applications as well as the underlying combinators provided by mainstream PBT frameworks.

The experimental result of verifying the monadic combinators for \autoref{sec:monad} is shown in \autoref{tab:qcheck}. These combinators are relatively simple and \name{} is able to verify them within $0.03$ seconds with small variance.

\begin{table}[h!]
\renewcommand{\arraystretch}{0.8}
\caption{ Experimental results for test generators of open-source repositories.  The columns follow the same format as in \autoref{tab:evaluation}, with an additional column showing the number of combinators (\#C) used by these generators. The verification time in the last column is given in seconds. }
\footnotesize
\bgroup
\setlength\tabcolsep{5pt}
\begin{tabular}{cccccc|ccc}
\toprule
 & \#Branch & Rec & \#LVar & \#UP & \#C & \#Query & (max. \#$\forall$,\#$\exists$) & total (avg. time) (s)\\
\midrule
\textsf{Tezos} & $7$ & $\checkmark$ & $40$ & $15$ & $3$ & $73$ & $(3, 35)$ & $10.75(0.15)$ \\
\textsf{Xen API} & $1$ &   & $21$ & $5$ & $3$ & $24$ & $(1, 16)$ & $6.51(0.27)$ \\
\textsf{Vellvm} & $10$ & $\checkmark$ & $41$ & $19$ & $4$ & $84$ & $(4, 19)$ & $2.05(0.02)$ \\
\textsf{Herdtools7} & $1$ &   & $32$ & $10$  & $3$ &  $90$ & $(2, 8)$ & $2.08(0.02)$ \\
\textsf{Zipperposition} & $7$ & $\checkmark$ & $71$ & $16$ & $4$ & $134$ & $(2, 49)$ & $15.81(0.12)$ \\
\bottomrule
\end{tabular}
\egroup
\label{tab:pbt-repos}
\end{table}

We also apply \name{} to identify coverage completeness violations
across five open-source repositories that use the \Code{QCheck}
framework to test their implementation. Each project deals with a
different real-world application domain: blockchain~\cite{Tezos},
virtualization~\cite{XenAPI}, compiler verification~\cite{Vellvm},
weak memory model testing~\cite{herdtools7}, and automated theorem
proving~\cite{zipperposition}. The details of these projects are as
follows:
\begin{enumerate}
\item \textbf{Tezos}. Tezos~\cite{Tezos} is a blockchain platform
  implemented in OCaml that supports decentralized assets and
  applications. Tezos writes its own generators to construct block
  trees, i.e., trees of block nodes representing the shape of a
  blockchain, where blocks are expected to have unique hash codes.
    \item \textbf{Xen API}. Xen API~\cite{XenAPI} is a management API
      for controlling the virtualization environments of then Xen
      Project Hypervisor~\cite{Xen}. We selected the file descriptor
      generator of the API, which produces random file descriptors
      with different attributes, e.g., file size, kind, read and write
      delay time. The generated test inputs are expected to be a
      subset of valid Unix file descriptors (e.g., not a symbolic
      link).
    \item \textbf{Vellvm}. Vellvm~\cite{Vellvm} is a formal
      verification framework for the LLVM IR, implemented in both
      OCaml and Coq/Rocq. We selected the generator for LLVM values
      used to provide as arguments of various LLVM functions under
      test. Given a specific LLVM type, this generator is expected to
      generate all well-typed LLVM values.
    \item \textbf{Herdtools7}. Herdtools7~\cite{herdtools7} is a
      collection of tools for modeling and analyzing concurrent
      programs with weak memory models. The generator \Code{Carpenter}
      in Herdtools7 produces random ASTs of the Arm Architecture
      Specification Language ~\cite{ASL}, where the generator is
      expected to generate every AST of a specified size.
    \item
      \textbf{Zipperposition}. Zipperposition~\cite{zipperposition} is
      an automated theorem prover that supports polymorphic types and
      higher-order unification. Test generators are used to produce
      valid first-order logic (e.g.,
      $\forall x{:}\alpha. \exists y{:}\alpha. x \not= y$) and
      lambda-free higher-order logic formulae.
\end{enumerate}

We used our polymorphic version of \name{} to check that these test
generators can cover all test inputs that satisfy the precondition of
the program under test. Thus, we type check generators against
coverage types that encode specifications extracted from README files,
documentation, and comments in the source code.  As shown in
\autoref{tab:pbt-repos}, these test generators are complex, including
1-10 control flow branches, 21-71 local variables, and 5-19
uninterpreted predicates. Most uninterpreted predicates are directly
converted from datatype constructors (e.g., $\Code{cons}$ for list
datatype).  Notably, \name{} reports that \emph{none} of these
generators are coverage complete with respect to the preconditions we
expected, with verification times ranging from $2.08$ to $15.81$
seconds. Due to the sophisticated nature of realistic test generators,
verification takes longer compared to previous benchmarks, though it
remains within acceptable limits. One observation is that the result
of type checking does not significantly affect the verification time,
which differs from standard refinement type checkers that can return
early when detecting safety violations in one control flow. This is
because \name{} cannot determine the coverage completeness of a
program without analyzing \emph{all }corresponding control flows, as
discussed in \autoref{sec:rules}.

In the rest of section, we highlight three representative categories
of coverage incompleteness detected by \name{} for these benchmarks,
examine the underlying causes, and suggest candidate fixes to address
these coverage missing.


\paragraph{Limited sampling range}
Manually written test input generators often rely on restricted
sampling ranges, resulting in coverage completeness violations. For
example, \textsf{Vellvm}~\cite{Vellvm} generates \Code{LLVM int32} and
\Code{LLVM int64} values only within the range $0$–$10000$, and the
generators in the \textsf{herdtools7} framework~\cite{herdtools7} also
restrict integer literals appearing in theorems to this
interval. These generators typically use QCheck’s built-in
$\Code{Gen.nat}$ combinator, which only produces natural numbers below
$10000$. While developers adopt this restriction under the assumption
that smaller values are more likely to detect bugs, such approaches
(theoretically) compromise completeness.  One solution is to run two
sets of tests, one with smaller values and one with the full range of
values.  Alternatively, it is straightforward to build a repaired
version of this generator by using QCheck’s \Code{frequency}
combinator to combine the original with the default, unconstrained
generator, adjusting the weights so that it is biased towards the
original input space while retaining the possibility of producing
other values:
\vspace{-0.4cm}
\begin{minted}[xleftmargin=5pt, numbersep=4pt, linenos = false, fontsize = \small, escapeinside=!!]{OCaml}
let nat = frequency 2 (function
                       | 0 -> (10, Gen.nat)
                       | _ -> (1, full_nat_gen))
\end{minted}

\paragraph{Fixed values}
\textsf{Zipperposition}~\cite{zipperposition} uses the \Code{oneof}
combinator to restrict variable names to one of three fixed options:
\Code{X}, \Code{Y}, or \Code{Z}. Similarly, the \textsf{Xen}
hypervisor~\cite{XenAPI} constrains file sizes to $9$ predetermined
lengths and only uses one of $4$ fixed float number (e.g., $0.001$,
$0.01$, $0.1$, and $0.4$) as file read and write delay times. While
such choices may reflect the developer's intuition about likely corner
cases, they also prevent the exploration of errors outside these
considerations. The repair strategies proposed earlier apply here as
well.

\begin{figure}
    \centering
    \begin{minted}[xleftmargin=5pt, numbersep=4pt, linenos = true, fontsize = \small, escapeinside=!!]{OCaml}
        type 'a tree =
        | Leaf of 'a
        | Node1 of ('a * 'a tree)
        | Node2 of ('a * 'a tree * 'a tree)

        let rec tree_gen (blocks: block list): block tree gen =
            match block with
            | [x] -> return (Leaf x)
            | x :: xs -> (
              ... (* the case producing Leaf *)
              ... (* the case producing Node1 *)
              int_bound (List.length xs - 1) >>= fun n ->
              let left, right = List.split_n n xs in
              tree_gen left >>= fun left ->
              tree_gen right >>= fun right ->
              return (Node2 (x, left, right))

        tree_gen (Block.set_to_list unique_nonempty_block_gen)
    \end{minted}
    \caption{A tree-like block chain generator.}
    \label{fig:tezos}
\end{figure}

\paragraph{Datatype transformations}
\name{} identified an interesting coverage completeness violation in
the \textsf{Tezos} blockchain project~\cite{Tezos}, where test input
generators construct block trees, i.e., trees of block nodes
representing the shape of the block chain. Instead of generating such
trees directly, Tezos implements a helper function $\Code{tree\_gen}$
(shown in \autoref{fig:tezos}) that builds trees from lists of
blocks. Specifically, $\Code{tree\_gen}$ accepts a non-empty,
duplicate-free list of blocks and returns a generator that produces a
tree with non-duplicate blocks. The code comments state that this
approach allows better control over sampling distributions and
uniqueness.  Thus, the expected coverage refinement type of
$\Code{tree\_gen}$ should be:
\begin{align*}
    \Code{tree\_gen} :\;&\Code{l}{:}\ort{\List[\Code{block}]}{\I{uniqList}(\vnu)}\sarr
    \\&\quad \Murt{\Tree[\Code{block}]}{\I{uniqTree}(\vnu) \land \forall u. \I{memTree}(\vnu, u)\iff\I{mem}(\Code{l}, u) }
\end{align*}\noindent
However, the actual implementation in Tezos fails to satisfy this
specification and is rejected by \name{}. The root of every generated
tree (line 16) is deterministically selected as the head of the input
list. Moreover, the use of $\Code{List.split\_n}$ partitions the input
list into subtlists that preserve element ordering, leading to the
left subtree always containing earlier elements in the list. Since
Tezos applies $\Code{tree\_gen}$ to sorted lists (line 18), the
resulting block trees consistently place the smallest element at the
root, with left subtree elements always being smaller than those in
the right subtree. The root cause of this error lies in the complexity
of maintaining coverage completeness through transformations between
inductive datatypes (e.g., from lists to trees).

One possible solution is to randomly permute the input list of
blocks. Another patch is to replace the list input with a \Code{set}:
\begin{align*}
    \Code{tree\_gen} :\;&\Code{s}{:}\ort{\Code{block\;set}}{\top}\sarr
    \\&\quad \Murt{\Tree[\Code{block}]}{\I{uniqTree}(\vnu) \land \forall u. \I{memTree}(\vnu, u)\iff\I{memSet}(\Code{s}, u) }
\end{align*}\noindent
A coverage complete implementation would then randomly select the root
node from this set and partition the set into two disjoint subsets
using a randomized split instead of relying on list indexing. We have
submitted this patch to the \textsf{Tezos} codebase.

\paragraph{Discussion}
As with any verification methodology, the question of whether the
completeness violations discovered by \name{} in our case studies are
true bugs or represent deliberate trade-offs is one that must
ultimately be answered by the authors of these generators. We
emphasize, however, that coverage types do enable developers to
rigorously specify the intended behaviors of an input generator and to
validate that an implementation exhibits those behaviors. The ability
to do so has concrete benefits for both the developers of input
generators and the test engineers that rely on them.

First, potential incompleteness bugs can help to surface
semantic mismatches between the test engineer's and system developer's
understandings of the precondition of the system under test. In the
case of the \textsf{Tezos} case study, for example, the test engineer
may have believed that a tree of distinct blocks is isomorphic to a
flattened, sorted list of blocks. By identifying this mismatch,
\name{} can provide actionable feedback to both parties, enabling them
to align their mental models of the system under test.

On the other hand, the coverage type of a generator also provides
useful documentation to its clients. As our case studies shows, 
developers may deliberately prioritize
efficient sampling over coverage completeness in practice, embedding \emph{ad hoc}
restrictions and choices in their implementations. When these
trade-offs are not explicitly documented, as is often the case, the
users of such generators may inadvertently fail to probe system
behaviors on relevant portions of the input space. A generator for a parser or
string manipulating program that
restricts variable names to \Code{a}, \Code{b}, and \Code{c} will fail to
discover errors that can only be triggered by
programs with more than three distinct variables. Similarly, only
considering small integers may fail to probe arithmetic overflow cases
like ${-}2^{31} / {-}1$ (the range of \Code{int32} is
${-}2^{31} \sim 2^{31}{-}1$). By codifying these sorts of trade-offs
as explicit type-level specifications, \name{} helps clients
understand and justify the use of a particular generator when testing a
system.

Finally, as \autoref{sec:synthesis} demonstrated, coverage types can
provide useful information when constructing new generators,
particularly when doing so automatically~\cite{LZ+25}. These types can
also be helpful when manually defining generators
\emph{compositionally} via monadic combinators, e.g., when a test
engineer wants to combine generators that sample from different
distributions. For example, the Xen API~\cite{XenAPI} includes the
file descriptor generator shown in \autoref{fig:fd-gen}, which applies
the \Code{oneof} combinator to generators for singleton lists and
general file descriptor lists. Currently, these generators are
informally described in the accompanying code comment; coverage
types provide a rigorous mechanism for capturing these sorts of
behaviors along with a principled characterization of how
they fit together.

\begin{figure}[ht!]
  \centering
  \begin{minted}[xleftmargin=5pt, numbersep=4pt, linenos = true, fontsize = \small, escapeinside=!!]{OCaml}
  (* generates 2 kinds of lists:
    - lists that contain only a single file kind
    - lists that contain multiple file kinds

    This is important for testing [select], because a single
      [Unix.S_REG] would cause it to return immediately,
      making it unlikely that we're actually testing
      the behaviour for other file descriptors.
  *)
  oneof [g1; g2]
  \end{minted}
  \caption{A file descriptor generator that uses oneof to combine two generators with different coverage guarantees.}
  \label{fig:fd-gen}
\end{figure}

\section{Related Work}
\label{sec:related}

\cbnewadding{ Property-based testing (PBT) is an automated testing
  approach that uses random tests to determine if a user-defined
  property is satisfied by the component under test. In a PBT setting,
  test inputs are expected to satisfy the preconditions of the
  functions under test and are equipped with a ``shrink'' procedure
  that automatically reduces failing inputs to minimal examples for
  reporting to users. First introduced for
  Haskell~\cite{claessen2011quickcheck} to test the behavioral
  correctness of simple programs over algebraic datatypes (ADTs), PBT
  has since been successfully applied to a number of application
  domains, including operating systems~\cite{XenAPI},
  compilers~\cite{NotUseless}, networks~\cite{networkErlang}, memory
  models~\cite{herdtools7}, web applications~\cite{Quickstrom}, and
  quantum computation~\cite{PBTQuantum}. Meanwhile, the properties in
  PBT have also been extended to richer logics. For instance,
  \citet{Quickstrom} allows users to provide specifications in linear
  temporal logic (LTL) to guide the testing of interactive
  applications. Property-based testing can also validate performance
  specifications, in addition to safety proeprties. For example,
  \citet{networkErlang} uses PropEr~\cite{propEr}, a PBT framework in
  Erlang, to verify that the energy consumption of sensor networks
  remains within a specific bound. }

\cbnewadding{ Although PBT test generators can be derived
  automatically for arbitrary datatypes, handwritten test generators
  remain widely used. Etna~\cite{Etna} evaluates the test generators
  of mainstream PBT frameworks in Coq/Rocq and Haskell, concluding
  that ``bespoke generators'' (user-defined test generators) perform
  well in most cases. However, a recent empirical
  study~\cite{PBTinPractice} analyzed the use of property-based
  testing (PBT) in industrial practice. One key finding is that
  writing test generators for complex precondition properties (e.g.,
  valid red-black trees) remains a significant challenge, even when
  PBT frameworks provide extensive combinators and monadic APIs.
  Developers described the process as ``tedious'' and ``high-effort'',
  particularly when balancing coverage and input quality (e.g.,
  generating inputs closer to realistic distributions). \name{}'s
  integration of PBT abstractions within an expressive refinement-type
  system tailored for coverage provides an automated validation
  mechanism to ensure generators satisfy rich safety and completeness
  constraints.
}

\cbnewadding{There has also been recent work that reduce user effort
  in writing high-quality test generators. Tyche~\cite{Tyche} enables
  better visualization of test generator quality, including valid and
  invalid test cases, duplicate test cases, and corresponding
  distribution. \citet{PBTCombinatorial} demonstrates that test
  generators for complex ADTs can be improved through combinatorial
  testing~\cite{CT}, where the domain of input ADT instances can be
  partitioned by combinations of datatype constructors. Several works
  have considered how to automatically synthesize sound and complete
  generators for a target precondition: the QuickChick PBT framework
  for Coq/Rocq, for example, is capable of deriving generators from
  preconditions expressed as inductive
  relations~\cite{LPP18,PEL+22}. \citet{LZ+25} present a technique for
  automatically repairing incomplete generators used
  coverage-type-guided program synthesis; when applied to a
  control-flow sketch, this technique is capable of synthesizing a
  coverage-complete generator for a given precondition. The recently
  proposed Palmedes~\cite{GPTSLP+25} tool uses deductive synthesis to
  generate sound and complete generators from precondition that are
  expressed in Lean using an extensible set of recursion schemes.
  Another set of works consider the more specific problem of how to
  automatically adjust the distribution of an input generator. The
  Dragen tool~\cite{MRH+18, PBTRichADT}, for example, refines the
  default input generators derived by the \Code{QuickCheck} framework
  by tuning input weights of \Code{frequency} combinators using
  additional structural information provided by users. Loaded
  Dice~\cite{TGGMPV+25} relaxes this restriction, allowing users to
  automatically tune the distribution of an arbitrary generator by
  first lowering it into a probabilistic DSL embedded in Julia and
  then optimizing those weights based on a user-provided objective
  function.}

A general limitation of PBT is that its effectiveness suffers when the
property of interest has a strict
precondition~\cite{Lampropoulos2018RandomTF}, because most of the
inputs produced by a purely random test generation strategy will be
simply discarded. As a result, there has been much recent interest on
improving the coverage of test generators with respect to a particular
precondition. Proposed solutions range from adopting ideas from
fuzzing~\cite{afl, crowbar} to intelligently mutate the outputs
produced by the generator~\cite{quickcheck-coverage-guided, PL+19}, to
focusing on generators for particular classes of inputs (e.g.,
well-typed programs)~\cite{FC+15, PCRH+11, CSmith}, to automatically
building complete-by-construction generators~\cite{CDP14, LG+17,
  LPP18}. While sharing broadly similar goals with these proposals,
our approach differs significantly in its framing of coverage in
purely type-theoretic terms. This fundamental change in perspective
allows us to statically and compositionally verify coverage properties
of a generator without the need for any form of instrumentation on, or
runtime monitoring of, the program under test (as in~\cite{crowbar,
  quickcheck-coverage-guided}).  Unlike other approaches
  that have also considered the verification of a generator's coverage
  properties~\cite{DH+03,DH+04,PH+15} using a mechanized proof
  assistant, our proposed type-based framing is highly-automated and
inherently  compositional.  Expressing coverage as part of a type system also
allows us to be agnostic to (a) how generators are constructed,
(b) the particulars of the application domain~\cite{FC+15, PCRH+11, CSmith, NotUseless},
and (c) the specific structure of the properties being
tested~\cite{LG+17,LPP18}.  \name's ability to specify and type-check
a complex coverage property depends only on whether we can express a
desired specification using available uninterpreted predicates.


A number of logics have been proposed for reasoning about
underapproximations of program behavior, including the recently
developed incorrectness logic (IL)~\cite{IncorrectnessLogic,RBD+20}, reverse Hoare
logic (RHL)~\cite{Reverse+Hoare+Logic}, and dynamic logic
(DL)~\cite{Pratt1976}. Both IL and RHL are formalisms similar to Hoare
logic, but support composable specifications that assert
underapproximate postconditions, with IL adding special
post-assertions for error states. IL was originally proposed as a way
of formalizing the conditions under which a particular program point
(say an error state) is guaranteed to be reachable, and has recently
been used in program analyses that discover memory
errors~\cite{LR+22}.  DL, in contrast, reinterprets Hoare logic as a
multi-modal logic equipped with operators for reasoning about the
existence of executions that end in a state satisfying some desired
postcondition.  This paper instead provides the first development that
interprets these notions in the context of a type system for a rich
functional language.  While our ideas are formulated in the context of
verifying coverage properties for test input generators, we believe
our framework can be equally adept in expressing type-based program
analyses for bug finding or compiler optimizations.

Our focus on reasoning about coverage properties of test input
generators distinguishes our approach, in obvious ways from other
refinement type-based testing solutions such as
\textsc{target}~\cite{SVJ11}.  Nonetheless, our setup follows the same
general verification playbook as Liquid Types~\cite{JV21,LiquidHaskell} --- our
underapproximate specifications are identical to their overapproximate
counterpart, except that we syntactically distinguish the return types
for functions to reflect their expected underapproximate (rather than
overapproximate) behavior.  An important consequence of this design is
that the burden of specifying and checking the coverage behavior of a
program is no greater than specifying its safety properties.

Another related line of work has explored how to reason about the
distribution of data produced by a function~\cite{ADD+17, BZS+19},
with a focus on ensuring that these distributions are free of unwanted
biases. These works have considered decision-making and
machine-learning applications, in which these sorts of fairness
properties can be naturally encoded as (probabilistic) formulas in
real arithmetic. In contrast, coverage types can only verify that a
generator has a nonzero probability of producing a particular
output. Extending our type system and its guarantees to provide
stronger fairness guarantees about the distribution of the sorts of
discrete data produced by test input generators is an exciting
direction for future work.

\section{Conclusion}
\label{sec:conc}

This paper adapts principles of underapproximate reasoning found in
recent work on Incorrectness Logic to the specification and automated
verification of test input generators used in modern property-based
testing systems.  Specifications are expressed in the language of
refinement types, augmented with coverage types, types that reflect
underapproximate constraints on program behavior.  A novel
bidirectional type-checking algorithm enables an expressive form of
inference over these types.
Our experimental results demonstrate that
our approach is capable of verifying both sophisticated hand-written
generators, as well as being able to successfully identify
type-correct (in an overapproximate sense) but coverage-incomplete
generators produced from a deductive refinement type-aware
synthesizer.
\cbnewadding{We also demonstrate how polymorphic extensions to our type system can be used to successfully detect coverage violations of complex test generators implemented using monadic combinators in real-world developments.}

\section*{Acknowledgements}

This work was supported in part by DARPA under the grants
HR0011-19-C-0073 and HR0011-22-90062, the Department of Defense under
grant FA8649-21-P-1511, and the NSF under grant CCF-2321680.

\newpage
\section{Data Availability}\label{sec:data}

An artifact containing this implementation, our benchmark suite, results and
corresponding Rocq proofs (i.e., our soundness theorems Theorem~\ref{theorem:type-sound}) is publicly available on
Zenodo~\cite{artifact}. 
The benchmark suite of our new case study is
available on GitHub:
https://github.com/zhezhouzz/CoverageType/tree/jfp (Commit c6fa15d)

\subsection*{Conflicts of Interest}

The authors have no conflicts of interest to report.

\bibliographystyle{JFPlike}
\bibliography{bibliography}

\ifdefined\showappendix
  \appendix
  \newpage
  \section{Operational Semantics}

\begin{figure}[h!]
 {\small
  {\normalsize
\begin{flalign*}
 &\text{\textbf{Operational Semantics }} & \fbox{$e \hookrightarrow e$}
\end{flalign*}
}
\\ \
\begin{prooftree}
\hypo{\I{op}\ \overline{v} \equiv v_y }
\infer1[\textsc{StAppOp}]{ 
\zlet{y}{\I{op}\ \overline{v}}{e} \hookrightarrow e[y\mapsto v_y]
}
\end{prooftree}
\\ \ \\ \ \\
\begin{prooftree}
\hypo{e_1 \hookrightarrow e_1'}
\infer1[\textsc{StLetE1}]{
\zlet{y}{e_1}{e_2} \hookrightarrow \zlet{y}{e_1'}{e_2}
}
\end{prooftree}
\quad
\begin{prooftree}
\hypo{}
\infer1[\textsc{StLetE2}]{
\zlet{y}{v}{e} \hookrightarrow e[y\mapsto v]
}
\end{prooftree}
\\ \ \\ \ \\
\begin{prooftree}
\hypo{}
\infer1[\textsc{StLetAppLam}]{ 
\zlet{y}{\zlam{x}{t}{e_1}\ v_x}{e_2} \hookrightarrow \zlet{y}{e_1[x\mapsto v_x]}{e_2}
}
\end{prooftree}
\\ \ \\ \ \\
\begin{prooftree}
\hypo{}
\infer1[\textsc{StLetAppFix}]{
\parbox{100mm}{\center
$\zlet{y}{\zfix{f}{t}{x}{t_x}{e_1}\ v_x}{e_2} \hookrightarrow$\quad
$\zlet{y}{(\zlam{f}{t}{e_1[x\mapsto v_x]}) \ (\zfix{f}{t}{x}{t_x}{e_1})}{e_2}$
}
}
\end{prooftree}
\\ \ \\ \ \\
\begin{prooftree}
\hypo{}
\infer1[\textsc{StMatch}]{ 
\match{d_i \ \overline{v_j}} \overline{d_i\ \overline{y_j} \to e_i} \hookrightarrow e_i[\overline{y_j \mapsto v_j}]
}
\end{prooftree}
}
\caption{Small Step Operational Semantics}
    \label{fig:semantics}
\end{figure}

The operational semantics of our core language is shown in \autoref{fig:semantics}, which is a standard small step semantics.
  \section{Basic Typing rules}

\begin{figure}[h!]
 {\small
 {\normalsize
\begin{flalign*}
 &\text{\textbf{Basic Typing }} & \fbox{$\Gamma \basicvdash e : t$}
\end{flalign*}
}
\\ \
\begin{prooftree}
\hypo{}
\infer1[\textsc{BtErr}]{ 
\Gamma \basicvdash \exn : t
}
\end{prooftree}
\quad
\begin{prooftree}
\hypo{}
\infer1[\textsc{BtConst}]{
\Gamma \basicvdash c : \S{Ty}(c)
}
\end{prooftree}
\quad
\begin{prooftree}
\hypo{}
\infer1[\textsc{BtOp}]{ 
\Gamma \basicvdash \I{op} : \S{Ty}(\I{op})
}
\end{prooftree}
\quad
\begin{prooftree}
\hypo{ \Gamma(x) = t }
\infer1[\textsc{BtVar}]{ 
\Gamma \basicvdash x : t
}
\end{prooftree}
\\ \ \\ \ \\
\begin{prooftree}
\hypo{\Gamma, x{:}t_1 \basicvdash e : t_2}
\infer1[\textsc{BtFun}]{ 
\Gamma \basicvdash \zlam{x}{t_1}{e} : t_1\sarr t_2
}
\end{prooftree}\quad
\begin{prooftree}
\hypo{\Gamma, f{:}t_1\sarr t_2 \basicvdash \zlam{x}{t_1}{e} : t_1\sarr t_2}
\infer1[\textsc{BtFix}]{ 
\Gamma \basicvdash \zfix{f}{t_1\sarr t_2}{x}{t_1}{e} : t_1\sarr t_2
}
\end{prooftree}
\\ \ \\ \ \\
\begin{prooftree}
\hypo{\emptyset \basicvdash e_1 : t_x \quad \Gamma, x{:}t_x \basicvdash e_2 : t}
\infer1[\textsc{BtLetE}]{ 
\Gamma \basicvdash \zlet{x}{e_1}{e_2} : t
}
\end{prooftree}
\quad
\begin{prooftree}
\hypo{\S{Ty}(\I{op}) = \overline{t_i}\sarr t_x \quad \Gamma \basicvdash v_i : t_i \quad \Gamma, x{:}t_x \basicvdash e : t}
\infer1[\textsc{BtAppOp}]{ 
\Gamma \basicvdash \zlet{x}{\I{op}\ \overline{v_i}}{e} : t
}
\end{prooftree}
\\ \ \\ \ \\
\begin{prooftree}
\hypo{\Gamma \basicvdash v_1 : t_2 \sarr t_x \quad \Gamma \basicvdash v_2 : t_2 \quad \Gamma, x{:}t_x \basicvdash e : t}
\infer1[\textsc{BtApp}]{ 
\Gamma \basicvdash \zlet{x}{v_1\ v_2}{e} : t
}
\end{prooftree}
\\ \ \\ \ \\
\begin{prooftree}
\hypo{\Gamma \basicvdash v : t_v \quad 
\forall i, \S{Ty}(d_i) = \overline{t_j} \sarr t_v \quad
\Gamma, \overline{y_j{:}t_j} \basicvdash e_i : t}
\infer1[\textsc{BtMatch}]{ 
\Gamma \basicvdash \match{v} \overline{d_i\ \overline{y_j} \to e_i} : t
}
\end{prooftree}
}
\caption{Basic Typing Rules}
    \label{fig:basic-type-rules}
\end{figure}

\begin{figure}[h!]
 {\small
 {\normalsize
\begin{flalign*}
 &\text{\textbf{Basic Qualifier Typing }} & \fbox{$\Gamma \basicvdash l: t  \quad \Gamma \basicvdash \phi: t$}
\end{flalign*}}
\\ \
\begin{prooftree}
\hypo{\S{Ty}(c) = t }
\infer1[\textsc{BtLitConst}]{
\Gamma \basicvdash c : t
}
\end{prooftree}
\quad
\begin{prooftree}
\hypo{ \Gamma(x) = t }
\infer1[\textsc{BtLitVar}]{
\Gamma \basicvdash x : t
}
\end{prooftree}
\quad
\begin{prooftree}
\hypo{ }
\infer1[\textsc{BtTop}]{
\Gamma \basicvdash \top : \Code{bool}
}
\end{prooftree}
\quad
\begin{prooftree}
\hypo{ }
\infer1[\textsc{BtBot}]{
\Gamma \basicvdash \bot : \Code{bool}
}
\end{prooftree}
\\ \ \\ \ \\
\begin{prooftree}
\hypo{\S{Ty}(\primop) = \overline{t_i}{\sarr}t \quad \forall i. \Gamma \basicvdash l_i : t_i }
\infer1[\textsc{BtLitOp}]{
\Gamma \basicvdash \primop\,\overline{l_i} : t
}
\end{prooftree}
\quad
\begin{prooftree}
\hypo{\S{Ty}(\I{uf}) = \overline{t_i}{\sarr}\Code{bool} \quad \forall i. \Gamma \basicvdash l_i : t_i }
\infer1[\textsc{BtLitMp}]{
\Gamma \basicvdash \I{uf}\,\overline{l_i} : t
}
\end{prooftree}
\\ \ \\ \ \\
\begin{prooftree}
\hypo{
\Gamma \basicvdash \phi_1 : \Code{bool} \quad
\Gamma \basicvdash \phi_2 : \Code{bool} }
\infer1[\textsc{BtAnd}]{
\Gamma \basicvdash \phi_1 \land \phi_2 : \Code{bool}
}
\end{prooftree}
\quad
\begin{prooftree}
\hypo{
\Gamma \basicvdash \phi_1 : \Code{bool} \quad
\Gamma \basicvdash \phi_2 : \Code{bool} }
\infer1[\textsc{BtOr}]{
\Gamma \basicvdash \phi_1 \lor \phi_2 : \Code{bool}
}
\end{prooftree}
\\ \ \\ \ \\
\begin{prooftree}
\hypo{
\Gamma, x{:}b \basicvdash \phi : \Code{bool} }
\infer1[\textsc{BtForall}]{
\Gamma \basicvdash \forall x{:}b. \phi : \Code{bool}
}
\end{prooftree}
\quad
\begin{prooftree}
\hypo{
\Gamma, x{:}b \basicvdash \phi : \Code{bool} }
\infer1[\textsc{BtExists}]{
\Gamma \basicvdash \exists x{:}b. \phi : \Code{bool}
}
\end{prooftree}
\quad
\begin{prooftree}
\hypo{
\Gamma \basicvdash \phi : \Code{bool} }
\infer1[\textsc{BtNeg}]{
\Gamma \basicvdash \neg \phi : \Code{bool}
}
\end{prooftree}
}
\caption{Basic Qualifier Typing Rules}
    \label{fig:basic-qualifier-type-rules}
\end{figure}

The basic typing rules of our core language and qualifiers are shown
in \autoref{fig:basic-type-rules} and
\autoref{fig:basic-qualifier-type-rules}. We use an auxiliary function
$\S{Ty}$ to provide a basic type for the primitives of our language,
e.g., constants, built-in operators, and data constructors.

\section{Type System Details}

The subset relation between the denotation of two refinement types $\tau_1$ and $\tau_2$ under a type context $\Gamma$ (written $\denotation{\tau_1}_{\Gamma} \subseteq \denotation{\tau_1}_{\Gamma}$) is:
\begin{alignat*}{2}
    \denotation{\tau_1}_{\emptyset} \subseteq \denotation{\tau_2}_{\emptyset}&\doteq \denotation{\tau_1} \subseteq  \denotation{\tau_2}
    &\\ \denotation{\tau_1}_{x{:}\tau_x, \Gamma} \subseteq \denotation{\tau_1}_{x{:}\tau_x, \Gamma} &\doteq 
    \forall v_x \in \denotation{\tau_x},\\
    &\qquad \denotation{\tau_1 [x\mapsto v_x]}_{\Gamma [x\mapsto v_x]} \subseteq 
    \denotation{\tau_2 [x\mapsto v_x]}_{\Gamma [x\mapsto v_x]}
    &\quad \text{ if $\tau
      \equiv \ort{b}{\phi}$} \\
       \denotation{\tau_1}_{x{:}\tau_x \Gamma} \subseteq \denotation{\tau_2}_{x{:}\tau_x \Gamma} 
      &\doteq \exists \hat{e}_x \in
    \denotation{\tau_x}, \forall v_x, \hat{e}_x \hookrightarrow^*
      v_x \implies  \\
    &\qquad \denotation{\tau_1
    [x\mapsto v_x]}_{\Gamma [x\mapsto v_x]} \subseteq \denotation{\tau_2
    [x\mapsto v_x]}_{\Gamma [x\mapsto v_x]}
    &\text{ otherwise}
\end{alignat*}
The way we interpret the type context $\Gamma$ here is the same as the definition of the type denotation under the type context, but we keep the denotation of $\tau_1$ and $\tau_2$ as the subset relation under the same interpretation of $\Gamma$, that is under the \emph{same} substitution $[x\mapsto v_x]$. This constraint is also required by other refinement type systems, which define the denotation of the type context $\Gamma$ as a set of substitutions, with the subset relation of the denotation of two types holding under the \emph{same} substitution. However, our type context is more complicated, since it has both under- and overapproximate types that are interpreted via existential and universal quantifiers, and cannot simply be denoted as a set of substitution. Thus, we define a subset relation over denotations under a type context to ensure the ame substitution is applied to both types.
  \section{Typing Algorithm Details}

\begin{figure}[h!]
    \centering
  \begin{minipage}{0.48\textwidth}
  \begin{algorithm}[H]
  \caption{Disjunction}
  \Procedure{$\disjop (\tau_1, \tau_2) := $}{
    \Match{$\tau_1, \tau_2$}{
      \Case{$\urt{b}{\phi_1}, \urt{b}{\phi_2}$}{
        \Return{$\urt{b}{\phi_1 \lor \phi_2}$}\;
      }
      \Case{$\ort{b}{\phi_1}, \ort{b}{\phi_2}$}{
        \Return{$\ort{b}{\phi_1 \lor \phi_2}$}\;
      }
      \Case{$a{:}\tau_{a_1} \sarr \tau_1, a{:}\tau_{a_2} \sarr \tau_2$}{
        $\tau_a \leftarrow \conjop (\tau_{a_1},\tau_{a_2})$\;
        \Return{$a{:}\tau_a \sarr \disjop (\tau_{1}, \tau_{2})$}\;
      }
    }
    }
  \end{algorithm}
  \end{minipage}
  \hfill
  \begin{minipage}{0.49\textwidth}
  \begin{algorithm}[H]
  \caption{Conjunction}
  \Procedure{$\conjop (\tau_1 ,\tau_2 ):= $}{
    \Match{$\tau_1, \tau_2$}{
      \Case{$\urt{b}{\phi_1}, \urt{b}{\phi_2}$}{
        \Return{$\urt{b}{\phi_1 \land \phi_2}$}\;
      }
      \Case{$\ort{b}{\phi_1}, \ort{b}{\phi_2}$}{
        \Return{$\ort{b}{\phi_1 \land \phi_2}$}\;
      }
      \Case{$a{:}\tau_{a_1} \sarr \tau_1, a{:}\tau_{a_2} \sarr \tau_2$}{
        $\tau_a \leftarrow \disjop (\tau_{a_1},\tau_{a_2})$\;
        \Return{$a{:}\tau_a \sarr \conjop(\tau_{1},\tau_{2})$}\;
      }
    }
    }
  \end{algorithm}
  \end{minipage}
  \caption{Type Disjunction and Conjunction}
    \label{algo:disj-conj}
  \end{figure}
  
  We implement the auxiliary $\disjop$ operation using the mutually
  recursive functions shown in \autoref{algo:disj-conj}. The disjunction
  of two underapproximate base types $\urt{b}{\nu = 1}$ and
  $\urt{b}{\nu = 2}$ is simply the disjunction of their qualifiers:
  $\urt{b}{\nu = 1 \lor \nu = 2}$ (lines 3-4), as is the disjunction of
  two overapproximate base types (lines 5-6).  We build the disjunction
  of two function types by taking the conjunction of their argument
  types and the disjunction of their return type (lines 7-9). We build
  the disjunction of a sequence of types by folding $\disjop$ over them:
  \begin{align*}
    \disjop(\tau_1, \tau_2, ..., \tau_{n-1}, \tau_n) \doteq \disjop(\tau_1,\disjop(\tau_2, ..., \disjop(\tau_{n-1}, \tau_n)))
  \end{align*}
  
  \begin{figure}[h!]
    \centering
  \begin{minipage}{0.49\textwidth}
  \begin{algorithm}[H]
  \caption{Exists}
  \Procedure{$\existsop (x, \urt{b}{\phi_x}, \tau) := $}{
    \Match{$\tau$}{
      \Case{$\urt{b}{\phi}$}{
        $\phi' \leftarrow \exists x. \phi_x[\nu\mapsto x] \land \phi$\;
        \Return{$\urt{b}{\phi'}$}\;
      }
      \Case{$\ort{b}{\phi}$}{
        $\phi' \leftarrow \exists x. \phi_x[\nu\mapsto x] \land \phi$\;
        \Return{$\ort{b}{\phi'}$}\;
      }
      \Case{$a{:}\tau_{a} \sarr \tau$}{
        $\tau_a' \leftarrow \forallop (x, \urt{b}{\phi_x}, \tau_{a})$\;
        $\tau' \leftarrow \existsop(x, \urt{b}{\phi_x}, \tau)$\;
        \Return{$a{:}\tau_a' \sarr \tau'$}\;
      }
    }
    }
  \end{algorithm}
  \end{minipage}
  \hfill
  \begin{minipage}{0.49\textwidth}
  \begin{algorithm}[H]
  \caption{Forall}
    \Procedure{$\forallop (x, \urt{b}{\phi_x}, \tau) := $}{
    \Match{$\tau$}{
      \Case{$\urt{b}{\phi}$}{
        $\phi' \leftarrow \forall x. \phi_x[\nu\mapsto x] \impl \phi$\;
        \Return{$\urt{b}{\phi'}$}\;
      }
      \Case{$\ort{b}{\phi}$}{
        $\phi' \leftarrow \forall x. \phi_x[\nu\mapsto x] \impl \phi$\;
        \Return{$\ort{b}{\phi'}$}\;
      }
      \Case{$a{:}\tau_{a} \sarr \tau$}{
        $\tau_a' \leftarrow \existsop (x, \urt{b}{\phi_x}, \tau_{a})$\;
        $\tau' \leftarrow \forallop(x, \urt{b}{\phi_x}, \tau)$\;
        \Return{$a{:}\tau_a' \sarr \tau'$}\;
      }
    }
    }
  \end{algorithm}
  \end{minipage}
  \caption{Embedding bindings into type qualifiers}
    \label{algo:exists-forall}
  \end{figure}
  
  The operation $\existsop(x, \urt{b}{\phi_x}, \tau)$ generalizes $\disjop$ by aggregating not just two types, but all instances of $\tau[x \mapsto v]$ for every $v$ such that $\phi_x$ holds.
  Thus, our algorithm also implements $\existsop$ using the pair of
  mutually recursive functions shown in \autoref{algo:exists-forall}. 
  To embed an underapproximate type binding into an under- or
  over-approximate base type, we simply add the appropriate quantifier
  into its type qualifier: thus, embedding the binding
  $x{:}\nuut{nat}{\nu > 0}$ into the type $\nuut{nat}{\nu = x + 1}$
  produces the type $\nuut{nat}{\exists x.  x > 0 \land \nu = x +
    1}$. 
  Bindings are
  embedded into a function type by contravariantly moving the binding
  into its argument type and covariantly into its result type. 
  To illustrate the necessity of embedding bindings contravariantly into function types, consider the following function, which returns normally only if the input parameter $y$ exceeds a random value $x$:
  \begin{minted}[fontsize = \small,xleftmargin=10pt, escapeinside=??]{ocaml}
    let x = 1 ?$\oplus$? 2 in
    fun (y: int) -> if y > x then x else err
  \end{minted}
  Within the type context $x{:}\urt{\Int}{\vnu = 1 \lor \vnu = 2}$, we can assign the function the type $y{:}\ort{\Int}{\vnu > x} \to \urt{\Int}{\vnu = x}$. After applying the $\existsop$ operation to embed the binding for $x$, we obtain:
  \begin{align*}
    &y{:}\ort{\Int}{\forall x.\ (x = 1 \lor x = 2) \impl \vnu > x}\ \to\ \urt{\Int}{\exists x.\ (x = 1 \lor x = 2) \land \vnu = x} \\
  \equiv\;\; & y{:}\ort{\Int}{\vnu > 2} \to \urt{\Int}{\vnu = 1 \lor \vnu = 2}
  \end{align*}
  In contrast, incorrectly treating the parameter position covariantly leads to unsoundness:
  \begin{align*}
    &y{:}\ort{\Int}{\exists x.\ (x = 1 \lor x = 2) \land \vnu > x}\ \to\ \urt{\Int}{\exists x.\ (x = 1 \lor x = 2) \land \vnu = x} \\
  \equiv\;\; & y{:}\ort{\Int}{\vnu > 1} \to \urt{\Int}{\vnu = 1 \lor \vnu = 2}
  \end{align*}
  Here, if $y = 2$, the function could only return $1$, violating the intended behavior.

\paragraph{SMT Query Encoding for data types} In order to reason over data types, we allow the user to specify refinement types with uninterpreted predicates (e.g., $\I{mem}$) and quantifiers ($e.g.,\forall u. \neg \mem(\nu,u)$). These uninterpreted predicates are encoded as uninterpreted functions with stratified sorts.  In order to ensure the query is an EPR sentence, we require that a normal refinement type (overapproximate types) can only use universal quantifiers. In addition, as shown in \autoref{fig:syntax}, we disallow nested uninterpreted predicate application (e.g., $\mem(\nu,\mem(\nu,u))$) and can only apply a uninterpreted predicate over constants $\mem(\nu,3)$ (it can be encoded as $\forall u. u = 3 \impl \mem(\nu,u)$).
  \section{Proofs}
\paragraph{Type Soundness}
The Rocq formalization of our core language, typing rules and the
proof of \autoref{theorem:type-sound} is publicly available on
Zenodo\cite{artifact}.

\paragraph{Soundness of Algorithmic Typing} We present the proof for \autoref{theorem:algo-sound} from \autoref{sec:algorithm}. The proof requires the following lemmas about the $\subquery$ subroutine, $\existsop$ and $\disjop$ functions.

\begin{lemma}\label{theorem:query}[Soundness of $\subquery$ subroutine] For all type context
  $\Gamma$ and coverage type $\urt{b}{\phi_1}$ and $\urt{b}{\phi_2}$, $\subquery(\Gamma, \urt{b}{\phi_1}, \urt{b}{\phi_2})$ implies $\Gamma \vdash \urt{b}{\phi_1} <: \urt{b}{\phi_2}$.
\end{lemma}

\begin{lemma}[The $\disjop$ Function implies disjunction judgement]\label{lemma:disj} For all type context
  $\Gamma$, type $\tau_1$ and $\tau_2$, $\Gamma \vdash \tau_1 \lor \tau_2 = \disjop(\tau_1, \tau_2)$.
\end{lemma}

\begin{lemma}[The $\existsop$ Function implies type judgement transformation]\label{lemma:eq-trans} For all type context
  $\Gamma$, $\Gamma'$, term $e$, and type $\tau$,
\begin{align*}
    \Gamma, \Gamma' \vdash e : \tau \implies \Gamma, \Gamma' \vdash e : \existsop(\Gamma',~\tau) \land \Gamma \vdash \existsop(\Gamma',~\tau)
\end{align*}
\end{lemma}



We also lift the subtyping relation to type contexts.
\begin{definition}[Subtyping relation over Type Contexts]\label{def:type-ctx-order} As in the subtyping relation between types, the subtyping relation between two type context $\Gamma_1 \sqsubseteq \Gamma_2$ means that if a term have type $\tau$ under one context, it should also have the same type in the second context.
\begin{align*}
    \Gamma_1 \sqsubseteq \Gamma_2 \doteq \forall \tau, \forall e, e \in \denotation{\tau}_{\Gamma_1} \implies e \in \denotation{\tau}_{\Gamma_2}
\end{align*}
\end{definition}

\begin{lemma}\label{theorem:subctx-trans}[Sub Type Context Implies Type Judgement Transformation] For two type context $\Gamma_1 \sqsubseteq \Gamma_2$, term $e$ and coverage type $\tau$,
\begin{align*}
    \Gamma_1 \vdash e : \tau \implies \Gamma_2 \vdash e : \tau
\end{align*}
\end{lemma}

Intuitively, modifying a type binding in a type context is equivalent to applying the subsumption rule before we introduce this binding into the type context. This subtype context relation allows us to prove the correctness of the typing algorithm, which lazily strengthens the types in the type context by need.

Now we can prove the soundness theorem of our typing algorithm with respect to our declarative type system. As the type synthesis rules are defined mutually recursively, we simultaneously prove both are correct:
\begin{theorem}\label{theorem:syn-check}[Soundness of the type synthesis and type check algorithm] For all type context
  $\Gamma$, term $e$ and coverage type $\tau$,
\begin{align*}
  &\Gamma \vdash e \typeinfer \tau \implies \Gamma \vdash e: \tau \\
  &\Gamma \vdash e \typecheck \tau \implies \Gamma \vdash e: \tau
\end{align*}
\end{theorem}
\begin{proof} We proceed by induction of the mutual recursive structure of $\Gamma \vdash e \typeinfer \tau$ and $\Gamma \vdash e \typecheck \tau$. In the cases for; synthesis and checking rules of rule \textsc{SynConst}, \textsc{SynOp}, \textsc{SynErr}, \textsc{SynVarBase}, \textsc{SynVarFun}, \textsc{ChkSub},\textsc{ChkFun}, and \textsc{ChkFix}, the coverage typing rules in \autoref{fig:type-rules} aligns exactly with these rules, thus the soundness is immediate in these cases.

In addition, the rule \textsc{SynAppOp} is similar to \textsc{SynAppBase}, but
has multiple arguments; the rule \textsc{SynLetE} is the same as
\textsc{SynAppFun} but has no application; the rule \textsc{SynMatch} is similar
with \textsc{ChkMatch}, thus we discuss one rule in each of these pairs while
the second follows in a similar fashion. Consequently, there are three
interesting cases, corresponding to the rules shown in
\autoref{fig:bi-type-rules}.

\begin{enumerate}[label=Case]
    \item \textsc{SynAppFun}: This rule can be treated as a combination of \textsc{TAppFun} and \textsc{TEq}. From the induction hypothesis and the precondition of \textsc{SynAppFun}, we know
    \begin{align*}
        &\Gamma \vdash v_1 : (a{:}\tau_a\sarr \tau_b)\sarr \tau_x
        &\text{since}\quad \Gamma \vdash v_1 \typeinfer (a{:}\tau_a\sarr \tau_b)\sarr \tau_x \\
        &\Gamma \vdash v_2 : a{:}\tau_a\sarr \tau_b
        &\text{since}\quad \Gamma \vdash v_2 \typecheck a{:}\tau_a\sarr \tau_b \\
        &\Gamma, x{:}\tau_x \vdash e : \tau
        &\text{since}\quad \Gamma, x{:}\tau_x \vdash e \typeinfer \tau
    \end{align*}
    For the $\tau' = \existsop(x{:}\tau_x,~\tau)$, according to Lemma \ref{lemma:eq-trans}, we know
    \begin{align*}
        \Gamma, x{:}\tau_x \vdash e : \tau' \land \Gamma \vdash \tau'
    \end{align*}
    Using the above conclusions, Since all the preconditions of \textsc{TAppFun} hold, applying the rule \textsc{TAppFun}, we have $\Gamma \vdash \zlet{x}{v_1\ v_2}{e} : \tau'$.
    \item \textsc{SynAppBase}: Notice that the value $v_2$ has the base type $t$, and can only be a constant or a variable, doing a case split on this:
    \begin{enumerate}
        \item If $v_2$ is a constant $c_2$, notice that $\phi[\nu \mapsto c_2]$ has to be true, otherwise the binding $a{:}\urt{b}{\nu = c_2 \land \phi}$ has the bottom type, and the type context that contains it is not well formed. Thus, using the well-formedness of the context, it follows that
        \begin{align*}
            \nu = c_2 \land \phi \equiv \nu = c_2
        \end{align*}
        Thus, again using the Induction Hypothesis on the antecedents of the rule we have:
        \begin{align*}
        &\Gamma \vdash v_1 : a{:}\ort{b}{\nu = c_2 \land \phi}\sarr\tau_x[a\mapsto c_2] &
        \\&\qquad \text{since } \Gamma \vdash v_1 \typeinfer a{:}\ort{b}{\nu = c_2 \land \phi}\sarr\tau_x \text{ and \textsc{TSub}}  \\
        &\Gamma \vdash c_2 : \urt{b}{\nu = c_2 \land \phi}
        \\&\qquad\text{ since \textsc{TConst} and } \nu = c_2 \equiv \nu = c_2 \land \phi \\
        &\Gamma, a{:}\urt{b}{\nu = c_2 \land \phi}, x{:}\tau_x[a\mapsto c_2]  \vdash e : \tau[a \mapsto c_2]
        \\&\qquad\text{since } \Gamma, a{:}\urt{b}{\nu = c_2 \land \phi}, x{:}\tau_x \vdash e \typeinfer \tau
    \end{align*}
    Since the variable $a$ is not free in the type judgment, we can remove it from the type context
    \begin{align*}
        &\Gamma, x{:}\tau_x[a\mapsto c_2]  \vdash e : \tau[a \mapsto c_2]
    \end{align*}
    According to the Lemma~\ref{lemma:eq-trans}, we know that
    \begin{align*}
        &\Gamma, x{:}\tau_x[a\mapsto c_2]  \vdash e : \existsop(x{:}\tau_x[a\mapsto c_2], \tau[a \mapsto c_2])
    \end{align*}
    The type $\existsop(x{:}\tau_x[a\mapsto c_2], \tau) $ is well formed under the type context $\Gamma$, and all preconditions of the rule \textsc{TApp} are satisfied, so we can conclude
    \begin{align*}
        &\Gamma \vdash e : \zlet{x}{v_1\ c_2}{e} : \existsop(x{:}\tau_x[a\mapsto c_2], \tau[a \mapsto c_2])
    \end{align*}
    Notice that, $\phi[\nu \mapsto c_2]$ is true, thus we have
    \begin{align*}
        &\existsop(x{:}\tau_x[a\mapsto c_2], \tau[a \mapsto c_2])\\
        \equiv\ &\existsop(a{:}\urt{b}{\nu = c_2}, x{:}\tau_x[a\mapsto a], \tau[a \mapsto a])\\
        \equiv\ &\existsop(a{:}\urt{b}{\nu = c_2 \land \phi}, x{:}\tau_x, \tau) \\
        \equiv\ &\tau'
    \end{align*}
    Thus, we can conclude $\Gamma \vdash e : \zlet{x}{v_1\ c_2}{e} : \tau'$.
    \item If $v_2$ is a variable $x_2$, we first construct a subcontext of $\Gamma$ where we modify the type of $x_2$ in the type context $\Gamma$. Since the variable $x_2$ has a type in the context $\Gamma$, then\footnote{We use the same way when the variable $x_2$ having a normal refinement type $\nuot{t_2}{\phi_2}$, thus we omitted this situation.}
    \begin{align*}
        \Gamma \equiv \Gamma_1, x_2{:}\nuut{t_2}{\phi_2}, \Gamma_2
    \end{align*}
    we build a type context $\Gamma^{*}$
    \begin{align*}
        \Gamma \equiv \Gamma_1, x_2{:}\nuut{t_2}{\phi_2 \land \phi}, \Gamma_2
    \end{align*}
    Intuitively, this new context gives us an assumption similar to the constant case above:
    \begin{align*}
        \nu = x_2 \land \phi \iff \nu = x_2
    \end{align*}
    In fact, the new context $\Gamma^{*}$ implies two subtyping relations over the context:
\begin{align*}
    \Gamma^{*} &\sqsubseteq \Gamma \\
    \Gamma, a{:}\urt{b}{\nu = x_2 \land \phi} &\sqsubseteq \Gamma^{*}, a{:}\urt{b}{\nu = x_2 \land \phi}
\end{align*}
    The first is obvious, since we only add a conjunction into the type of $x_2$. On the other hand, $\Gamma, a{:}\urt{b}{\nu = x_2 \land \phi}$ is a subtype of $\Gamma^{*}, a{:}\urt{b}{\nu = x_2 \land \phi}$ in reverse, since we strengthen the coverage type of $x_2$ in the last binding $a{:}\urt{b}{\nu = x_2 \land \phi}$.
    Then, according to the second subtype context relation, we have
    \begin{align*}
        &\Gamma^{*} \vdash v_1 : a{:}\ort{b}{\nu = x_2 \land \phi}\sarr\tau_x
        &\text{since}\quad \Gamma \vdash v_1 \typeinfer a{:}\ort{b}{\phi}\sarr\tau_x \text{ and \textsc{TSub}}
    \end{align*}
    Based on the fact $\nu = x_2 \land \phi \iff \nu = x_2 $, we have
    \begin{align*}
        &\Gamma^{*} \vdash v_2 : \urt{b}{\nu = x_2 \land \phi}
        &\text{According to the rule \textsc{TVar}}
    \end{align*}
    According to the second subtype context relation, we have
    \begin{align*}
        &\Gamma^{*}, a{:}\urt{b}{\nu = x_2 \land \phi}, x{:}\tau_x[a\mapsto x_2]  \vdash e : \tau[a \mapsto x_2]
        \\&\qquad \text{since } \Gamma, a{:}\urt{b}{\nu = x_2 \land \phi}, x{:}\tau_x \vdash e \typeinfer \tau
    \end{align*}
    Again, since the variable $a$ is not free, we can also remove it. Moreover, according to the typing rule \textsc{TApp} and the Lemma~\ref{lemma:eq-trans}, we know
    \begin{align*}
        &\Gamma^{*} \vdash \zlet{x}{v_1\ x_2}{e} : \existsop(x{:}\tau_x[a \mapsto x_2], \tau[a \mapsto x_2])
    \end{align*}
    Again, we have
    \begin{align*}
        &\existsop(x{:}\tau_x[a\mapsto x_2], \tau[a \mapsto x_2])\\
        \equiv\ &\existsop(a{:}\urt{b}{\nu = x_2}, x{:}\tau_x[a\mapsto a], \tau[a \mapsto a])\\
        \equiv\ &\existsop(a{:}\urt{b}{\nu = x_2 \land \phi}, x{:}\tau_x, \tau)\\
        \equiv\ &\tau'
    \end{align*}
    Then we have
    \begin{align*}
        &\Gamma^{*} \vdash \zlet{x}{v_1\ x_2}{e} :  \tau'
    \end{align*}
    Finally, by combining Lemma~\ref{theorem:subctx-trans} and $\Gamma^{*} \sqsubseteq \Gamma$, we have
    \begin{align*}
        &\Gamma \vdash \zlet{x}{v_1\ x_2}{e} : \tau'
    \end{align*}
        \end{enumerate}
    \item \textsc{ChkMatch}: The rule is a combination of \textsc{TMatch} and \textsc{TMerge}.
    For the $i^{th}$ branch of the pattern matching branch, we have the following judgment after unfolding $\Gamma_i'$
    \begin{align*}
        &\Gamma,  \overline{{y}{:}\urt{b_y}{\theta_y}}, a{:}\urt{b}{\nu = v_a \land \psi_i} \vdash e_i : \tau_i &\text{ since $\Gamma,  \Gamma_i'\vdash e_i \typeinfer \tau_i$}
    \end{align*}
    Similarly to the approach we used for the \textsc{SynAppBase} case, since $v_a$ is a value of base type, it can only be a constant or a variable. Then we can derive the following judgement without the variable $a$:
    \begin{align*}
        \Gamma,  \overline{{y}{:}\urt{b_y}{\theta_y}} \vdash e_i : \existsop(\overline{{y}{:}\urt{b_y}{\theta_y}}, \tau_i[a\mapsto v_a]) \equiv \tau_i'
    \end{align*}
    According to the rule \textsc{TMatch}, we have the following judgement for all branches
    \begin{align*}
        \Gamma \vdash \match{v_a} \overline{d_i\ \overline{y} \to e_i} : \tau_i'
    \end{align*}
    Then according to the Lemma~\ref{lemma:disj}, we have
    \begin{align*}
        \Gamma \vdash \match{v_a} \overline{d_i\ \overline{y} \to e_i} : \disjop(\overline{\tau_i'})
    \end{align*}
    Finally, according to \textsc{TSub}, for a type $\tau'$ that $\Gamma \vdash \disjop(\overline{\tau_i'}) <: \tau'$, we have
        \begin{align*}
        \Gamma \vdash \match{v_a} \overline{d_i\ \overline{y} \to e_i} : \tau'
    \end{align*}
    which is exactly what we needed to prove for this case.
\end{enumerate}
\end{proof}

\paragraph{Completeness of Algorithmic Typing} We present the proof for \autoref{theorem:algo-complete} from \autoref{sec:algorithm}. The theorem assumes``an oracle for all formulas produced by the Query subroutine'', which can be stated as the following lemma.

\begin{lemma}\label{theorem:query-complete}[An oracle of $\subquery$ subroutine exists] For all type context
  $\Gamma$ and coverage type $\urt{b}{\phi_1}$ and $\urt{b}{\phi_2}$, $\Gamma \vdash \urt{b}{\phi_1} <: \urt{b}{\phi_2}$ iff $\subquery(\Gamma, \urt{b}{\phi_1}, \urt{b}{\phi_2})$.
\end{lemma}

With the assumption above, we introduce the following lemmas about the $\subquery$ subroutine, $\disjop$ and $\existsop$ functions as we did in the soundness proof.

\begin{lemma}[$\subquery$ subroutine implies propositional equality]\label{lemma:subquery-eq} For all type context
  $\Gamma$, type $\urt{b}{\phi_1}$, $\urt{b}{\phi_2}$
  \begin{align*}
      \subquery(\Gamma, \urt{b}{\phi_1}, \urt{b}{\phi_2}) \land \subquery(\Gamma, \urt{b}{\phi_2}, \urt{b}{\phi_1}) \implies \phi_1 = \phi_2
  \end{align*}
\end{lemma}

\begin{lemma}[Disjunction judgement can be simulated by the $\disjop$ Function]\label{lemma:disj-complete} For all type context
  $\Gamma$, type $\tau_1$, $\tau_2$, $\tau_3$, $\Gamma \vdash \tau_1 \lor \tau_2 = \tau_3 \implies \Gamma \vdash \tau_3 <: \disjop(\tau_1, \tau_2) \land \Gamma \vdash \disjop(\tau_1, \tau_2) <: \tau_3$.
\end{lemma}

\begin{lemma}[$\existsop$ Function is identical when well-fromed]\label{lemma:existsop-wf} For all type context $\Gamma$, $\Gamma'$ and type $\tau$, $\Gamma \vdash \tau \implies \forall e, \Gamma, \Gamma' \vdash e \typeinfer \tau \iff  \Gamma, \Gamma' \vdash e \typeinfer \existsop(\Gamma', \tau)$.
\end{lemma}

We also have the corresponding lemma about the subtyping judgement.

\begin{lemma}[Subtyping judgement iff the $\existsop$ Function]\label{lemma:existsop-complete} For all type context $\Gamma$, and type $\tau_1$, $\tau_2$, $\Gamma \vdash \tau_1 <: \tau_2 \iff \emptyset \vdash \existsop( \Gamma, \tau_1) <: \existsop (\Gamma, \tau_2)$.
\end{lemma}

Now we can prove the completeness theorem of our typing algorithm with respect to our declarative type system.
\begin{theorem} [Relative completeness of typing algorithm] For all type context
  $\Gamma$, term $e$ and coverage type $\tau$, $\Gamma \vdash e: \tau \implies \Gamma \vdash e \typecheck \tau$.
\end{theorem}
\begin{proof} We proceed by induction of $\Gamma \vdash e : \tau$. In the cases for typing rules of rule \textsc{TErr}, \textsc{TConst}, \textsc{TOp}, \textsc{TVarBase}, \textsc{TVarFun}, \textsc{TFun}, and \textsc{TFix}, the coverage typing synthesis rules in \autoref{fig:bi-type-rules} aligns exactly with these rules. By applying the rule \textsc{ChkSub} to shift from the typing synthesis judgement to the typing check judgement, the completeness is immediate in these cases. For the same reason, in the case for the rule \textsc{TSub}, the completeness also holds.

\begin{enumerate}[label=Case]
\item \textsc{TLetE}, \textsc{TAppOp}, \textsc{TAppFun}, \textsc{TApp}, \textsc{TMatch}: The coverage typing synthesis rules in \autoref{fig:bi-type-rules} aligns similar rules (\textsc{SynLetE}, \textsc{SynAppOp}, \textsc{SynAppFun}, \textsc{SynAppBase}, \textsc{SynMatch}) in these cases, which synthesis the type $\existsop(\Gamma', \tau)$ instead of $\tau$. This difference can be fixed by the Lemma~\ref{lemma:existsop-wf} and the precondition that $\tau$ is well formed under type context $\Gamma$.
\item \textsc{TEq}:
The key idea of this case is to use the auxiliary term $\zlet{x}{e}{x}$ and the rule \textsc{SynLetE} to simulate the type judgment transformation. Notice that the auxiliary term $\zlet{x}{e}{x}$ is equivalent to $e$ with respect to the operational semantics, that is,
\begin{align*}
    \forall v, e \hookrightarrow^* v \iff \zlet{x}{e}{x} \hookrightarrow^* v
\end{align*}
It also implies that
\begin{align*}
    \forall \Gamma\ e\ \tau, e \in \denotation{e}_{\Gamma} \iff \denotation{\zlet{x}{e}{x}}_{\Gamma}
\end{align*}
Thus, the goal of this case can be simplified as
\begin{align*}
    \forall \Gamma\ e\ \tau_1\ \tau_2, \Gamma \vdash e \typeinfer \tau_1 \land \Gamma \vdash \tau_1 <: \tau_2 \land \Gamma \vdash \tau_2 <: \tau_1 \implies \Gamma \vdash \zlet{x}{e}{x} \typeinfer \tau_2
\end{align*}
According to the Lemma~\ref{lemma:existsop-complete}, we know
\begin{align*}
    \emptyset \vdash \existsop( \Gamma, \tau_1) <: \existsop (\Gamma, \tau_2) \land \emptyset \vdash \existsop( \Gamma, \tau_1) <: \existsop (\Gamma, \tau_2)
\end{align*}
According to the Lemma~\ref{theorem:query-complete} and Lemma~\ref{lemma:subquery-eq}, we know $\existsop( \Gamma, \tau_1) = \existsop( \Gamma, \tau_2)$.
On the other hand, according to the rule \textsc{SynVarBase} (or, \textsc{SynVarFun}) and the rule \textsc{SynLetE}, we can infer the type of the auxiliary term $\zlet{x}{e}{x}$ as $\existsop( \Gamma, \tau_1)$, thus we know
\begin{align*}
    \Gamma \vdash \zlet{x}{e}{x} \typeinfer \existsop(\Gamma, \tau_2)
\end{align*}
The according to Lemma~\ref{lemma:existsop-complete} and the type $\existsop(\Gamma, \tau_2)$ has no free variable, we know
\begin{align*}
    \Gamma \vdash \zlet{x}{e}{x} \typeinfer \tau_2
\end{align*}
which is exactly what we needed to prove for this case.
\item \textsc{TMerge}: Similarly to the approach we used in the case \textsc{TEq}, the key idea is to use the auxiliary term $e \oplus e$ (non-deterministic choice between two $e$) and the rule \textsc{SynMatch} to simulate the typing rule \textsc{TMerge}.  Notice that the auxiliary term $e \oplus e$ is equivalent to $e$ with respect to the operational semantics, which implies that
\begin{align*}
    \forall \Gamma\ e\ \tau, e \in \denotation{e}_{\Gamma} \iff \denotation{e \oplus e}_{\Gamma}
\end{align*}
Thus, the goal of this case can be simplified as
\begin{align*}
    \forall \Gamma\ e\ \tau_1\ \tau_2\ \tau_3, \Gamma \vdash e \typeinfer \tau_1 \land \vdash e \typeinfer \tau_2 \land \Gamma \vdash \tau_1 \lor \tau_2 = \tau_3 \implies \Gamma \vdash e \oplus e \typeinfer \tau_3
\end{align*}
With the rule \textsc{SynMatch}, we can infer the type of the term $e \oplus e$ as $\disjop(\tau_1, \tau_2)$. On the other hand, according to the Lemma~\ref{lemma:disj-complete}, we know
\begin{align*}
    \Gamma \vdash \tau_3 <: \disjop(\tau_1, \tau_2) \land \Gamma \vdash \disjop(\tau_1, \tau_2) <: \tau_3
\end{align*}
Finally, it falls back to the same situation of the case \textsc{TEq}, obviously can be proved in the same way.
\end{enumerate}
\end{proof}

\paragraph{Soundness of Extended Algorithmic Typing} We present the proof for \autoref{theorem:extended-algo-sound} from \autoref{sec:poly}. Besides the lemmas about the $\subquery$ subroutine, $\existsop$ and $\disjop$ functions which are required by the proof of \autoref{theorem:algo-sound}, this proof additionally requires the following lemma about the $\instop$ subroutine.

\begin{lemma}\label{theorem:inst}[Soundness of $\instop$ subroutine] For all type context
  $\Gamma$ and coverage type $\tau_1$, $\tau_2$ and $\tau_3$, $\instop(\Gamma, \tau_1, \tau_2) = \tau_3$ implies $\Gamma \vdash \tau_1 <: \tau_3$.
\end{lemma}

Similar with the proof for \autoref{theorem:extended-algo-sound}, we prove the soundness theorem of our extended typing algorithm with respect to our declarative type system. As the type synthesis rules are defined mutually recursively, we simultaneously prove both are correct:
\begin{theorem}\label{theorem:disjop}[Soundness of the type synthesis and type check algorithm] For all type context
  $\Gamma$, term $e$ and coverage type $\tau$,
\begin{align*}
  &\Gamma \vdash e \typeinfer \tau \implies \Gamma \vdash e: \tau \\
  &\Gamma \vdash e \typecheck \tau \implies \Gamma \vdash e: \tau
\end{align*}
\end{theorem}
\begin{proof} We proceed by induction of the mutual recursive structure of $\Gamma \vdash e \typeinfer \tau$ and $\Gamma \vdash e \typecheck \tau$ and focus on the extended rules shown in \autoref{fig:extended-bi-type-rules}. In addition, the rule \textsc{ChkPolyType} and \textsc{ChkPolyPred} aligns exactly to declarative tying rule \textsc{TPolyType} and \textsc{TPolyPred} shown in \autoref{fig:extended-typing}, thus the soundness is immediate in thse cases. Consequently, we now provide the proof of two modified rules, \textsc{SynAppFun} and \textsc{SynAppBase}.

\begin{enumerate}[label=Case]
    \item \textsc{SynAppFun}: This rule can be treated as a combination of \textsc{TAppFun} and \textsc{TEq}. From the induction hypothesis and the precondition of \textsc{SynAppFun}, we know
    \begin{align*}
        &\Gamma \vdash v_1 : \tau_1
        &\text{since}\quad \Gamma \vdash v_1 \typeinfer \tau_1 \\
        &\Gamma \vdash v_1 : a{:}\tau_2'\sarr\tau_x
        &\text{since}\quad \instop(\Gamma, \tau_1, \tau_2) = a{:}\tau_2'\sarr\tau_x \text{ and Lemma~\ref{theorem:inst}}\\
        &\Gamma \vdash v_1 : a{:}\tau_2\sarr \tau_x
        &\text{since}\quad \Gamma \vdash \tau_2 <: \tau_2' \text{ and \textsc{TSub}}\\
        &\Gamma \vdash v_2 : a{:}\tau_2
        &\text{since}\quad \Gamma \vdash v_2 \typeinfer a{:}\tau_2 \\
        &\Gamma, x{:}\tau_x \vdash e : \tau
        &\text{since}\quad \Gamma, x{:}\tau_x \vdash e \typeinfer \tau
    \end{align*}
    For the $\tau' = \existsop(x{:}\tau_x,~\tau)$, according to Lemma \ref{lemma:eq-trans}, we know
    \begin{align*}
        \Gamma, x{:}\tau_x \vdash e : \tau' \land \Gamma \vdash \tau'
    \end{align*}
    Using the above conclusions, Since all the preconditions of \textsc{TAppFun} hold, applying the rule \textsc{TAppFun}, we have $\Gamma \vdash \zlet{x}{v_1\ v_2}{e} : \tau'$.
    \item \textsc{SynAppBase}: From the induction hypothesis and the precondition of \textsc{SynAppBase}, we know
    \begin{align*}
        &\Gamma \vdash v_1 : \tau_1
        \\&\qquad\text{since}\quad \Gamma \vdash v_1 \typeinfer \tau_1 \\
        &\Gamma \vdash v_1 : a{:}\ort{b}{\phi}\sarr\tau_x
        \\&\qquad\text{since}\quad \instop(\Gamma, \tau_1, \urt{b}{\phi_2}) = a{:}\ort{b}{\phi}\sarr\tau_x \text{ and Lemma~\ref{theorem:inst}}
    \end{align*}\noindent
    which leads the same precondition of original rule of \textsc{SynAppBase} as shown in \autoref{fig:bi-type-rules}, thus the proof is the same as in the proof of \autoref{theorem:algo-sound}.
\end{enumerate}
\end{proof}

\paragraph{Completeness of Extended Algorithmic Typing} We present the proof for \autoref{theorem:extended-algo-complete} from \autoref{sec:poly}. The theorem besides assumes``an oracle for all formulas produced by the Query subroutine'', also assumes ``$\instop$ subroutine always succeeds`` which can be stated as the following lemma.

\begin{lemma}\label{theorem:instop-complete}[$\instop$ subroutine always succeeds] For all type context $\Gamma$ and coverage type $\tau_1$ and $\tau_2$, if there exists a type $x{:}\tau_2'\sarr\tau_\Code{ret}$ such that
$\Gamma \vdash a{:}\tau_2\sarr\tau_\Code{ret} <: \tau_1$ and $\tau_2' <: \tau_2'$, then $\instop(\Gamma, \tau_1, \tau_2) = x{:}\tau_2'\sarr\tau_\Code{ret}$.
\end{lemma}

Now we can prove the completeness theorem of our typing algorithm with respect to our declarative type system.
\begin{theorem} [Relative completeness of typing algorithm] For all type context
  $\Gamma$, term $e$ and coverage type $\tau$, $\Gamma \vdash e: \tau \implies \Gamma \vdash e \typecheck \tau$.
\end{theorem}
\begin{proof} We proceed by induction of $\Gamma \vdash e : \tau$ and focus on the extended rules shown in \autoref{fig:extended-bi-type-rules}. In addition, the rule \textsc{TPolyType} and \textsc{TPolyPred} aligns exactly to rule \textsc{ChkPolyType} and \textsc{ChkPolyPred}, thus the completeness also holds. The remaining proof for the modified rules \textsc{ChkAppFun} and \textsc{ChkAppBase} can be simplified to showing that all terms type-checkable under the original rules remain type-checkable under the modified versions.
\begin{enumerate}[label=Case]
    \item \textsc{ChkAppFun}: According to \autoref{theorem:algo-complete}, we just needs to show the precondition of original rule \textsc{ChkAppFun} can implies the precondition of modified rule \textsc{ChkAppFun}. We know
    \begin{align*}
        &\Gamma \vdash v_1 \typeinfer \tau_2\sarr\tau_x
        &\text{precondition of original \textsc{ChkAppFun}}\\
        &\instop(\Gamma, \tau_2\sarr\tau_x, \tau_2) = \tau_2\sarr\tau_x
        &\text{since Lemma~\ref{theorem:instop-complete}}\\
        &\Gamma \vdash v_2 \typecheck \tau_2
        &\text{precondition of original \textsc{ChkAppFun}}\\
        &\Gamma \vdash v_2 \typeinfer \tau_2
        &\text{since \textsc{CheckSub}}\\
        &\Gamma \vdash \tau_2 <: \tau_2
        &\text{since soundness of subtyping}
    \end{align*}\noindent
    Then all preconditions of modified rule \textsc{ChkAppFun} hold, then the completeness also holds.
    \item \textsc{ChkAppBase}: According to \autoref{theorem:algo-complete}, we just needs to show the precondition of original rule \textsc{ChkAppBase} can implies the precondition of modified rule \textsc{ChkAppBase}. We know
    \begin{align*}
        &\Gamma \vdash v_1 \typeinfer a{:}\ort{b}{\phi}\sarr\tau_x
        \\&\qquad\text{precondition of original \textsc{ChkAppBase}}\\
        &\Gamma \vdash v_2 \typeinfer \urt{b}{\vnu = v_2}
        \\&\qquad\text{since rule \textsc{SynVarBase}}\\
        &\instop(\Gamma, a{:}\ort{b}{\phi}\sarr\tau_x, \urt{b}{\vnu = v_2}) = a{:}\ort{b}{\phi \land \vnu = v_2}\sarr\tau_x
    \end{align*}\noindent
    Notice that $a{:}\ort{b}{\phi \land \vnu = v_2}\sarr\tau_x$ is equal to $a{:}\ort{b}{\phi \land \vnu = v_2 \land \vnu = v_2}\sarr\tau_x$, then all preconditions of modified rule \textsc{ChkAppFun} hold, then the completeness also holds.
\end{enumerate}
\end{proof}

\else
\fi


\label{lastpage01}

\end{document}